%% file: rsub-v3.tex
\documentclass[%
reprint, 
 amsmath,amssymb,
 aps,
]{revtex4-1}

\usepackage[dvipdfmx]{graphicx}
\usepackage{dcolumn}
\usepackage{bm}
\usepackage{color}

\newcommand{\trento}{T\raisebox{-.5ex}{R}ENTo}
\makeatletter
\newcommand{\colorcaption}[2][]{%
  \begingroup%
  \renewcommand{\@caption@fignum@sep}{ (color online).  }%
  \caption[#1]{#2}%
  \endgroup%
}
\makeatother

\begin{document}


\title{Temperature dependence of transport coefficients of QCD 
\\in high-energy heavy-ion collisions}

\author{Kazuhisa Okamoto$^1$}
\email{okamoto@hken.phys.nagoya-u.ac.jp}
\author{Chiho Nonaka$^{1,2}$}%
 \email{nonaka@hken.phys.nagoya-u.ac.jp}
\affiliation{%
$^1$Department of Physics, Nagoya University, Nagoya 464-8602, Japan \\
$^2$Kobayashi Maskawa Institute, Nagoya University, Nagoya 464-8602, Japan
 }%

\date{\today}

\begin{abstract}
Using our developed new relativistic viscous hydrodynamics code, we investigate the temperature 
dependence of shear and bulk viscosities from comparison with the ALICE data: single particle spectra and 
collective flows of Pb+Pb $\sqrt{s_{\rm NN}}=2.76$ TeV collisions at the Large Hadron Collider.
We find that from the comprehensive analyses of centrality dependence of 
single particle spectra and collective flows, we can extract detailed information on 
the quark-gluon plasma bulk property, without the information being smeared by the final state interactions. 
\end{abstract}

\pacs{24.10.Nz, 25.75.-q, 47.75.+f, 12.38.Mh}
\maketitle

\section{Introduction}
Since the success of production of strongly interacting quark-gluon plasma (QGP) at 
the Relativistic Heavy Ion Collider (RHIC) \cite{QGP-RHIC},  
a relativistic viscous hydrodynamic model has been widely used for the description of space-time evolution of 
the hot and dense matter created after collisions. 
Now at RHIC as well as at the Large Hadron Collider (LHC) high-energy heavy-ion collisions are 
performed and many experimental data are reported. 
Because the relativistic viscous hydrodynamic equation has close a relation to an 
equation of state (EoS) and transport coefficients of the QCD matter, 
analyses of experimental data at RHIC and the LHC based on a relativistic viscous hydrodynamic 
model can provide an insight into the detailed information on the QGP bulk property. 

The recent development of a lattice QCD calculation for the EoS at vanishing chemical 
potential is remarkable. Two groups, the 
Wuppertal-Budapest and HotQCD Collaborations, report almost the same (pseudo)critical 
temperatures, $T_c=155 \pm 6$ MeV \cite{Borsanyi:2010cj} and 
$T_c=154 \pm 9$ MeV \cite{Bazavov:2011nk},  respectively. 
On the other hand, the evaluation of the shear viscosity to entropy density ratio $\eta/s$ 
of the hadronic phase and the 
QGP phase is investigated based on the Boltzmann equation \cite{Prakash:1993bt, Arnold:2003zc}. 
Due to the existence of the Kovtun-Son-Starinets (KSS) bound, the lower bound of $\eta/s$ 
\cite{Kovtun:2004de}, $\eta/s$  takes the minimum around the critical 
temperature \cite{Csernai:2006zz}. 
On the other hand, the behavior of temperature dependence of the bulk viscosity to entropy 
density ratio is not clear.  
For example, $\zeta/s$ of a massive pion gas decreases with temperature \cite{FernandezFraile:2008vu, Lu:2011df} below the 
critical temperature. Also, above the critical temperature, $\zeta/s$  of quark-gluon matter  \cite{Karsch:2007jc} or gluon plasma \cite{Chen:2011km} decreases with temperature.
There is not a conclusive understanding of quantitative information on the transport 
coefficients of QCD matter.  

Therefore phenomenological analyses of  transport coefficients from comparison with experimental data 
at RHIC and the LHC are indispensable \cite{Romatschke:2007mq, Luzum:2008cw, Dusling:2009df, Shen:2010uy, Song:2010mg, Gale:2012rq, Xu:2016hmp, Weller:2017tsr}.  
It turns out that the value of $\eta/s$ at the LHC is larger than that  at RHIC, which 
suggests a temperature-dependent $\eta/s$. 
At RHIC elliptic flow $v_2$ is sensitive to the $\eta/s$ of the hadronic phase, whereas at the LHC 
it depends on $\eta/s$ of both the hadronic phase and the QGP phase \cite{Niemi:2011ix, Niemi:2012ry, Molnar:2014zha, Niemi:2015qia}. 
Simultaneous analyses of $v_2$ at RHIC and the LHC give a constraint on the temperature 
dependence of $\eta/s$  \cite{Niemi:2011ix, Niemi:2012ry, Molnar:2014zha, Niemi:2015qia}. 
Even at one collision energy, we can explore 
the temperature dependence of shear and bulk viscosities from 
centrality and/or rapidity dependence of observables.  
At peripheral collisions the effect of $\eta/s$ of the hadronic phase is dominant,  
compared with that of the QGP phase \cite{Niemi:2012ry, Niemi:2015qia}. 
At the forward rapidity where the temperature becomes small, the behavior of 
$\eta/s$ in the hadronic phase affects the elliptic flow \cite{Molnar:2014zha, Denicol:2015nhu}. 

Now not only the shear viscosity but also the bulk viscosity
are included in relativistic viscous hydrodynamic simulations \cite{Denicol:2009am, Monnai:2009ad, Song:2009rh, Bozek:2009dw, Bozek:2012qs, Noronha-Hostler:2013gga, Noronha-Hostler:2014dqa, Rose:2014fba,Ryu:2015vwa, Bernhard:2016tnd, Ryu:2017qzn}.
Generally bulk viscosity reduces the growth of radial flow in hydrodynamic expansion. 
In computations with IP-Glasma initial condition, finite bulk viscosity 
is important for explanation of experimental data, for example, mean $p_T$ 
\cite{Rose:2014fba, Ryu:2015vwa, Ryu:2017qzn}. 
However, the evaluation of the effect of bulk viscosity in the calculation of particle distribution 
in the Cooper-Frye formula is not fixed yet. 
Furthermore, from the application of Bayesian analyses to a model-to-data comparison, 
the temperature dependences of shear and bulk viscosities are investigated \cite{Bernhard:2016tnd}. 
The results support that $\eta/s$ takes the minimum value around the critical temperature and increases 
with temperature in the  QGP phase, and $\zeta/s$ takes the maximum value around the critical temperature.

Furthermore, to achieve the quantitative analyses of the transport coefficients of QCD matter 
from comparison with high statistics and high precision experimental data, 
we  need to perform numerical calculations for relativistic viscous hydrodynamics with high accuracy. 
We have developed a new relativistic viscous hydrodynamics code optimized in the 
Milne coordinates \cite{Okamoto:2017ukz}. The code is constructed based on  a Riemann solver with the two shock approximation \cite{Takamoto:2011wi,Okamoto:2016pbc}. 
It is stable even with small numerical viscosity \cite{Akamatsu:2013wyk}. 

Using our newly developed hydrodynamics code 
we investigate the temperature dependence of shear and bulk viscosities from 
comprehensive analyses of centrality and rapidity dependence of particle distributions and 
higher flow harmonics at the LHC. 

This paper is organized as follows. We begin in Sec. \ref{sec:hydro} by showing the relativistic viscous hydrodynamic equations and the numerical algorithm for solving them briefly. 
In Sec. \ref{sec:model} we explain the phenomenological model: initial conditions, equation of states used in the 
relativistic viscous hydrodynamic equations, 
freezeout process, and final state interactions based on UrQMD.  
We discuss the temperature dependence of shear and bulk viscosities in Sec. \ref{sec:viscous}. 
In Sec. \ref{sec:results}, we show our numerical results of particle 
distributions and collective flows at the LHC.
We end in Sec. \ref{sec:summary} with our conclusions.

\section{Relativistic viscous hydrodynamic equation and algorithm \label{sec:hydro}}
In a hydrodynamic model,  we numerically solve the relativistic viscous hydrodynamic equation 
which is based on the conservation equations,  $T^{\mu\nu}_{\;\;\; ;\mu} =0$, 
where $T^{\mu\nu}$ is the energy-momentum tensor. 
In the Landau frame, the energy-momentum tensor of the viscous fluid is decomposed as 
$T^{\mu\nu} =  eu^\mu u^\nu - (p+ \Pi) \Delta^{\mu\nu} + \pi^{\mu\nu},   \label{eq:tensor2} $
where $\Pi$ is the bulk pressure and $\pi^{\mu\nu}$ is the shear tensor \cite{Landau1987}.   
The relativistic extension of Navier-Stokes theory in a nonrelativistic fluid  usually is not an easy task because of a problem of acausality and instability  \cite{Hiscock:1983zz,Hiscock:1985zz, Pu:2009fj}. 
The problem can be resolved by introducing the second-order terms of the viscous tensor and the derivative of fluid variables into the hydrodynamic equations \cite{Stewart1977,Israel:1979wp}. 
Recently turned out that the original Israel-Stewart theory \cite{Stewart1977,Israel:1979wp} 
does not reproduce the results 
of the kinetic equation quantitatively \cite{Huovinen:2008te,Bouras:2010hm, Takamoto:2010ji, El:2009vj}.  
Here we use the relativistic viscous hydrodynamic equation derived from the 
Boltzmann equation based on the method of moments \cite{Denicol:2012cn,Denicol:2014vaa}. 
The relaxation equations for the bulk viscous pressure $\Pi$ and the shear-stress tensor $\pi^{\mu \nu}$ 
read
\begin{align} \tau_\Pi  \dot{\Pi} + \Pi  &= - \zeta\theta - \delta_{\Pi\Pi} \Pi \theta + \lambda_{\Pi \pi} \pi^{\mu\nu}\sigma_{\mu\nu}, \label{eq:IS1} \\
 \tau_\pi \dot{\pi}^{\langle \mu\nu\rangle } + \pi^{\mu\nu} &= 2\eta\sigma^{\mu\nu} - \delta_{\pi\pi}\pi^{\mu\nu}\theta
 + \varphi_7 \pi_\alpha^{\langle \mu} \pi^{\nu\rangle \alpha} \nonumber \\
&\quad - \tau_{\pi\pi} \pi_\alpha^{\langle \mu}\sigma^{\nu\rangle \alpha} + \lambda_{\pi\Pi} \Pi \sigma^{\mu\nu},  \label{eq:IS2}
\end{align}
where  $\tau_\Pi$ and $\tau_\pi$ are the relaxation times and  $\delta_{\Pi\Pi}$, $\lambda_{\Pi\pi}$, $\delta_{\pi\pi}$, $\varphi_7$, $\tau_{\tau\tau}$, and $\lambda_{\pi\Pi}$ are the transport coefficients. 
To analyze high-energy heavy-ion collisions where strong 
longitudinal expansion exists we perform numerical computation 
in the Milne coordinates \cite{Bjorken:1983}. 
For details, see Refs.~\cite{Okamoto:2016pbc, Okamoto:2017ukz}. 

In our algorithm \cite{Okamoto:2017ukz}, we split the conservation equation 
into two parts, an ideal part and a viscous part using the Strang splitting method \cite{Strang:1968}. 
It is also applied to evaluate the constitutive equations of the viscous tensors Eqs.~\eqref{eq:IS1} and \eqref{eq:IS2}. 
We decompose them into the following three parts, the convection equations, 
the relaxation equations, and the equations with source terms. 
In numerical simulation of relativistic hydrodynamic equation, a time-step size $\Delta \tau$ is usually 
determined by the Courant-Friedrichs-Lewy (CFL) condition. 
However in the relativistic dissipative hydrodynamics,   
one needs to determine the value of $\Delta \tau$ carefully.  
To save computational cost,  we use the piecewise exact solution (PES) method  
\cite{Takamoto:2011wi}, instead of using a simple explicit scheme. 
If, however, the relaxation times are larger than $\Delta\tau$ determined by the CFL condition, 
the PES method is not applied. 
We have checked  the energy and momentum conservation in a one-dimensional expansion of high-energy 
heavy-ion collisions \cite{Okamoto:2016pbc} and the correctness of our code in the following test problems: the viscous Bjorken flow for one-dimensional expansion and the Israel-Stewart theory in Gubser flow regime for the three-dimensional 
calculation \cite{Okamoto:2017ukz}. 

\section{Model \label{sec:model}}
For an initial condition of our hydrodynamic model, we use \trento\ \cite{Moreland:2014oya,Ke:2016jrd}. 
In the parametric model \trento,  the initial entropy density $s(\bm{x},\eta_s)$ is given by a function 
on the transverse plane at midrapidity  $f(\bm{x})$  and a rapidity-dependent function $g(\bm{x},\eta_s)$; 
$s(\bm{x},\eta_s) \propto f(\bm{x})\times g(\bm{x},\eta_s)$. 
The function $f(\bm x)$ is given by 
\begin{align} f(\bm x) \propto \left( \frac{\tilde{T}_A^p + \tilde{T}_B^p}{2} \right)^{1/p},
\end{align}
where $\tilde{T}_A$ is the nucleus thickness function expressed by proton thickness function $T_p$, 
\begin{align} & \tilde{T}_A(\bm x) = \sum^{N_{\rm part}}_{i=1} w_i T_p(\bm x - \bm x_i), \\
 & T_p(\bm x) = \frac{1}{2\pi w^2}{\rm exp} \left( - \frac{\bm x^2}{2w^2}\right).
\end{align}
$w_i$ is random weight for introduction of a negative binomial distribution of produced particles and 
$\omega$ is Gaussian nucleon width. 
A parameter $p$ in the function $f(\bm{x})$ can interpolate among different types of entropy schemes such as the wounded nucleon model ($p=1$), KLN model ($p\sim -0.67$), and IP-Glasma and EKRT ($p \sim 0)$ \cite{Ke:2016jrd}.
Throughout our calculations we fix the parameter $p$ to $p=0$, which is suggested by the Bayesian analyses 
\cite{Ke:2016jrd}. 
The initial distribution in the rapidity direction is described by $g(\bm x, \eta_s)= g(\bm x, y) dy/d\eta_s$, where 
$g(\bm x, y)$ is constructed by the inverse Fourier transform of its cumulant-generating function, 
\begin{eqnarray} g(\bm x, y) &  = & \mathcal{F}^{-1}\left \{ \tilde{g}(\bm x, k)\right \} \nonumber \\
& = & \mathcal{F}^{-1} \left\{ {\rm exp}\left(i\mu k - \frac{1}{2}\sigma^2 k^2-\frac{1}{6}\gamma \sigma^3 k^3  \right) \right\}. 
\end{eqnarray}
The first three cumulants $\mu$, $\sigma$, and $\gamma$ of the local rapidity 
distribution are parametrized by three corresponding coefficients 
$\mu_0$, $\sigma_0$, and $\gamma_0$ which characterize the
rapidity distribution's shift, width, and skewness, respectively. 
$dy/d\eta_s$ is determined by the Jacobian parameter $J$. 
The initial entropy density contains the following parameters other than the parameter $p$: 
normalization $N$, nucleon width $w$, multiplicity fluctuation shape $k$, rapidity distributions' shift $\mu_0$,  
width $\sigma_0$, and skewness $\gamma_0$, and Jacobian parameter $J$. 
We set initial flows and initial values of viscous tensors to be vanishing at $\tau_0=0.6$ fm. 

We use a realistic parametrized EoS \cite{Bluhm:2013yga} 
based on continuum-extrapolated lattice QCD results in 
the physical quark mass limit \cite{Borsanyi:2012vn},   
which is also combined with a hadron resonance gas model 
\cite{Karsch:2003zq,Tawfik:2004sw} at low temperature. 
In the parametrization, the sound velocity takes the minimum value at $T_c\sim167$ MeV \cite{Bluhm:2013yga}. 

At the switching temperature $T_{\rm SW}=150$ MeV the hydrodynamic expansion terminates and 
the UrQMD model starts for the description of space-time evolution \cite{Hirano:2005xf, Nonaka:2006yn}.
We sample produced particles from the fluid, 
using the Cooper-Frye formula \cite{Cooper:1974mv}
\begin{equation}
E\frac{dN_i}{d^3p}= \frac{g_i}{2 \pi} \int _\Sigma f_i(x,p) p^\mu d^3 \sigma_\mu, 
\end{equation}
where $i$ is an index over particle species, $f_i$ is the distribution function, and 
$d^3 \sigma_\mu$ is a volume element of the isothermal hypersurface $\Sigma$ defined by 
$T_{SW}$. We introduce the shear viscous correction to the distribution function based on Ref.~\cite{Pratt:2010jt}. 
We neglect the bulk viscosity correction, because ambiguity in the estimate exists \cite{Noronha-Hostler:2013gga}. 
Both bulk viscosity in the hydrodynamic expansion and the bulk viscous correction to the distribution function 
reduce the mean transverse momentum \cite{Monnai:2009ad,Song:2009rh, Dusling:2011fd,Bozek:2012qs, Noronha-Hostler:2013gga,Rose:2014fba,Ryu:2015vwa}.  
We find the particlization hypersurface based on Ref.~\cite{Huovinen:2012is}, 
checking the temperature of volume elements of the fluid at each time step. 
The sampled particle distribution is used for an initial state for UrQMD \cite{Bass:1998ca, Bleicher:1999xi}.  
In UrQMD all produced hadrons move along classical trajectories, including their scattering, resonance formations,
and decay processes until interactions among them stop.

\section{Temperature-dependent transport coefficients \label{sec:viscous}}
One of the pioneer works of analyses of temperature dependent $\eta/s$ was done in Ref.~\cite{Niemi:2011ix}. 
More comprehensive analyses were performed in Refs.~\cite{Molnar:2014zha, Niemi:2015qia}. 
In Ref.~\cite{Denicol:2015nhu} they authors showed that a temperature-independent $\eta/s$ is 
disfavored from comparison with experimental data at RHIC, and event-by-event flow 
as a function of rapidity is useful for constraining the 
temperature dependence of shear viscosity. 
The first analyses of experimental data with both shear and bulk viscosities of the hadronic phase were carried out  in Ref.~\cite{Bozek:2009dw}. 
The effect of bulk viscosity on higher flow harmonics was also 
investigated \cite{Noronha-Hostler:2013gga,Noronha-Hostler:2014dqa}. 
In addition, it was found that inclusion of bulk viscosity is preferable for a better description of the data of ultracentral 
relativistic heavy-ion collisions at the LHC \cite{Rose:2014fba}. 
\begin{figure}[t]
 \centering
 \includegraphics[width=7.5cm]{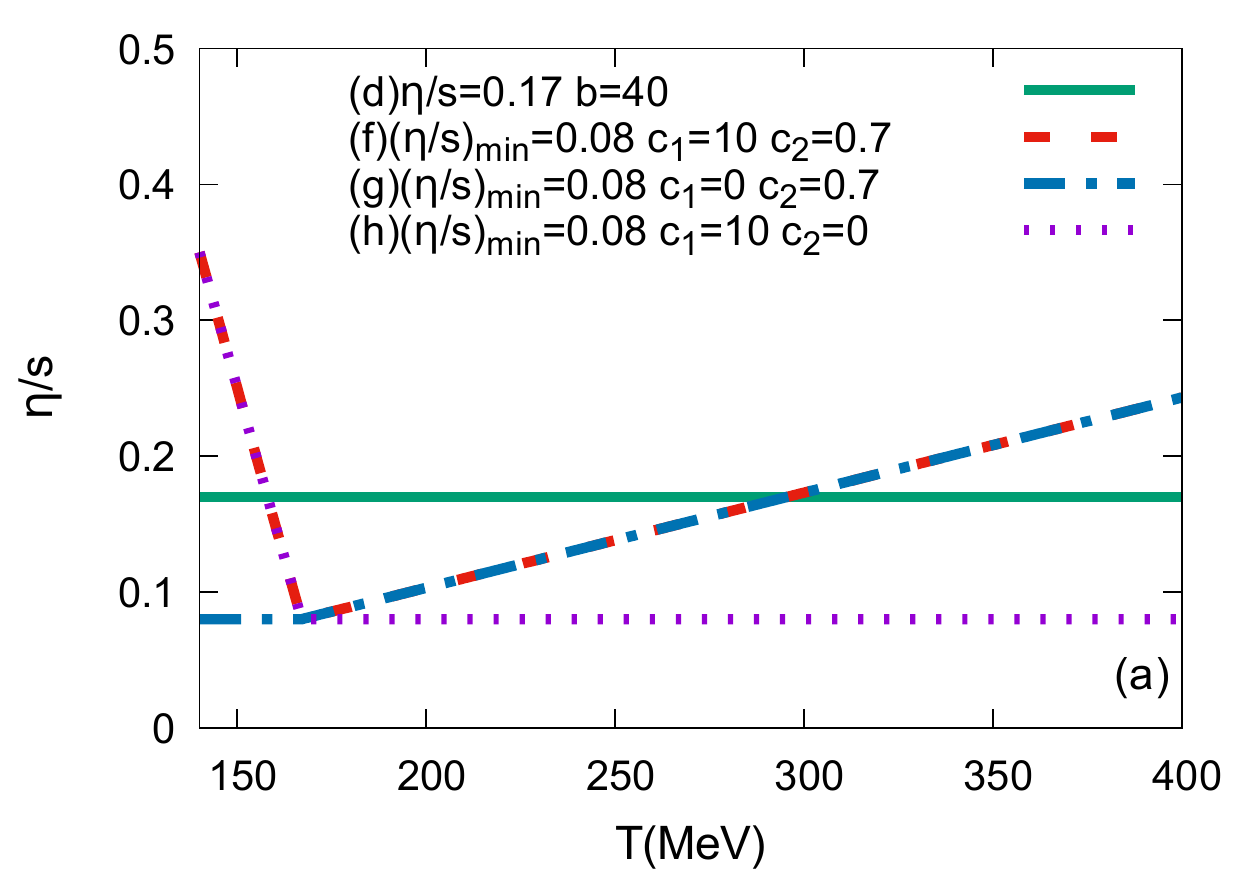}
 \centering
 \includegraphics[width=7.5cm]{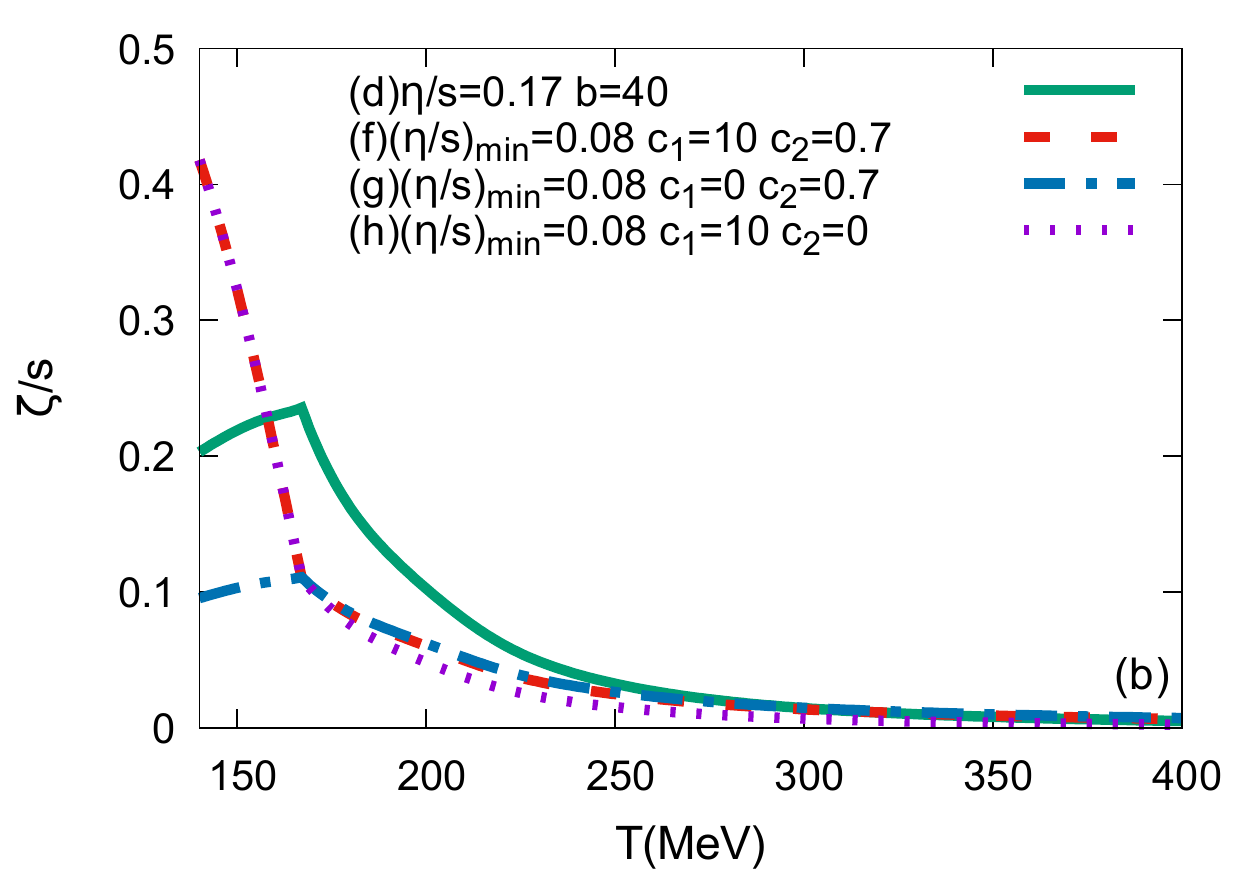}
 \colorcaption{The temperature dependence of the shear viscosity to entropy density ratio $\eta/s$ (a) 
 and the bulk viscosity to entropy density ratio  $\zeta/s$ (b).\label{fig:viscosity} } 
\end{figure}

Here we use the same parametrization as that 
in Refs.~\cite{Weinberg:1971mx,Denicol:2014vaa} for bulk viscosity, 
\begin{equation}
\frac{\zeta}{s} = b \frac{\eta}{s}  \left ( \frac{1}{3} - c_s^2\right)^2, 
\label{eq:bulk}
\end{equation}
where $c_s$ is the sound velocity and $b$ is a parameter. 
We parametrize $\eta/s(T)$ \cite{Niemi:2015qia, Denicol:2015nhu},  
\begin{equation}
\frac{\eta}{s}(T) = \left(\frac{\eta}{s}\right)_{min} + c_1(T_c - T)\theta(T_c-T) 
+ c_2(T - T_c)\theta(T-T_c),
\end{equation}
where $c_1$ and $c_2$ are parameters and $T_c=167$ MeV. 
We compare our calculated results with experimental data,  
in the case of the following parameter sets (Fig.~\ref{fig:viscosity}): 
(d) $\eta/s=0.17, b=40$;  
(f) $(\eta/s)_{\rm min}=0.08, c_1=10, c_2=0.7, b=40$; 
(g) $(\eta/s)_{\rm min}=0.08, c_1=0, c_2=0.7, b=40$; and 
(h) $(\eta/s)_{\rm min}=0.08, c_1=10, c_2=0, b=40$. 
The values of $c_1$ and $c_2$  are given in GeV$^{-1}$. 
Because the speed of sound takes the minimum value at $T_c$ \cite{Bluhm:2013yga},  
the temperature dependence of bulk viscosity has a sharp peak at $T_c$ in the case of (d).

\section{Numerical Results\label{sec:results}}

\subsection{Parameters in computations}
In the next subsections we explore transport coefficients of QGP through quantitative analyses of the 
LHC 
data, such as one-particle disaributions and collective flows in Pb+Pb $\sqrt{s_{NN}}=2.76$ TeV collisions. 
First in Sec. \ref{subsec:shear} we discuss an appropriate value of constant 
shear viscosity without including bulk viscosity 
and then in the Sec. \ref{subsec:bulk} we consider the effect of bulk viscosity on physical observables. 
In Sec. \ref{subsec:temp-shear} we investigate the temperature dependence of the shear viscosity.
Finally in Sec. \ref{subsec:FSI}, we discuss the effect of
final state interactions to $p_T$ spectra and collective flows. 
We determine parameters in the initial condition \trento, using pseudorapidity distributions of charged hadrons 
at central collision \cite{Abbas:2013bpa, Adam:2015kda}. 
Our parameters are the same as those in Ref.~\cite{Ke:2016jrd}, except for 
the normalization $N$ and $\sigma_0$. 

We fix the centrality based on initial entropy densities. 
First we produce initial entropy distributions of a certain number of events with minimum bias, using \trento. 
Then we arrange the events in decreasing order of total entropy $\frac{dS}{dy}|_{y=0}$.  
For example, for the centrality 0-10 \% we pick up the largest 10 \% events from entire events. 
To save computational time, we perform numerical calculation only for our focusing centralities. 
For the following analysis of experimental data, we prepare 2000 minimum bias events using \trento. 

\begin{figure}[t]
 \centering
 \includegraphics[width=7.5cm]{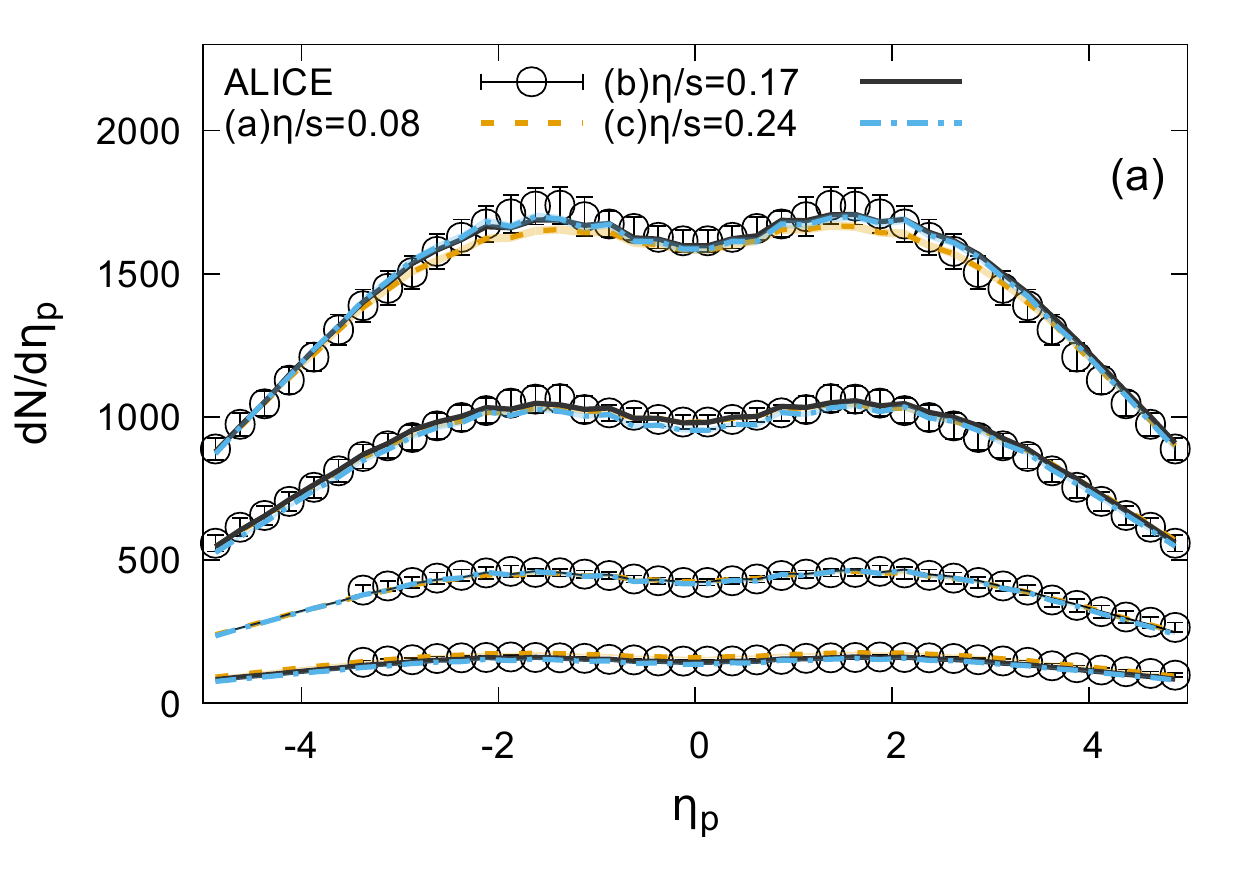}
\centering
 \includegraphics[width=7.5cm]{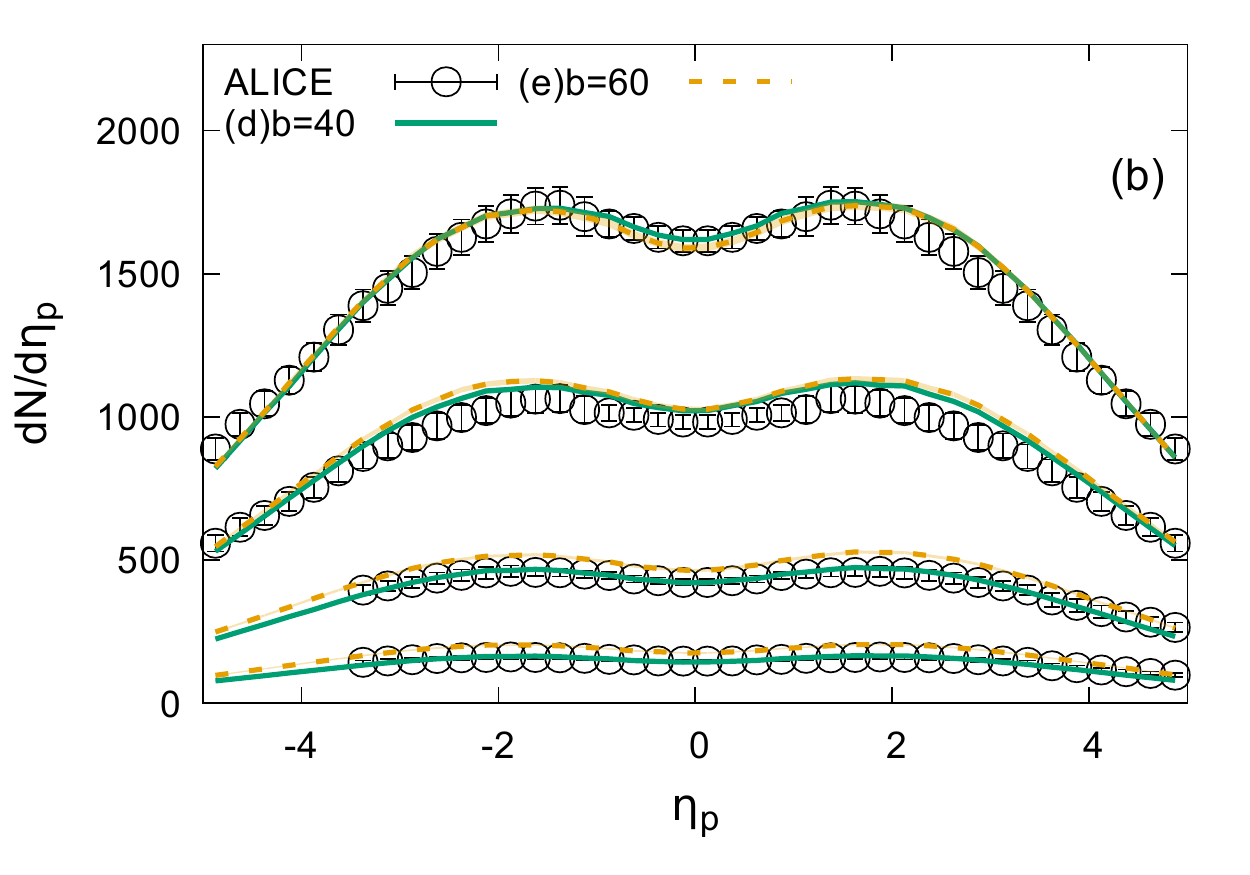}
\centering
 \includegraphics[width=7.5cm]{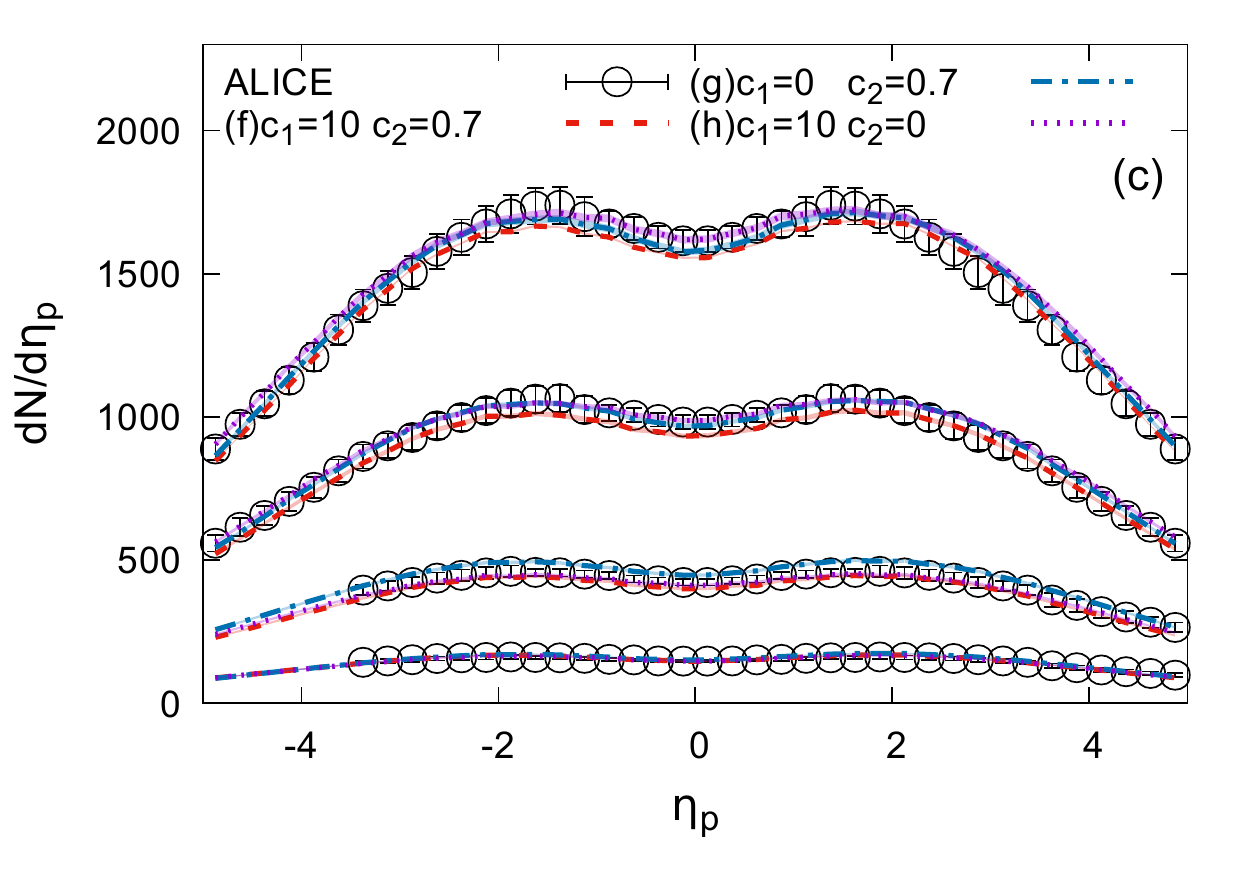}
 \colorcaption{The pseudorapidity distributions of charged hadrons in 0-5 $\%$, 10-20 $\%$, 30-40 $\%$, and 50-60 $\%$ 
 centralities together with the ALICE data (open circles) \cite{Abbas:2013bpa, Adam:2015kda}. 
 In the top panel, the pseudorapidity distributions were obtained 
 in the case of $\eta/s=0.08$, 0.17, and 0.24 at vanishing bulk viscosity (a).  
 The middle panel shows the computational results  with shear and bulk viscosities (b). 
 In the bottom panel, temperature dependence of $\eta/s$ is included (c). 
 See the text for details. 
 \label{fig:Ncheta} } 
\end{figure}

In the top panel of Fig.~\ref{fig:Ncheta} we show pseudorapidity distributions 
of charged particles in 0-5 $\%$, 10-20 $\%$, 30-40 $\%$, and 50-60 $\%$  centralities in the case of 
$\eta/s=0.08$, 0.17, and 0.24, neglecting the bulk viscosity.  
The middle panel shows the computational results with finite bulk viscosities and $\eta/s=0.17$. 
In the bottom panel we include the temperature dependence of $\eta/s$. 
For all the cases we reproduce centrality dependence of pseudorapidity distributions of charged particles very well (Fig. \ref{fig:Ncheta}). 
In Table \ref{tab:normalization} we list the values of $N$ and $\sigma_0$ which are used in Sec. 
\ref{subsec:shear} [constant shear viscosity: (a) $\eta/s=0.08$, (b) $\eta/s=0.17$, and (c) $\eta/s=0.24$],  
Sec. \ref{subsec:bulk} [finite bulk viscosity: (d) b=40 and (e) b=60], and Sec. \ref{subsec:temp-shear} [temperature-dependent 
shear viscosity: (f) $c_1=10, c_2=0.7$, (g) $c_1=0, c_2=0.7$, and (h) $c_1=10, c_2=0$]. 
In the case of (d) and (e), we need to choose smaller $\sigma_0$ 
because finite bulk viscosity increases the width of pseudorapidity distributions. 
We carry out the numerical computation with spatial grid sizes $\Delta x=\Delta y = 0.2$ fm, $\Delta \eta_s=0.3$ and 
time step $\Delta \tau= 0.5 \Delta x$ fm. 
\begin{table}[h]
\caption{The values of normalization $N$ and $\sigma_0$ in \trento.\label{tab:normalization}}
\begin{tabular}{|c|ccc|cc|ccc|}
\hline 
 & (a) &  (b)&(c)&(d)&(e)&(f)&(g)&(h)\\ \hline 
 $N$ & 116 & 110 & 105 & 94 & 88 & 94 & 98 & 101  \\
 $\sigma_0$ & 2.9 & 2.9 & 2.9 & 2.7 & 2.7& 2.9 & 2.9& 2.9 \\
\hline 
\end{tabular}
\end{table}

\subsection{Constant shear viscosity with vanishing bulk viscosity \label{subsec:shear}}
\begin{figure*}[t]
 \centering
 \includegraphics[width=14.3cm]{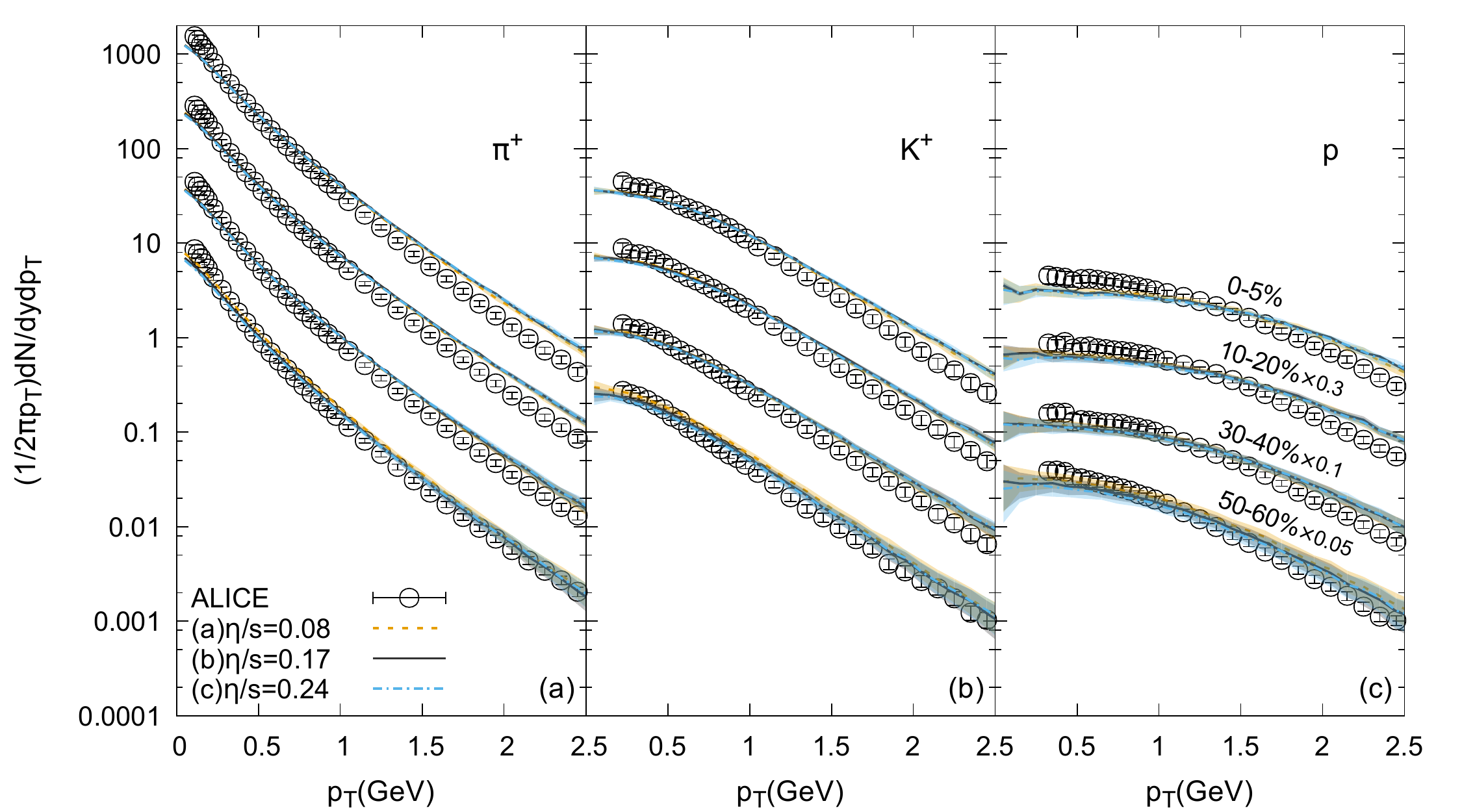}
 \colorcaption{
 The $p_T$ distributions for $\pi^+$ (a), $K^+$ (b), and $p$ (c) in 0-5 \%, 
 10-20 \%, 30-40 \%, and 50-60 \% centralities, 
together with the ALICE data (the open circles)  \cite{Abelev:2013vea}. 
The orange dashed line, the black solid line, and the blue dashed-dotted line 
stand for $\eta/s=0.08$, 0.17, and 0.24, respectively. \label{fig:pT_shear} } 
\end{figure*}
\begin{figure}[t]
 \centering
 \includegraphics[width=7.5cm]{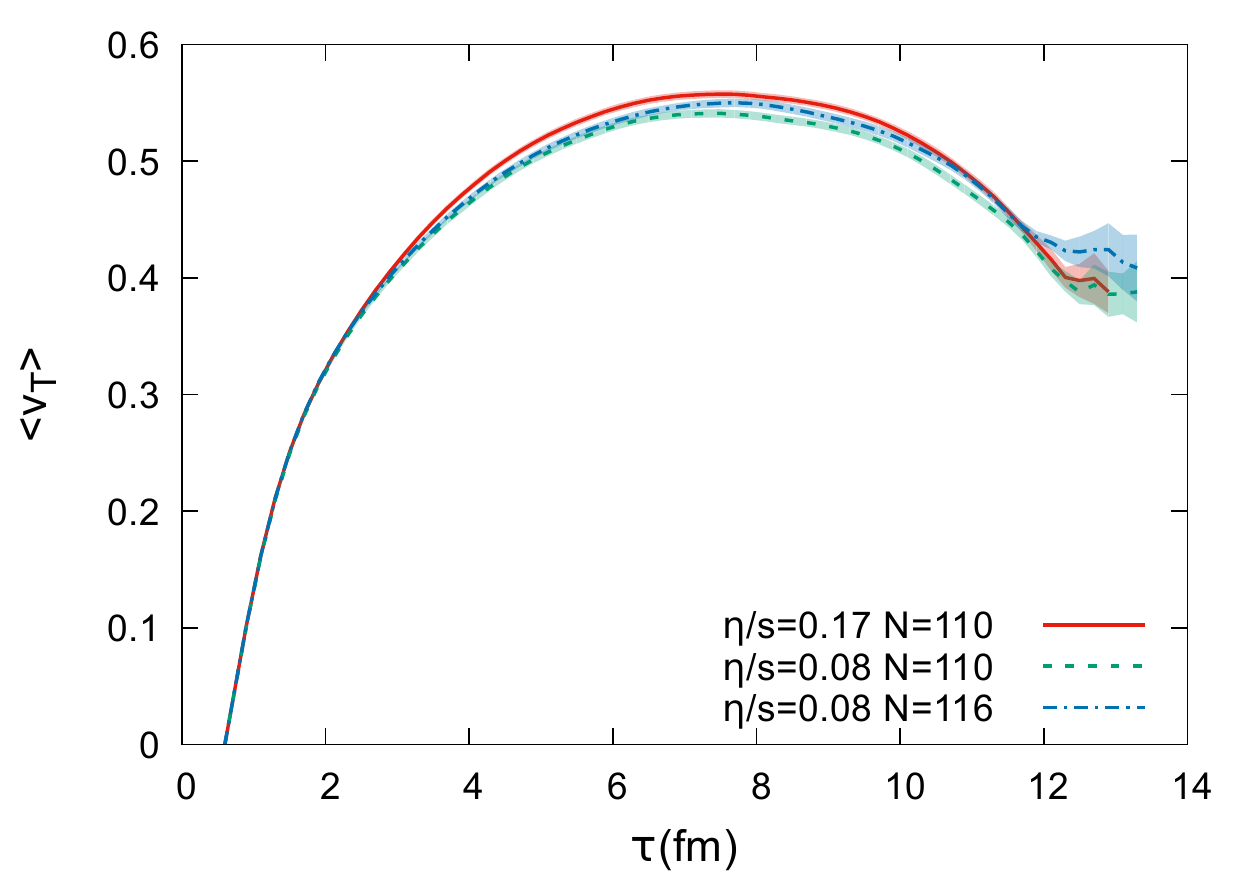}
 \colorcaption{The time evolution of the spatial averaged radial flow $v_T=\sqrt{v_x^2 + v_y^2}$ of fluid cells whose temperatures 
 are  $T>150$ MeV at $\eta_s=0$ in 0-5 \% centrality.  
 The values are taken from the average over 20 events. 
 The solid line, the dashed line and the dashed-dotted line 
 stand for $\eta/s=0.17$ with $N=110$, $\eta/s=0.08$ with $N=110$, and 
 $\eta/s=0.08$ with $N=116$, respectively. 
 \label{fig:vT-shear}} 
\end{figure}
\begin{figure}[t]
 \centering
 \includegraphics[width=7.5cm]{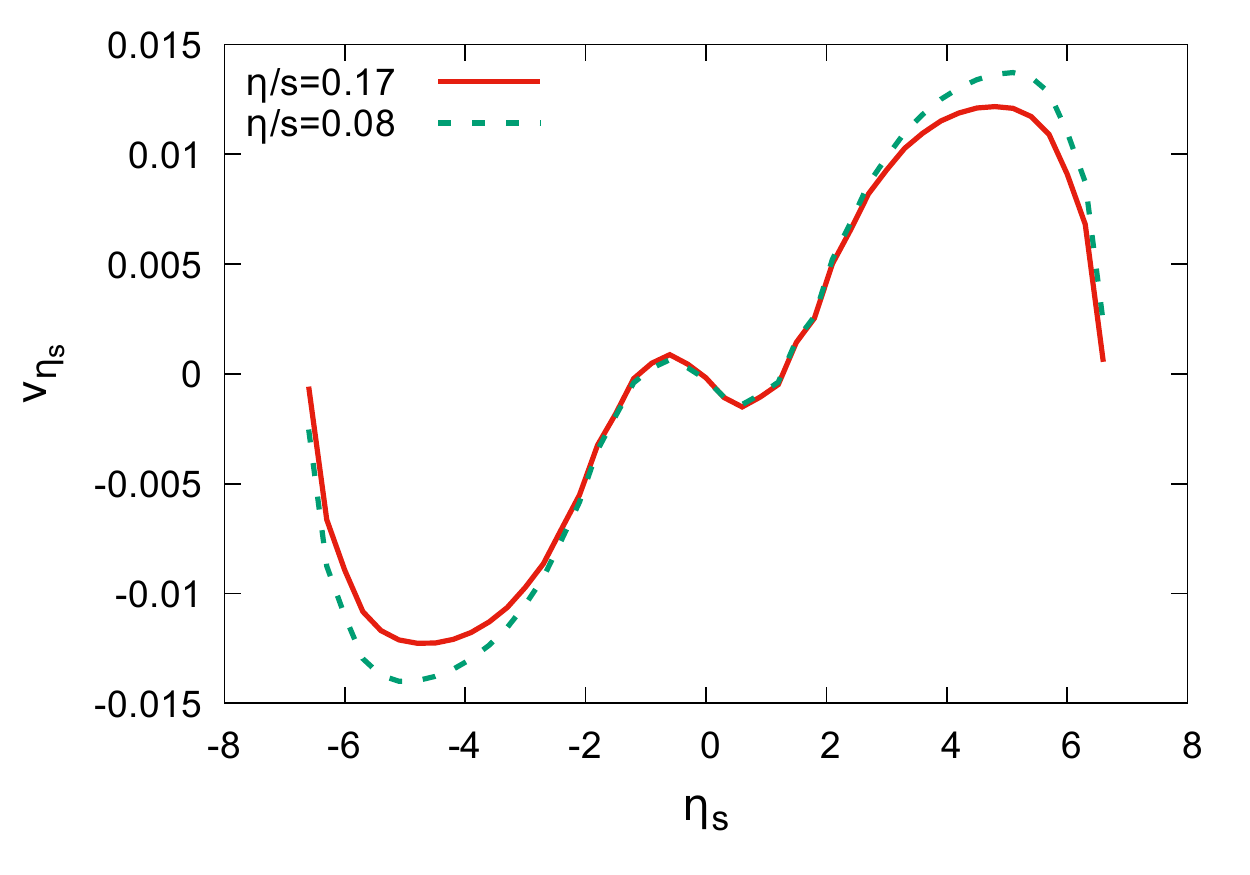}
 \colorcaption{The longitudinal flow $v_{\eta_s}$ as a function of $\eta_s$ ($x=y=0$ fm) at $\tau=7$ fm 
 in 0-5 \% centrality.  The solid line stands for $\eta/s=0.17$ and the dashed line stands for $\eta/s=0.08$. 
 Both of them are computed with the normalization $N=110$ for one event.
 \label{fig:vl-shear}} 
\end{figure}

First we extract a suitable value of the shear viscosity $\eta/s$ from comparison with experimental data 
of $p_T$ distributions and collective flows $v_2$ and $v_3$, neglecting bulk viscosity. 
In Fig.~\ref{fig:pT_shear} the $p_T$ distributions for $\pi^+$, $K^+$, and $p$ are shown, together with 
the ALICE data \cite{Abelev:2013vea}. The differences among calculated results of $p_T$ spectra with 
$\eta/s=0.08$, 0.17, and 0.24 are very small, which suggests that $p_T$ spectra themselves are not sensitive 
to the value of $\eta/s$. For all cases we reproduce experimental data reasonably well, 
though  in 0-5 \% and 10-20 \% centralities we observe a deviation from experimental data above $p_T \sim 1.5 $  GeV. 

\begin{figure*}[t!]
\begin{minipage}{0.5\hsize}
 \centering\hspace{1cm}
\includegraphics[width=7.3cm]{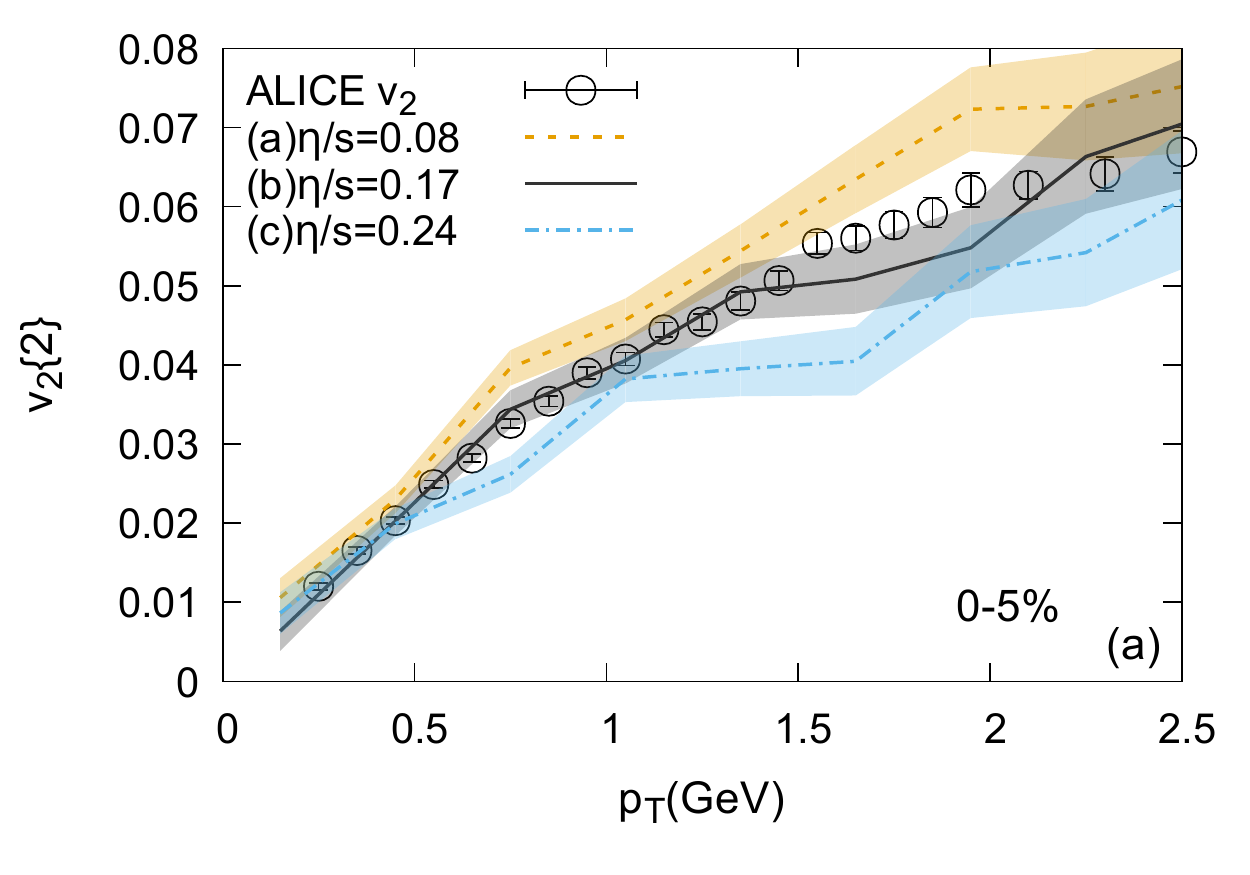}
 \end{minipage}
 \begin{minipage}{0.49\hsize}
 \centering\hspace{-1cm}
 \includegraphics[width=7.3cm]{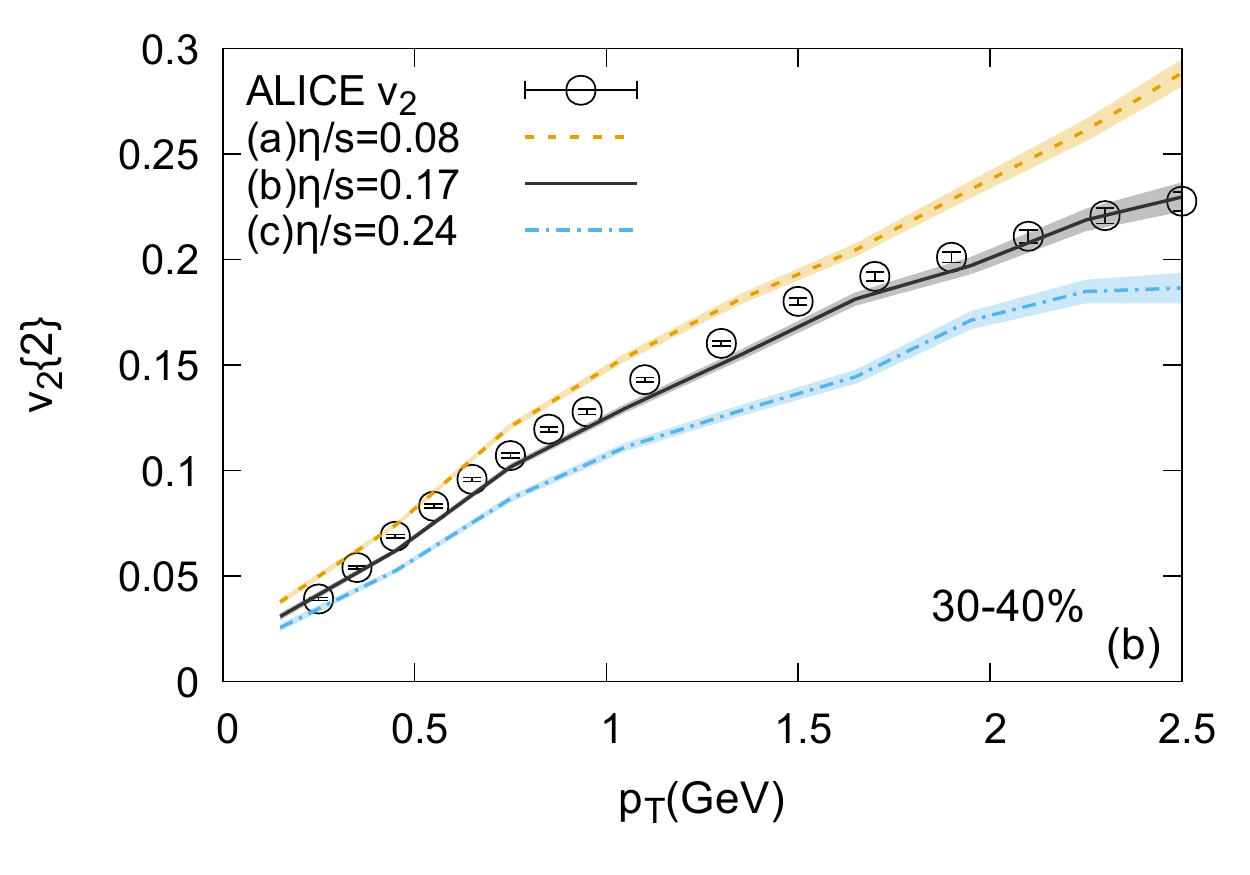}
\end{minipage}
\begin{minipage}{0.5\hsize}
 \centering\hspace{1cm}
 \includegraphics[width=7.3cm]{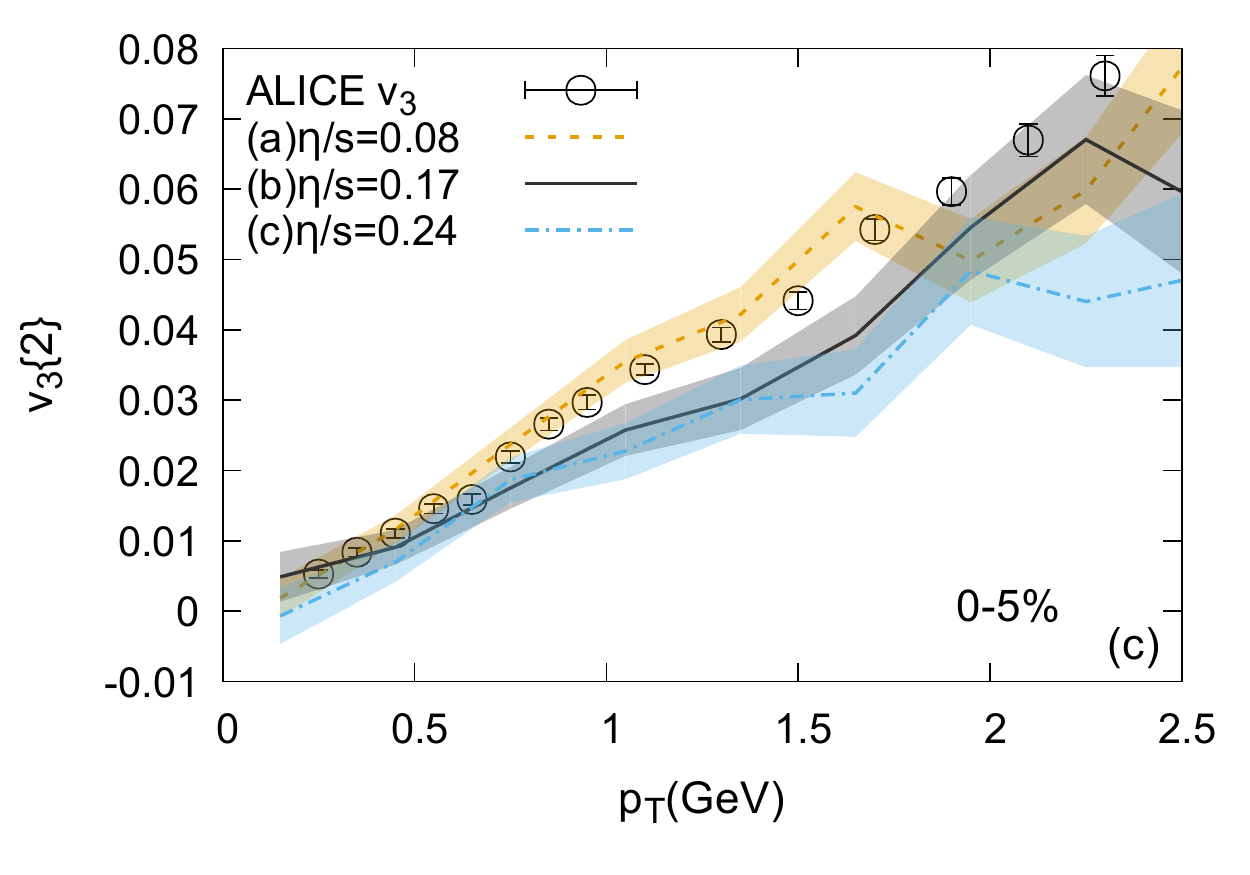}
 \end{minipage}
 \begin{minipage}{0.49\hsize}
 \centering\hspace{-1cm}
 \includegraphics[width=7.3cm]{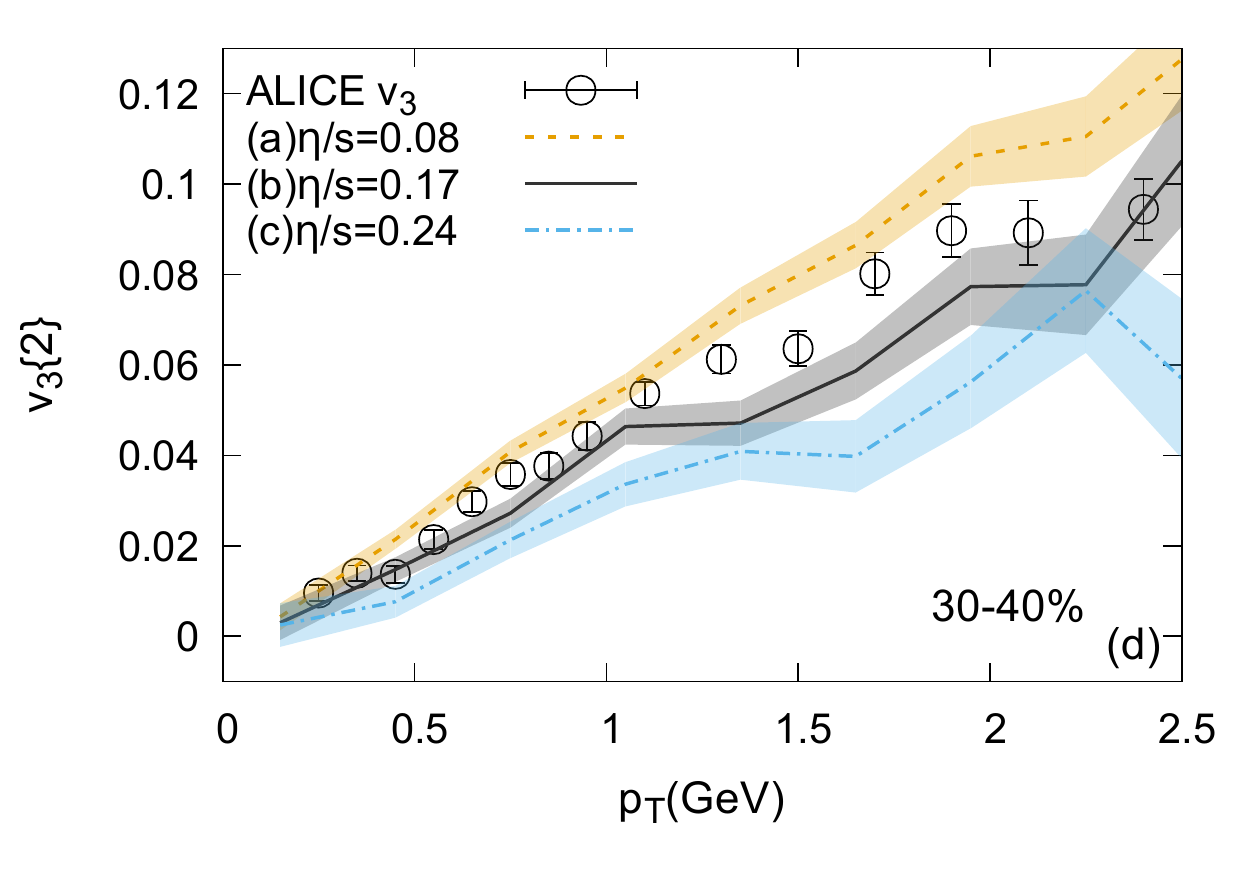}
\end{minipage}
 \colorcaption{
The elliptic and triangular flows of charged hadrons 
 as a function of $p_T$ in 0-5 $\%$  ((a) for $v_2$ and (c) for $v_3$) and 30-40 $\%$ centralities ((b) for $v_2$ and (d) for $v_3$) , 
 together with the ALICE data (the open circles) \cite{ALICE:2011ab}. The orange dashed line, black solid line, and the 
 blue dashed-dotted line stand for $\eta/s=0.08$, 0.17, and 0.24, respectively. \label{fig:vn-shear}
} 
\end{figure*}
\begin{figure*}[t!]
 \centering
 \includegraphics[width=14cm]{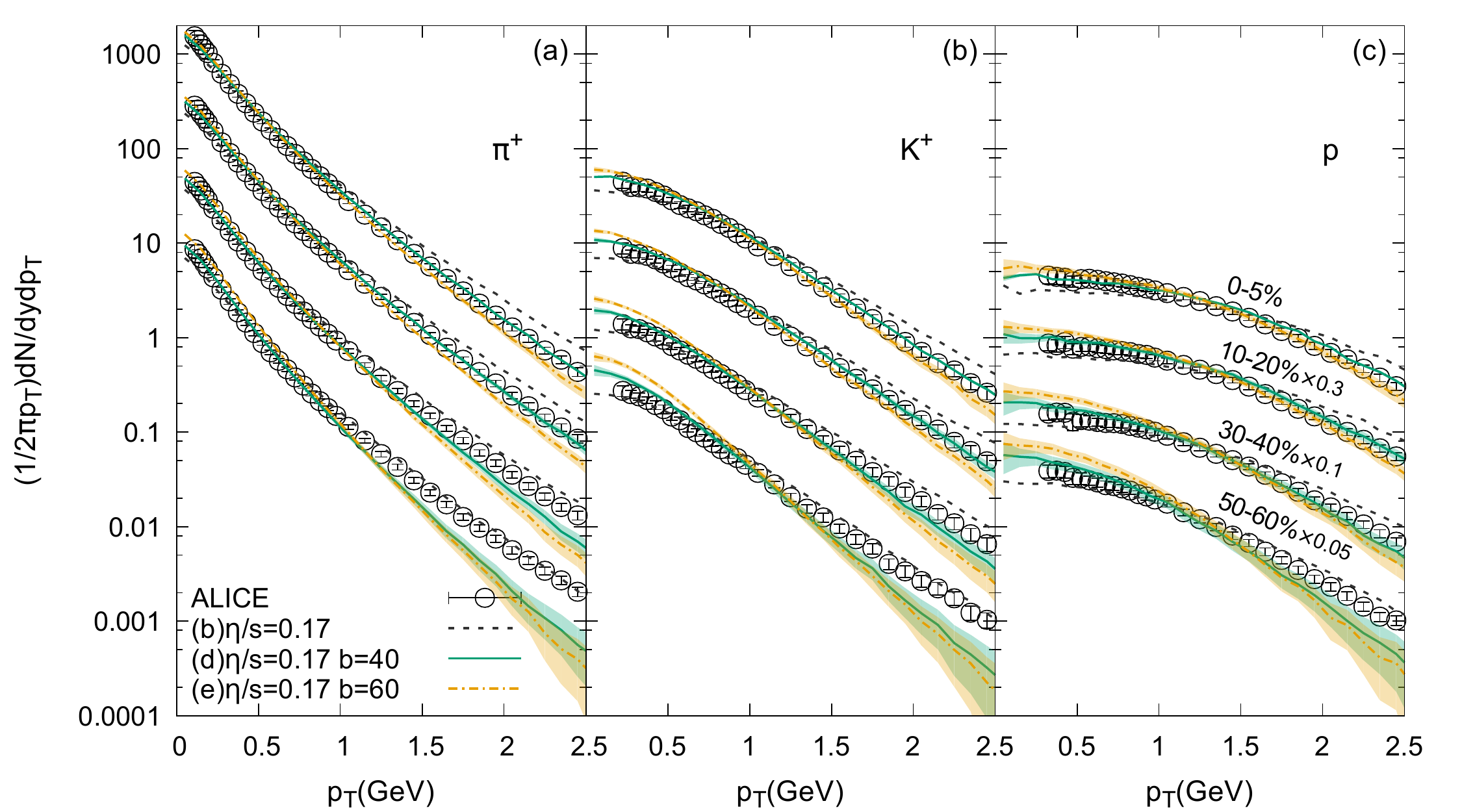}
 \colorcaption{
 The $p_T$ distributions for $\pi^+$ (a), $K^+$ (b), and $p$ (c) in 0-5 $\%$, 10-20 $\%$, 30-40 $\%$, 
and 50-60 $\%$ centralities, together with the ALICE data (the open circles)  \cite{Abelev:2013vea}. 
The black dashed line, the green solid line, and the yellow dashed-dotted line 
stand for vanishing bulk viscosity, $b=40$, and $b=60$, respectively. \label{fig:pT_bulk} 
} 
\end{figure*}

To understand the effect of the shear viscosity on the $p_T$ distributions, 
we show the time evolution of the spatial averaged radial flow $v_T=\sqrt{v_x^2 + v_y^2}$ of fluid cells whose temperatures 
 are  $T>150$ MeV at $\eta_s=0$ in 0-5 \% centrality in 
 Fig.~\ref{fig:vT-shear}. 
Keeping the normalization $N=110$, we change the value of $\eta/s$  from $\eta/s=0.17$ to $\eta/s=0.08$. 
With the larger  $\eta/s$ the growth of radial flow becomes larger.  
If we change $N$ from $N=110$ to $N=116$ in the case of $\eta/s=0.08$ 
to reproduce the rapidity distributions, the difference between $\eta/s=0.17$ and $\eta/s=0.08$ 
becomes very small. 
In addition, the average life time of the fluid for $\eta/s=0.17$  is the shortest, since 
its radial flow is the largest among the three cases.

Figure \ref{fig:vl-shear} shows the longitudinal flow as a function of $\eta_s$ at 
$\tau=7$ fm in 0-5 \% centrality. 
In the calculation, we use the same normalization $N=110$ for $\eta/s=0.08$ and $0.17$. 
Near midrapidity $|\eta_s| < 1$ the sign of $v_{\eta_s}$ is changed. 
The structure comes from the initial condition of pressure distribution of \trento \ 
in the longitudinal direction: there is a small dent at midrapidity and two bumps beside it. 
At early times, the initially produced partons are massless and free-streaming in the $z$ direction, $z/t=p_z/|\bf{p}|$. 
Under this assumption, 
we can estimate $\eta_s = \frac{1}{2} \ln \left( \frac{t+z}{t-z} \right)  \sim  
\eta_p = \frac{1}{2} \ln \left ( \frac{ |{\bm p}| + p_z}{ |{\bm p}| - p_z} \right ) $, which 
suggests that the dent in pseudo-rapidity distributions also appears in space-time 
rapidity distributions.
Around midrapidity $v_{\eta_s}$ becomes smaller than Bjorken's flow, 
whereas, in $|\eta_s| > 2$, $v_{\eta_s}$ is larger than Bjorken's flow due to acceleration from the 
pressure gradient. 
Around $|\eta_s|\sim 4$ the difference between two cases becomes large, which 
means that in the case of larger shear viscosity the growth of longitudinal 
flow at the early stage of expansion 
transforms into the larger radial flow. 
As a result, transverse flow is larger with the larger shear viscosity.

Figure \ref{fig:vn-shear} shows the elliptic flow $v_2$ and triangular flow $v_3$ 
of charged hadrons as a function of $p_T$ in 0-5 $\%$ and 30-40 $\%$ centralities. 
We compute the flow harmonics $v_n$ from the two-particle cumulant, 
using the $Q$-cumulant method \cite{Bilandzic:2010jr}. 
The same $p_T$ and rapidity cuts as those of the ALICE data \cite{ALICE:2011ab} are applied. 
In contrast to the $p_T$ distribution Fig.~\ref{fig:pT_shear}, there is clear 
$\eta/s$ dependence in behavior of collective flows: 
The larger $\eta/s$ is, the smaller $v_2$ and $v_3$ are. 
The existence of shear viscosity suppresses the growth of anisotropy of the 
flow on the transverse plane. 
Our result suggests that a suitable $\eta/s$ can be chosen between $\eta/s=0.08$ and $\eta/s=0.17$, which does 
not contradict the results of Ref.~\cite{Ke:2016jrd}.

Here we give a short summary for the constant shear viscosity. 
The $p_T$ spectra of $\pi^+$, $K^+$, and $p$ are insensitive to the value of $\eta/s$ and in 0-5 \% and 
10-20 \% centralities our computed $p_T$ spectra overestimate above $p_T > 1.5$ GeV, 
which suggests that the mean $p_T$ is larger. 
We find the clear $\eta/s$ dependence in $v_2$ and $v_3$;
{\it i.e.}, the larger $\eta/s$ is, the smaller $v_2$ and $v_3$ of charged hadrons are.

\begin{figure}[t]
  \centering
 \includegraphics[width=7.3cm]{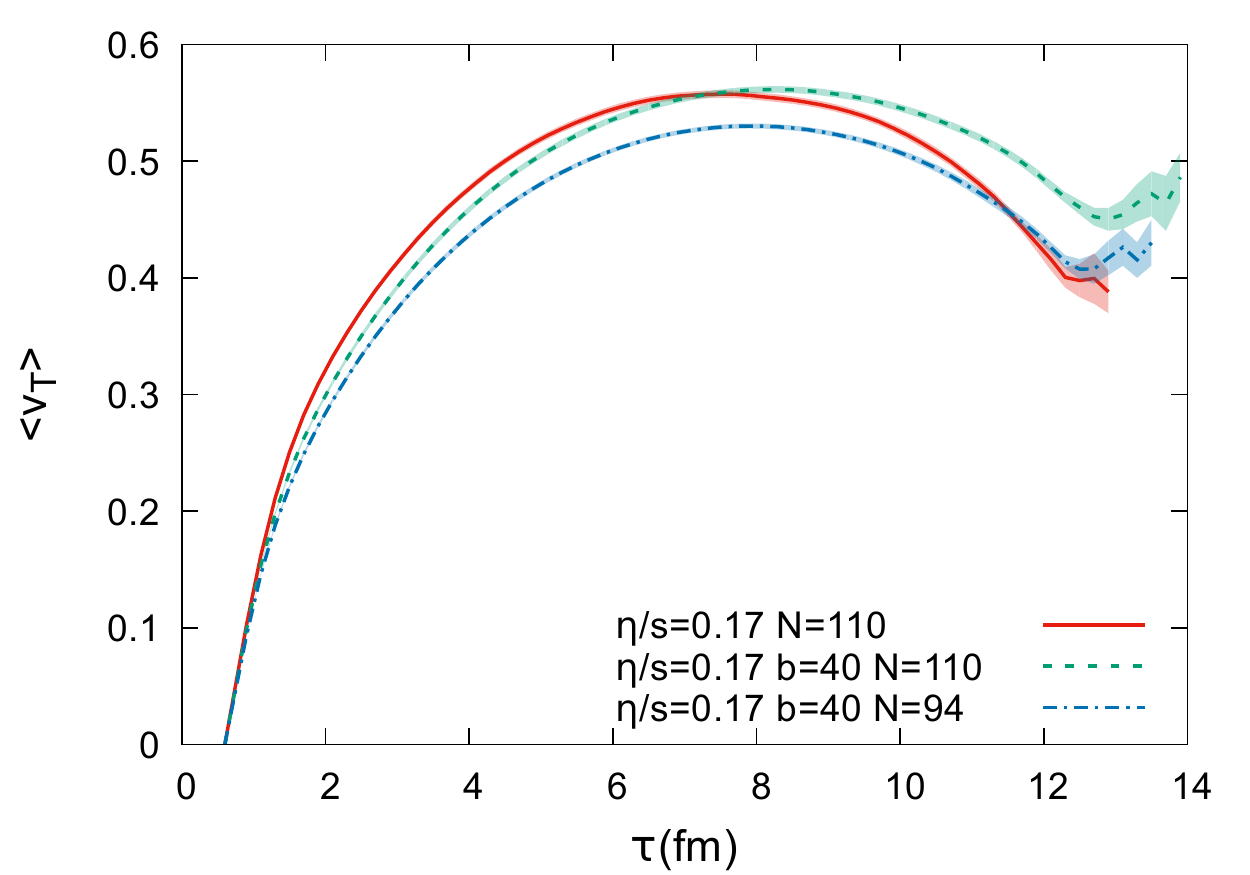}
 \colorcaption{
 The time evolution of the spatial averaged radial flow $v_T=\sqrt{v_x^2 + v_y^2}$ of fluid cells whose temperatures 
 are  $T>150$ MeV at $\eta_s=0$ in 0-5 \% centrality. 
 The values are taken from the average over 20 events. 
 The red solid line, the green dashed line, and the blue dashed-dotted line 
 stand for $\zeta/s=0$ with $N=110$, $b=40$ with $N=110$, and 
 $b=40$ with $N=94$, respectively.
 For all cases we fix the shear viscosity to $\eta/s=0.17$. 
  \label{fig:vT-bulk} 
}
\end{figure}

\subsection{Effect of bulk viscosity \label{subsec:bulk}} 
\begin{figure}[t!]
 \centering
 \includegraphics[width=7.1cm]{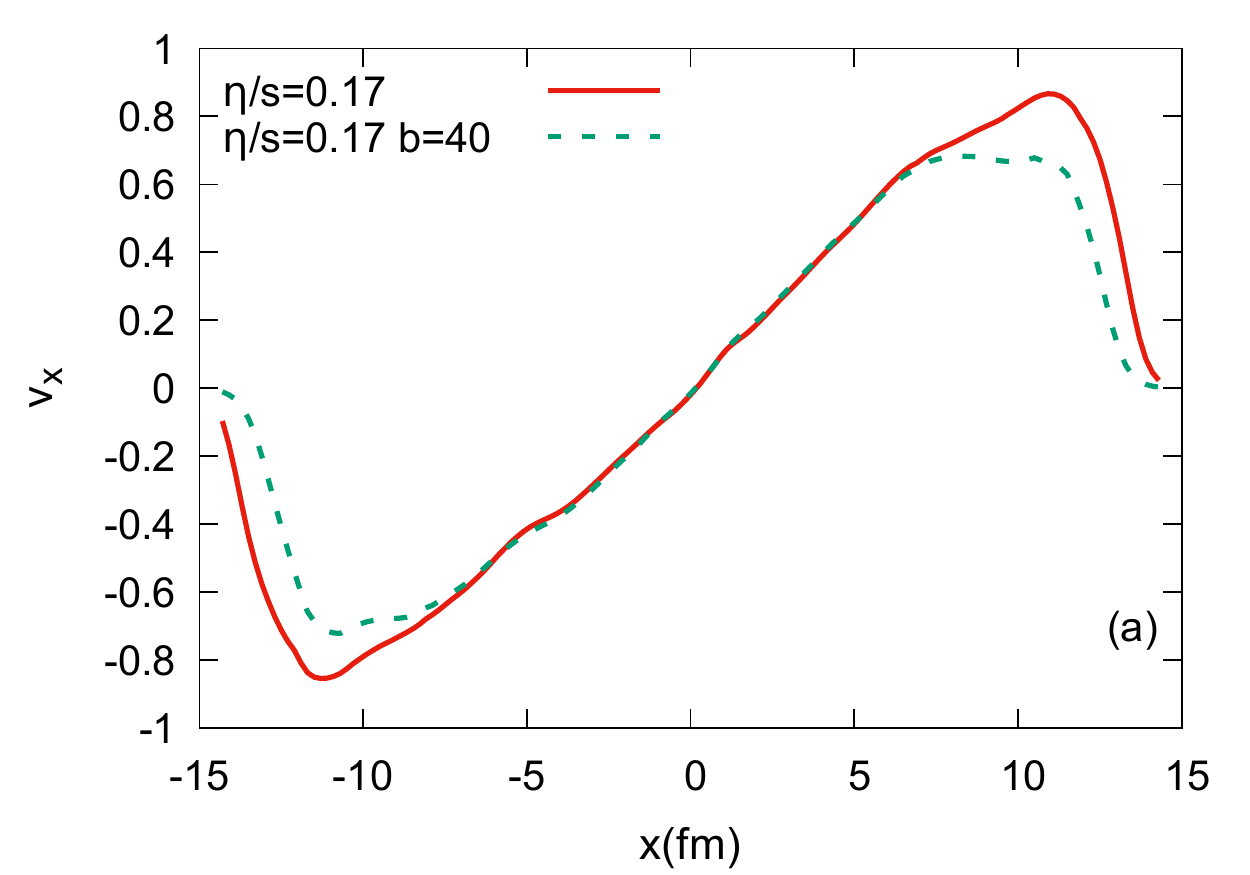}
  \centering
 \includegraphics[width=7.1cm]{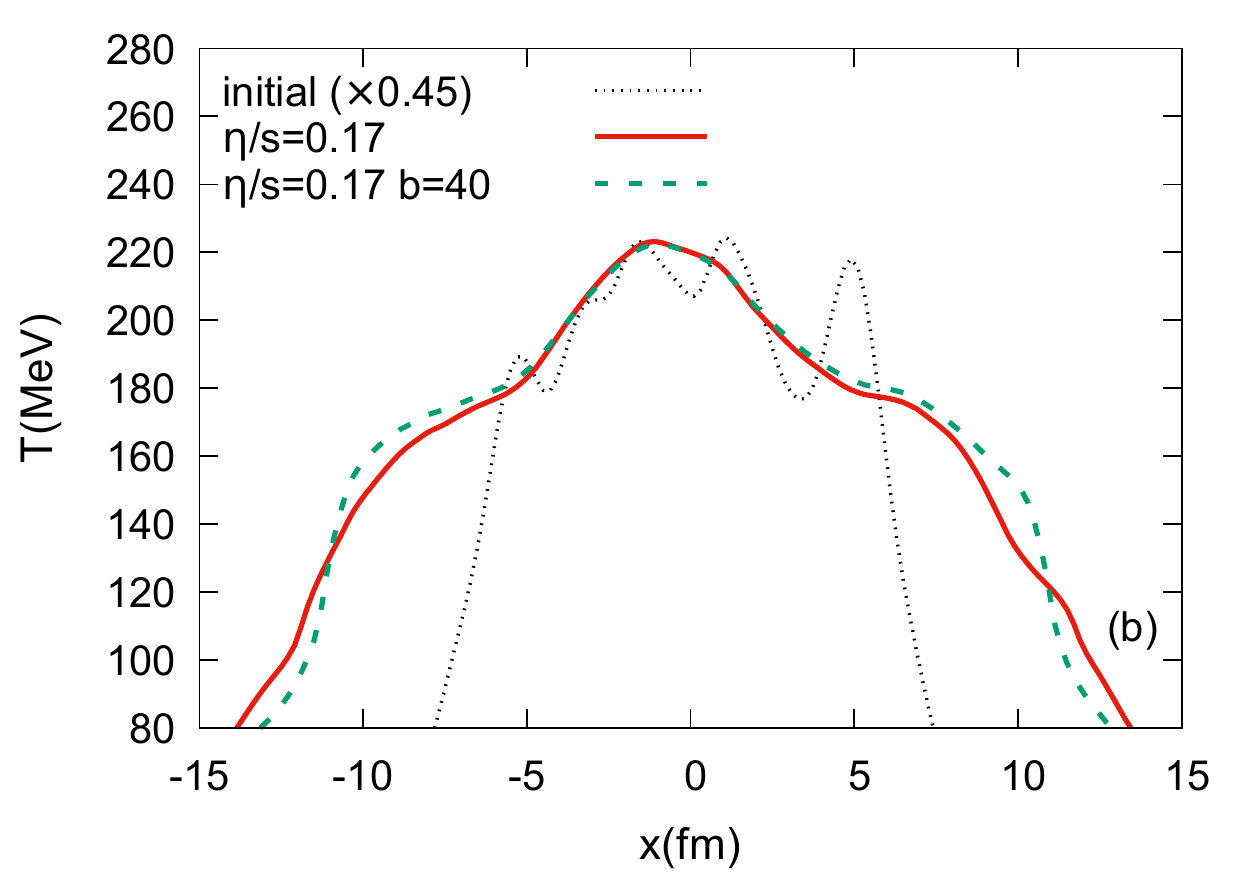}
 \colorcaption{
 The $v_x$ (a) and temperature (b) distributions 
 as a function of $x$ fm ($y=0$ fm and $\eta_s=0$) at $\tau=7$ fm 
 in 0-5 \% centrality.  The red solid line stands for $\eta/s=0.17$ with vanishing bulk viscosity 
 and the green dashed line stands for $\eta/s=0.17$ with $b=40$. 
 Both of them are computed with the normalization $N=110$ for one event. 
  \label{fig:vx-bulk}
}
\end{figure}
\begin{figure}[t!]
  \centering
 \includegraphics[width=7.3cm]{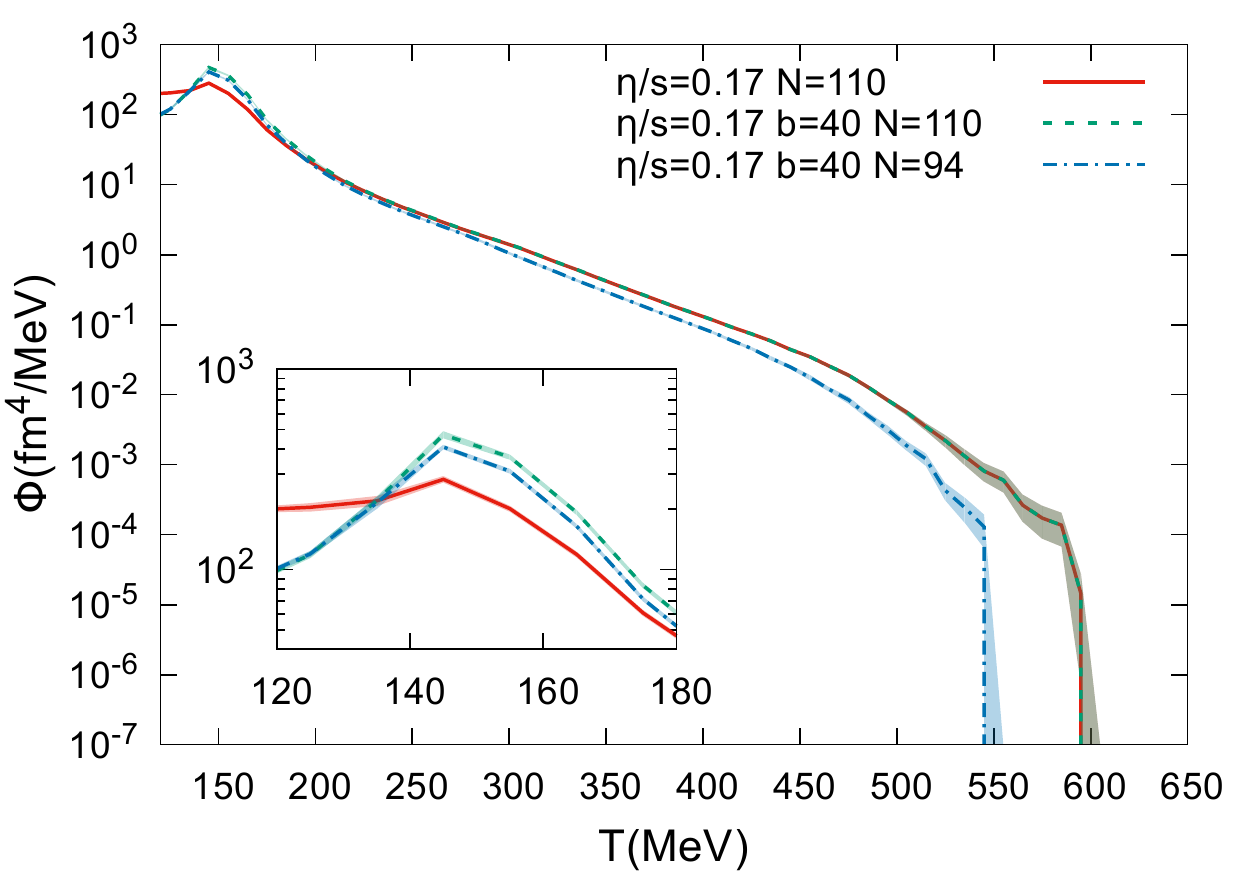}
  \colorcaption{
 The profile functions in 0-5\% centrality.
 The red solid line, the green dashed line, and the blue dashed-dotted line 
 stand for $\zeta/s=0$ with $N=110$, $b=40$ with $N=110$, and 
 $b=40$ with $N=94$, respectively. All calculations are done with $\eta/s=0.17$.
  The values are taken from the average over ten events. 
  \label{fig:profile_ft}} 
\end{figure}
\begin{figure*}[t]
\begin{minipage}{0.5\hsize}
 \centering \hspace{1cm}
 \includegraphics[width=7.3cm]{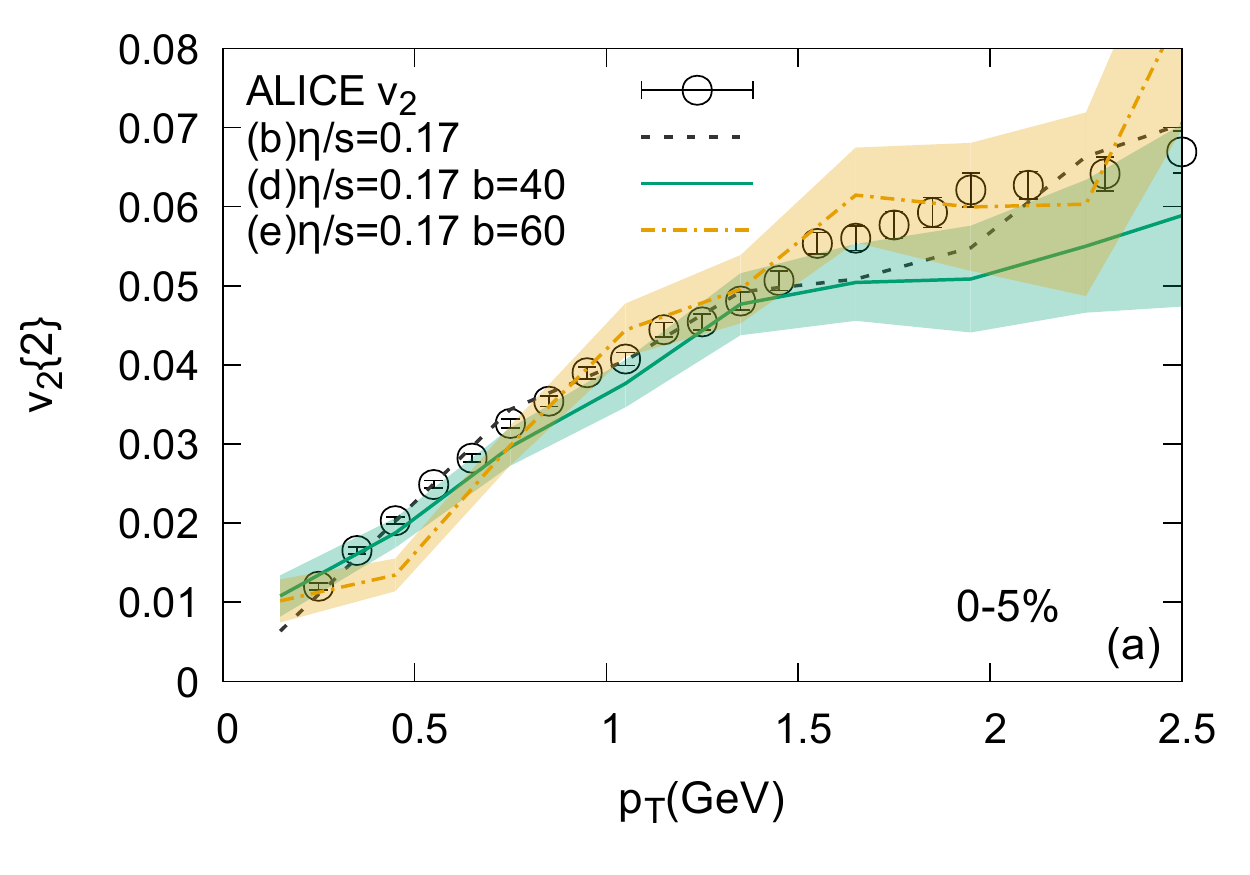}
 \end{minipage}
 \begin{minipage}{0.49\hsize}
 \centering \hspace{-1cm}
 \includegraphics[width=7.3cm]{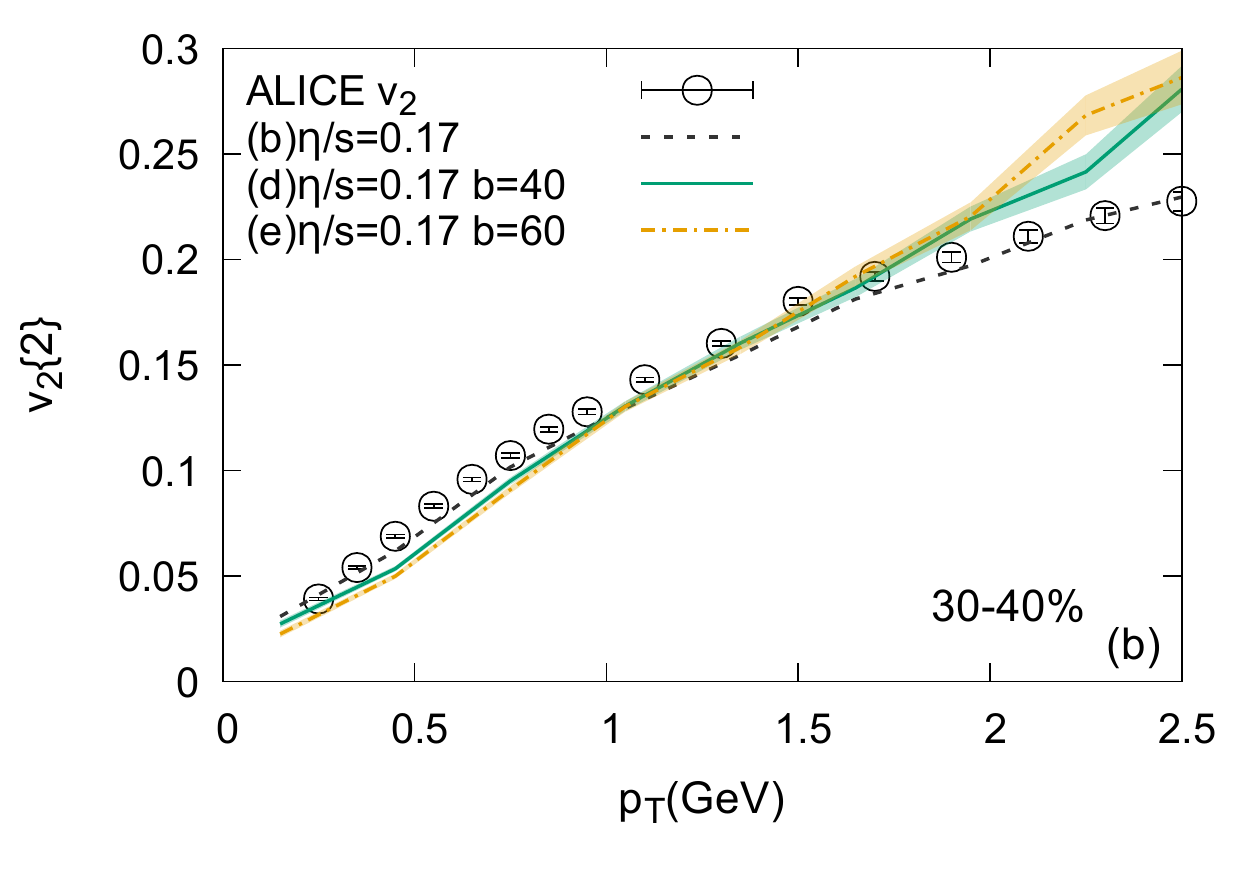}
\end{minipage}
\begin{minipage}{0.5\hsize}
 \centering \hspace{1cm}
 \includegraphics[width=7.3cm]{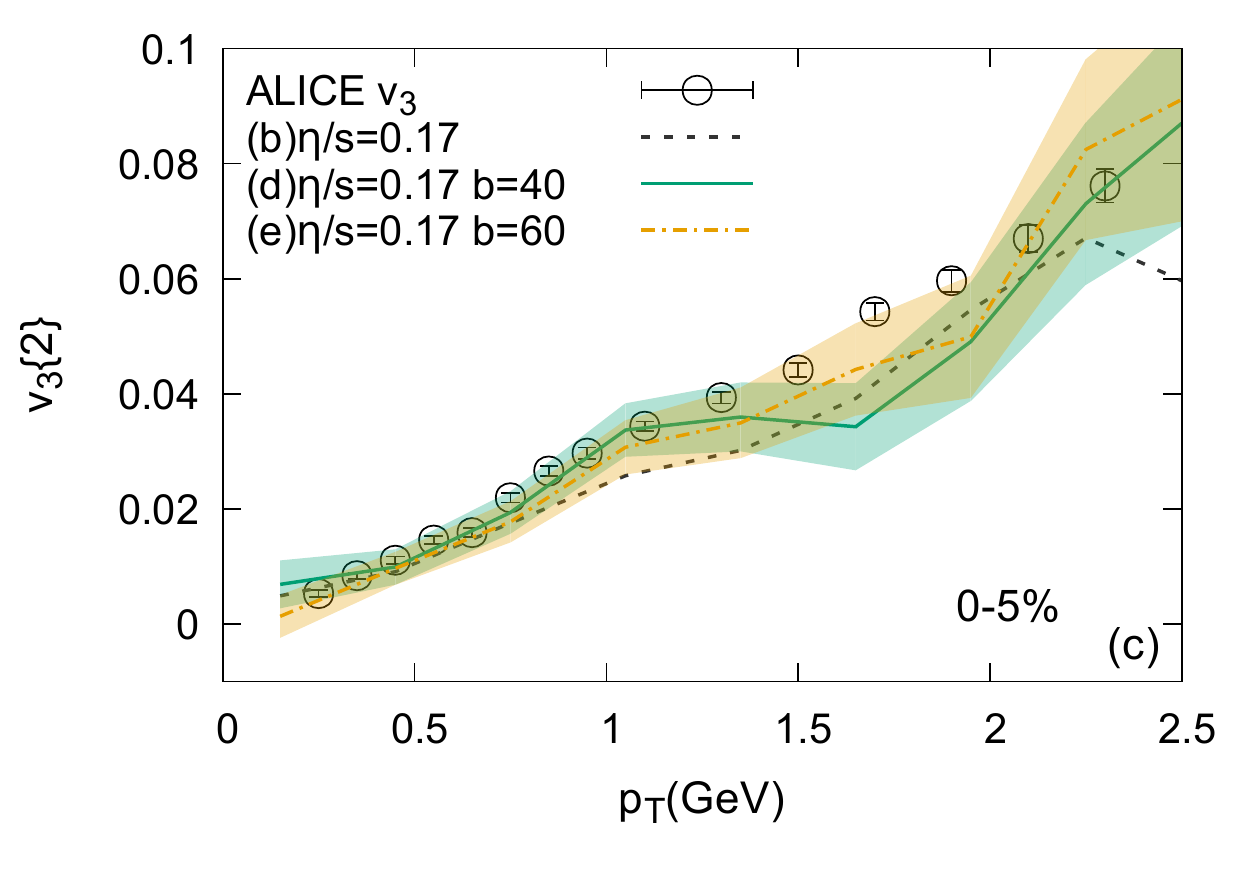}
 \end{minipage}
 \begin{minipage}{0.49\hsize}
 \centering \hspace{-1cm}
 \includegraphics[width=7.3cm]{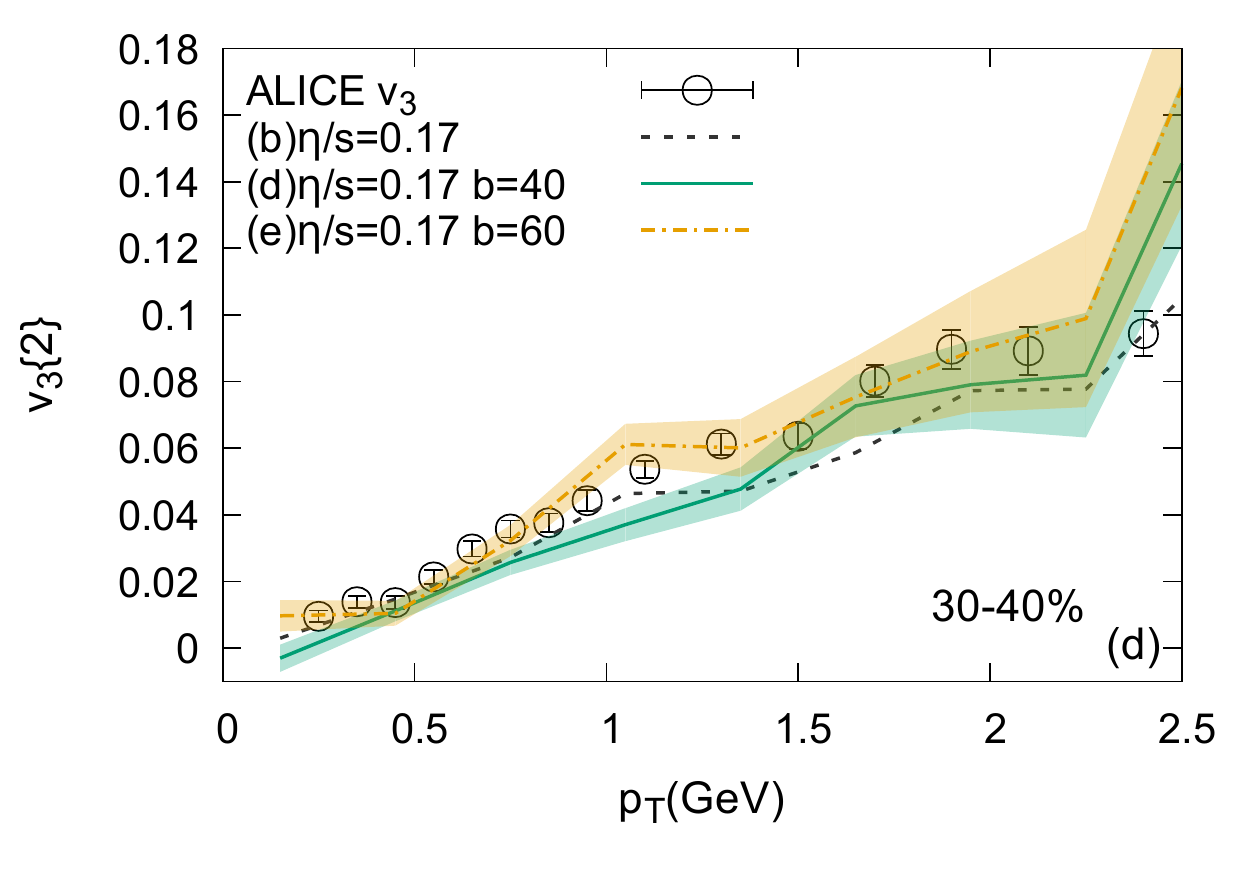}
\end{minipage}
 \colorcaption{
 The elliptic and triangular flows of charged hadrons 
 as a function of $p_T$ in 0-5 $\%$ [(a) for $v_2$ and (c) for $v_3$] and 30-40 $\%$ 
 [(b) for $v_2$ and (d) for $v_3$] centralities, 
 together with the ALICE data (the open circles) \cite{ALICE:2011ab}. 
 The black dashed line, the green solid line, and the 
 dashed-dotted line stand for $\eta/s=0.17$ without bulk viscosity, 
$\eta/s=0.17$ with $b=40$, and $\eta/s=0.17$ with $b=60$, 
respectively. \label{fig:vn-bulk}}
\end{figure*}

Next we investigate how the effect of bulk viscosity appears in $p_T$ spectra and 
collective flows $v_2$ and $v_3$. 
We introduce the bulk viscosity through 
Eq.~(\ref{eq:bulk}), fixing the value of shear viscosity to $\eta/s=0.17$. 

Figure \ref{fig:pT_bulk} shows the transverse momentum distributions 
for $\pi^+$, $K^+$, and $p$ in 0-5 $\%$, 10-20 $\%$, 30-40 $\%$, 
and 50-60 $\%$ centralities. The slopes of $p_T$ spectra of $\pi^+$, $K^+$, 
and $p$ are steeper, if the value of bulk viscosity is larger. 
The growth of the transverse flow becomes small due to the existence of bulk viscosity. 

\begin{figure*}[t!]
 \centering
 \includegraphics[width=14cm]{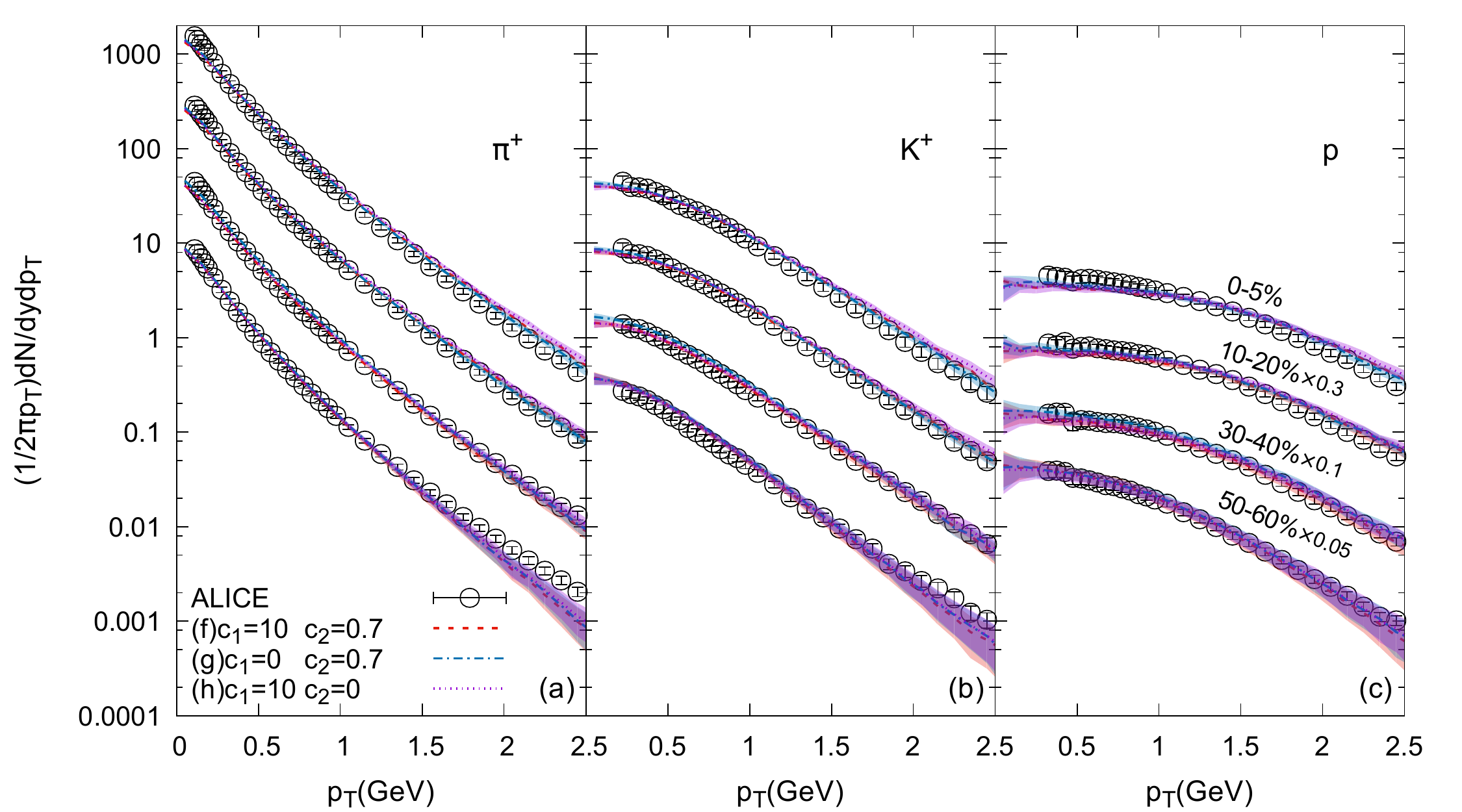}
 \colorcaption{
 The $p_T$ distributions for $\pi^+$ (a), $K^+$ (b), and $p$ (c) in 0-5 $\%$, 10-20 $\%$, 30-40 $\%$, 
and 50-60 $\%$ centralities together with the ALICE data (the open circles)  \cite{Abelev:2013vea}. 
The red dashed line, the blue dashed-dotted line, and the purple dotted line 
stand for (f) $c_1=10$, $c_2=0.7$,  (g) $c_1=0$, $c_2=0.7$, and  (h) $c_1=10$, $c_2=0.7$, respectively. 
See the details in the text. 
\label{fig:pT_shearT} 
}
\end{figure*}
\begin{figure}[t]
 \centering
 \includegraphics[width=7.3cm]{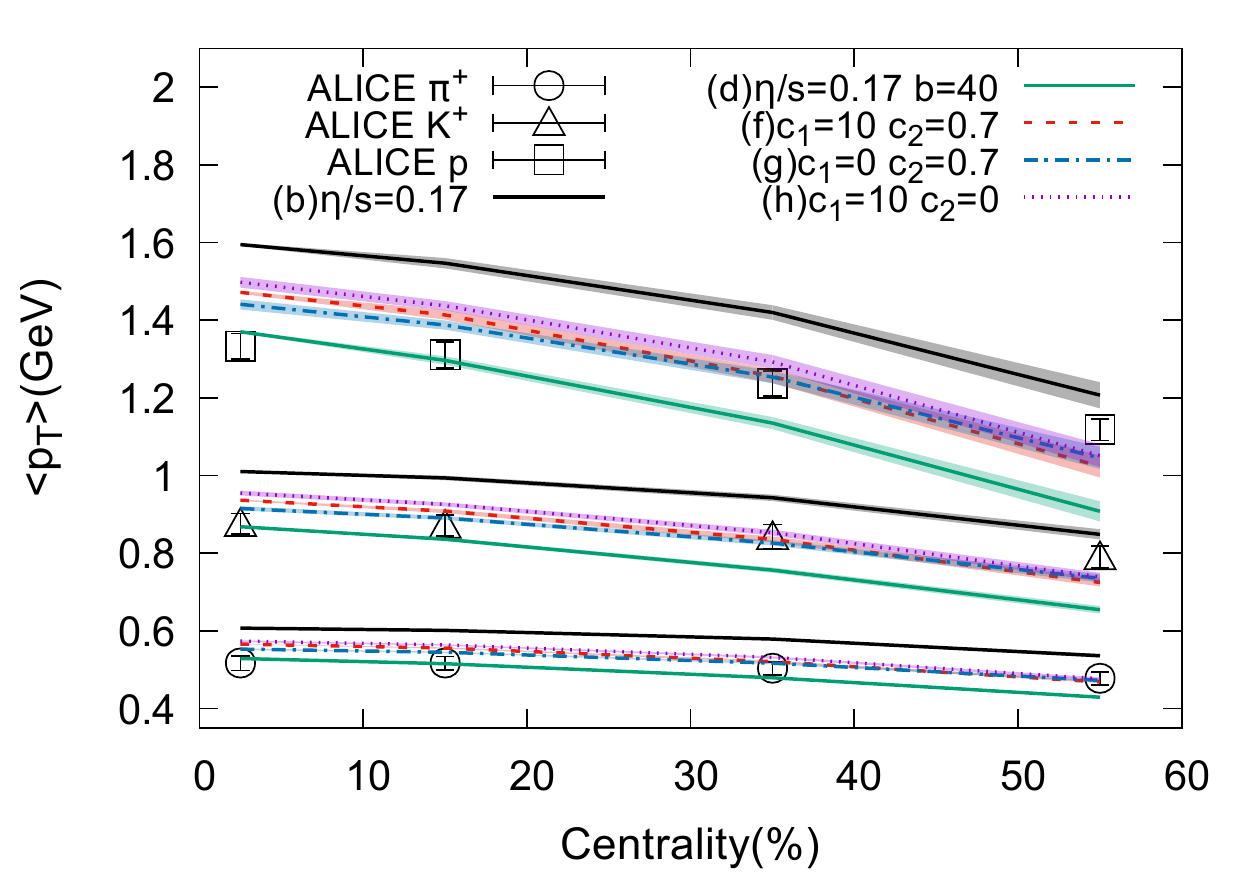}
 \colorcaption{The mean $p_T$ of $\pi^+$ (open circles), $K^+$ (open triangles), 
 and $p$ (open squares) as a function of centrality, together with the ALICE data \cite{Abelev:2013vea}. 
 We evaluate the mean $p_T$ for the whole region of calculated $p_T$, $p_T>0$ GeV. 
\label{fig:meanpt} }
 \end{figure}

To understand the detailed feature of the behavior of $p_T$ spectra, 
we investigate the time evolution of the transverse flow in Fig.~\ref{fig:vT-bulk}. 
From comparison between the red solid line ($\eta/s=0.17$, $N=110)$ and the 
green dashed line ($\eta/s=0.17$, $b=40$, $N=110$) we can see the bulk 
viscosity effect.  
The existence of bulk viscosity delays the growth of the transverse flow until around $\tau=7.0$ fm. 
After $\tau=7.0$ fm, the $\langle v_T \rangle$ of the green dashed line is larger than that of 
the red solid line, in turn. 
On the other hand, if we change the normalization $N$ from $N=110$ to 
$N=94$ to reproduce the rapidity distributions in Fig.~\ref{fig:Ncheta}, almost all time of 
expansion $\langle v_T \rangle$ of  the blue dashed-dotted line ($\eta/s=0.17$, $b=40$, $N=94)$ 
is smaller than that of the red solid line ($\eta/s=0.17$, $\zeta/s=0$). 
Therefore the slope of the $p_T$ distribution with larger bulk viscosity becomes steeper 
than that without bulk viscosity in Fig.~\ref{fig:pT_bulk}.

Furthermore we investigate the bulk viscosity effect in 
the $v_x$ and temperature distributions as a function of $x$ ($y=0$ fm and $\eta_s=0$) at $\tau=7$ fm 
 in 0-5 \% centrality in Fig.~\ref{fig:vx-bulk}. 
 In our parametrization of the bulk viscosity Eq.~(\ref{eq:bulk}), it becomes large below $T<200$ MeV (Fig.~\ref{fig:viscosity}). 
As a result, around $|x| \sim 10$ fm growth of $v_x$ is suppressed and diminution of  
temperature is smaller, in the case of finite bulk viscosity. 
The fraction of fluid elements whose temperature is around the critical temperature 
becomes large. 
On the other hand, 
around $x\sim0$ fm where the temperature is above 200 MeV 
 the bulk viscosity does not affect $v_x$ and temperature distributions.

To understand the bulk viscosity effect in the temperature distributions,  we evaluate the 
profile function \cite{SHURYAK1978150} in Fig.~\ref{fig:profile_ft}, 
\begin{align} \varPhi(T) = \int d^4 x \delta( T(x) - T). 
\end{align}
Here we can see that in the case of finite bulk viscosity the profile functions around the 
critical temperature are enhanced. 
The enhancement may affect physical observables such as  the photon 
and lepton-pair production in the medium.

In Fig.~\ref{fig:vn-bulk} we show the elliptic and triangular flows of 
charged hadrons as a function of $p_T$ in 0-5 \% and 30-40 $\%$ centralities. 
Here we display calculated results of $v_2$ and $v_3$ for $b=40$ and $60$. 
In spite of our limited statistics,  we find the following tendency. 
At low $p_T$, the elliptic flow is smaller with larger bulk viscosity. 
On the other hand, above $p_T\sim 2$  GeV it is larger with larger bulk 
viscosity. 
Because the slope of $p_T$ spectra is steeper with larger bulk viscosity (Fig.~\ref{fig:pT_bulk}), the anisotropy of flow is smaller (larger) at low (high) $p_T$. 
For the triangular flow, we do not find clear dependence of bulk viscosity in 0-5 \% centrality, 
however, in 30-40 \% centrality we observe enhancement of $v_3$ with larger bulk viscosity. 

For the finite bulk viscosity, we obtain the following results. 
The slope of $p_T$ spectra of $\pi^+$, $K^+$, and $p$ becomes steep in the finite bulk viscosity, 
which suggests a small mean $p_T$. The elliptic flow $v_2$ becomes small at low $p_T$, whereas 
above $p_T > 2$ GeV it becomes large. The triangular flow $v_3$ is enhanced for the larger bulk viscosity in 
30-40 \% centrality. 
Furthermore we find the bulk viscosity effect as the enhancement of the profile function around 
the critical temperature which may affect physical observables. 

\begin{figure}[t]
 \centering
 \includegraphics[width=7.5cm]{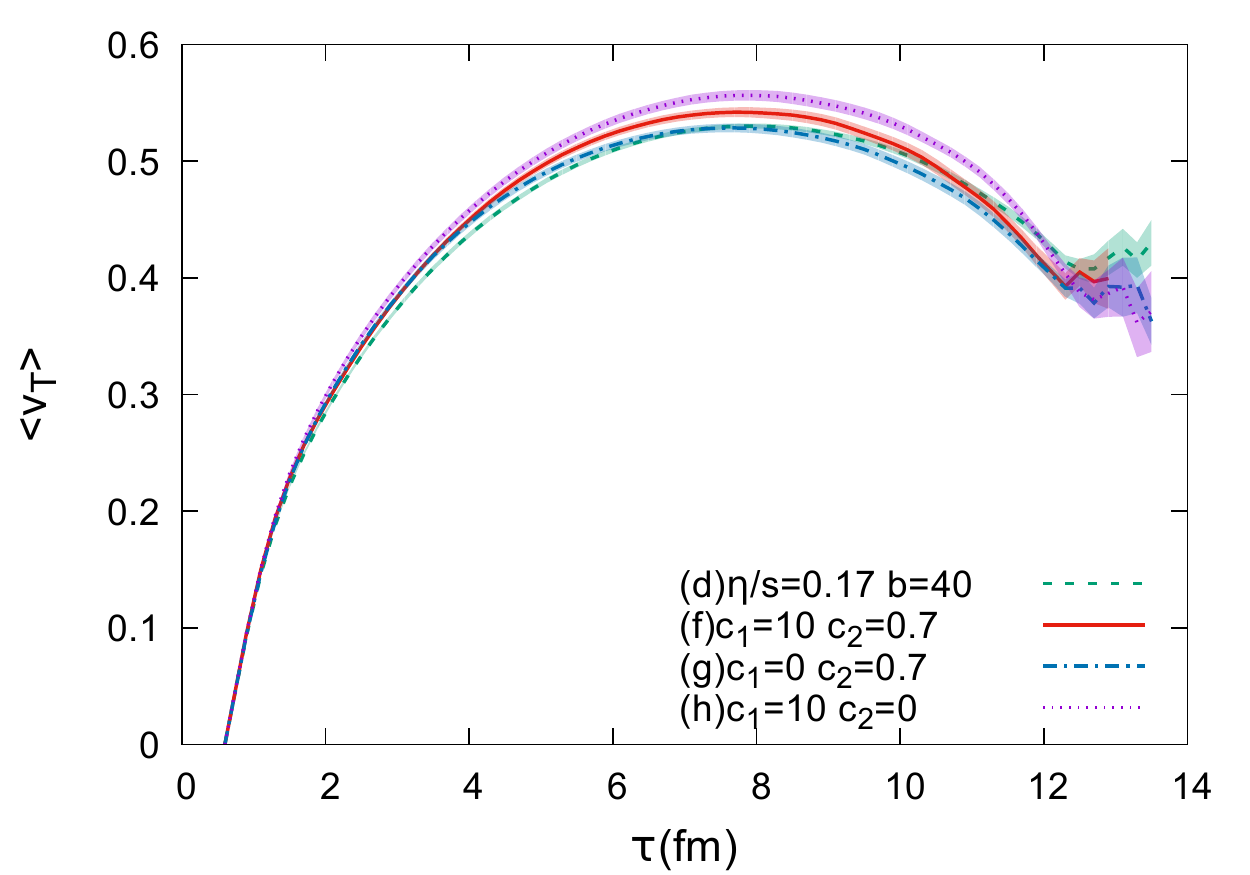}
 \colorcaption{The time evolution of the spatial averaged radial flow $v_T=\sqrt{v_x^2 + v_y^2}$ of fluid cells whose temperatures 
 are  $T>150$ MeV at $\eta_s=0$ in 0-5 \% centrality, for cases (d), (f), (g) and (h).  
 The values are taken from the average over 20 events. 
 \label{fig:vT-etaT}} 
\end{figure}

\subsection{Temperature-dependent $\eta/s(T)$ and $\zeta/s(T)$ \label{subsec:temp-shear}}

\begin{figure*}[t]
\begin{minipage}{0.5\hsize}
 \centering \hspace{1cm}
 \includegraphics[width=7.5cm]{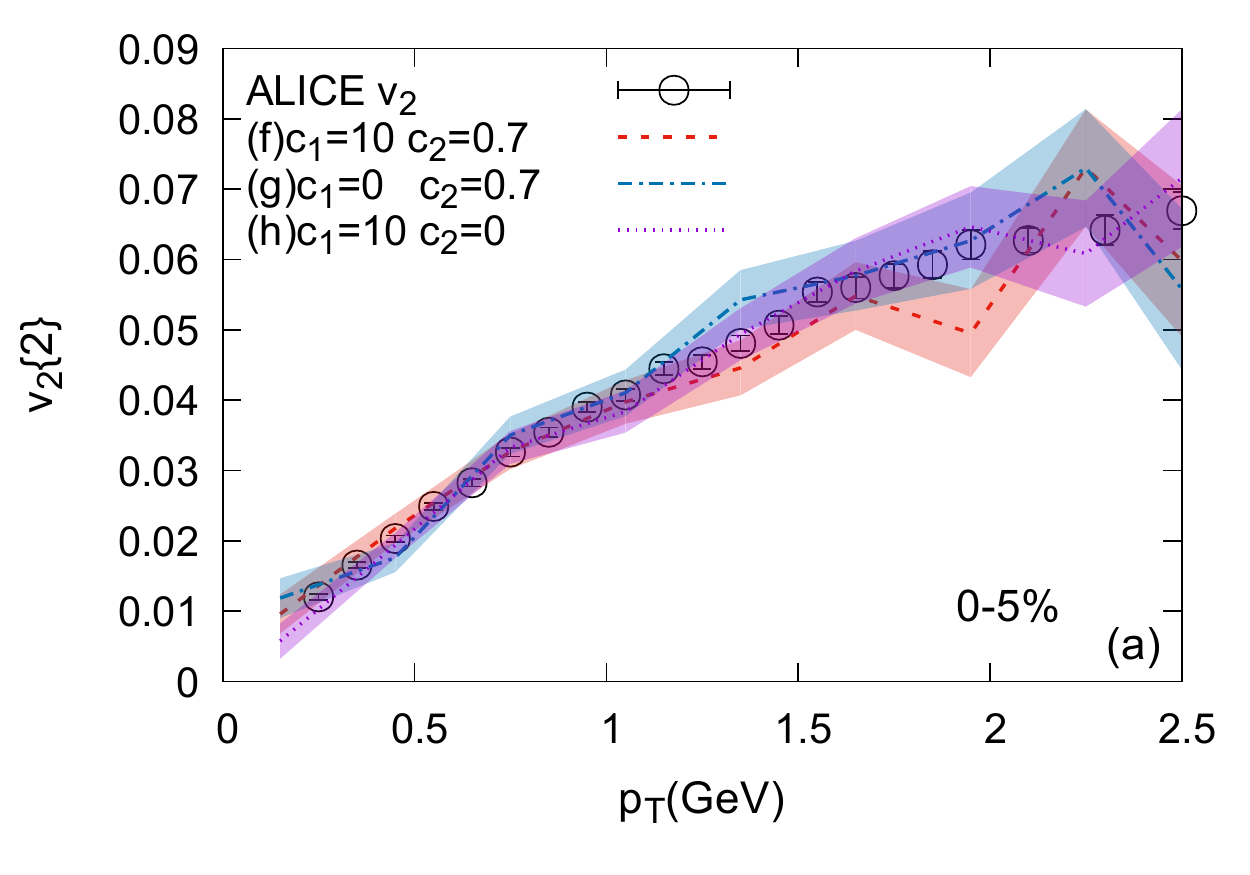}
 \end{minipage}
 \begin{minipage}{0.49\hsize}
 \centering \hspace{-1cm}
 \includegraphics[width=7.5cm]{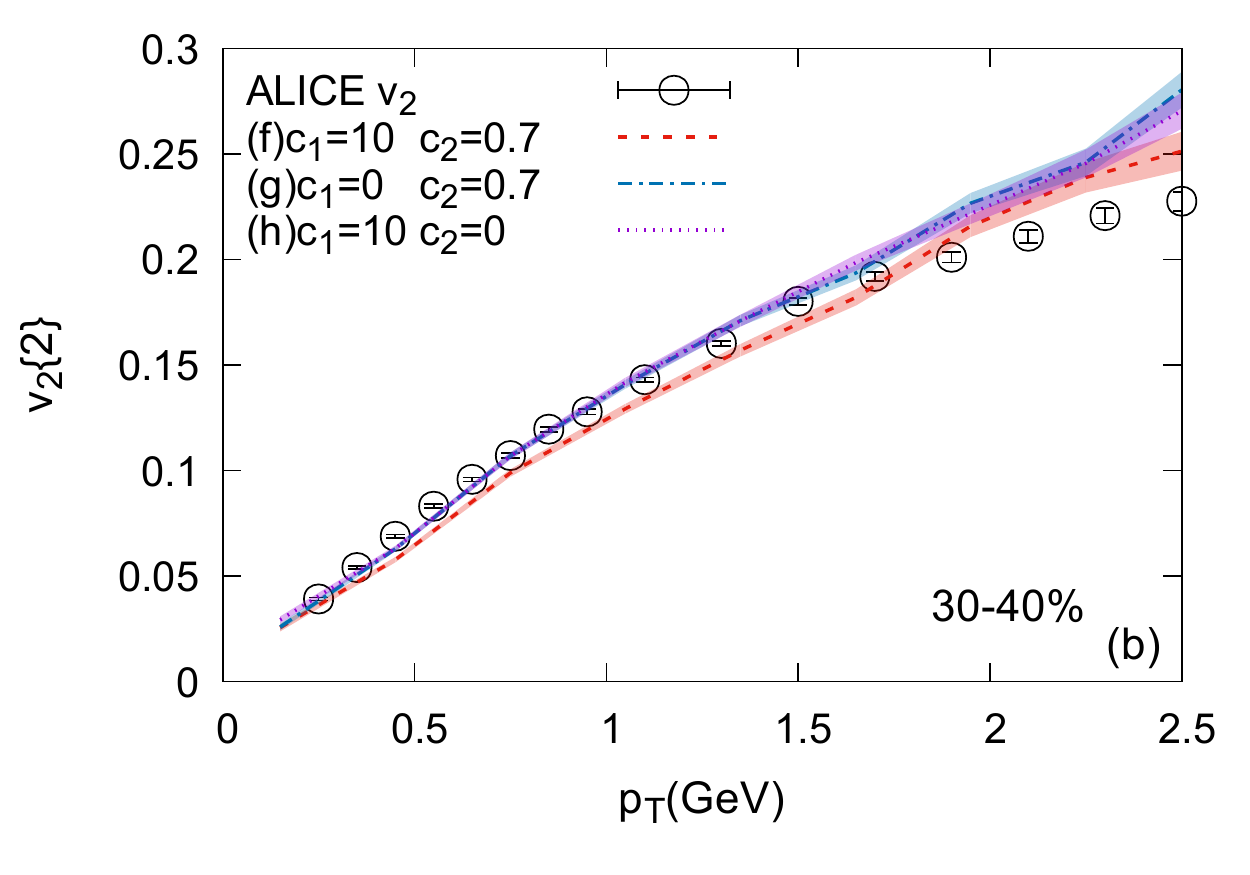}
\end{minipage}
\begin{minipage}{0.5\hsize}
 \centering \hspace{1cm}
 \includegraphics[width=7.5cm]{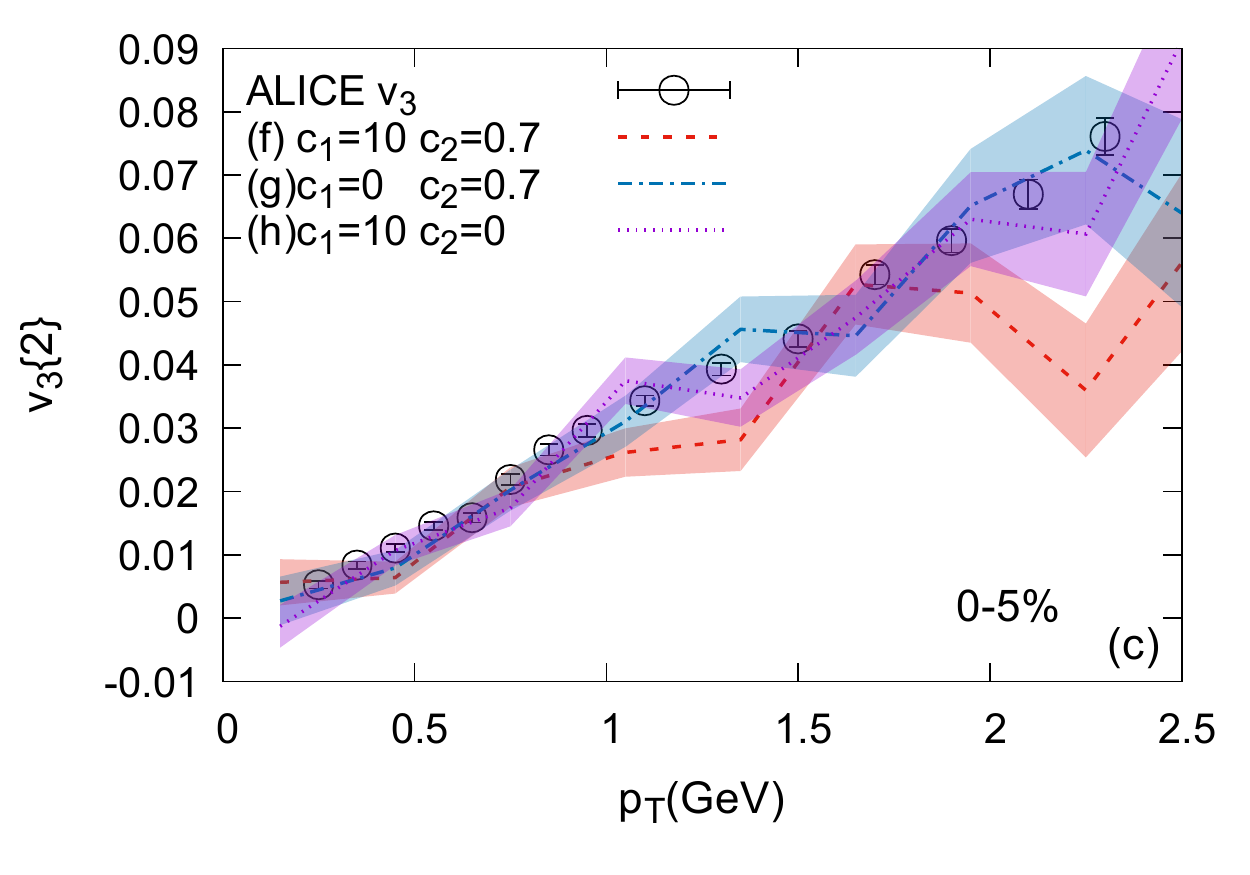}
 \end{minipage}
 \begin{minipage}{0.49\hsize}
 \centering \hspace{-1cm}
 \includegraphics[width=7.5cm]{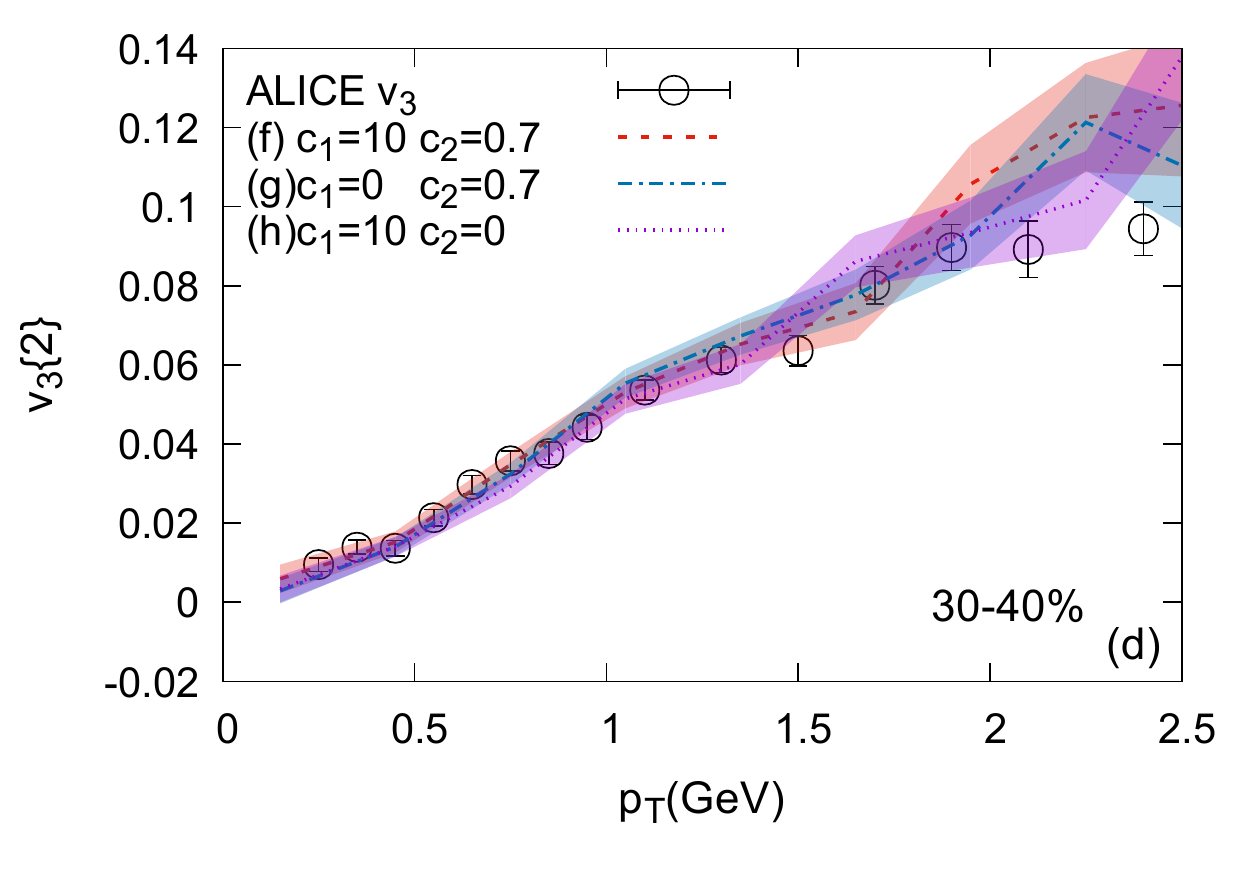}
\end{minipage}
 \colorcaption{
 The elliptic and triangular flows of charged hadrons 
 as a function of $p_T$ in 0-5 $\%$ [(a) for $v_2$ and (c) for $v_3$] and 30-40 $\%$[(b) for $v_2$ and (d) for $v_3$]  centralities, 
 together with the ALICE data (the open circles) \cite{ALICE:2011ab}. 
 The red dashed line, the blue dashed, and purple dotted line 
 stand for 
 (f) $c_1=10$, $c_2=0.7$,  (g) $c_1=0$, $c_2=0.7$, and  (h) $c_1=10$, $c_2=0.7$, respectively.
\label{fig:vn-shearT}}
\end{figure*}
\begin{figure}[h]
 \includegraphics[width=7.5cm]{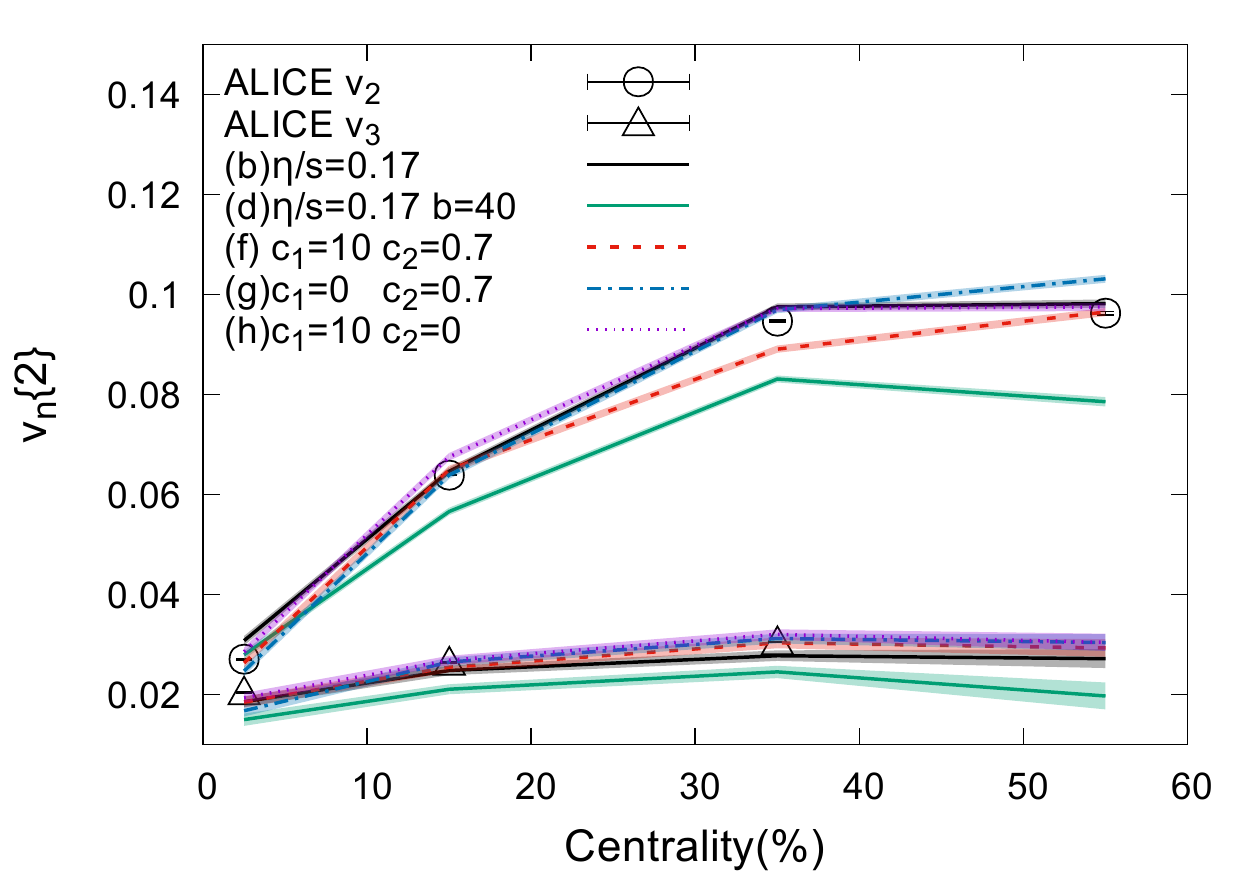}
 \colorcaption{
The integrated $v_2$ and $v_3$ of charged hadrons 
 as a function of centrality in the cases of (b) $\eta/s=0.17$ (the black solid line), 
 (d)$\eta/s=0.17, b=40$ (the green solid line), 
 (f) $c_1=10, c_2=0.7$ (the red dashed line), 
 (g) $c_1=0, c_2=0.7$ (the blue dashed-dotted line), and 
 (h) $c_1=10, c_2=0$ (the purple dotted line).  
 The open circles and squares stand for the $v_2$ and $v_3$ of the ALICE data, respectively \cite{ALICE:2011ab}. 
\label{fig:v2_centrality}} 
\end{figure}

We investigate the temperature dependence of shear and bulk viscosities from 
comparison with the ALICE data. 
Because the bulk viscous correction to the distribution function is neglected, 
there should be uncertainties in determination of the value of $b$ itself. 
Here we use $b=40$ as one of the possible values for $b$. 
We shall discuss the consequences of different temperature-dependent $\eta/s(T)$ and $\zeta/s(T)$ models. 
First in Fig.~\ref{fig:pT_shearT} we show the $p_T$ distributions for 
$\pi^+$, $K^+$, and $p$ in 0-5 $\%$, 10-20 $\%$, 30-40 $\%$, and 50-60 $\%$ centralities, 
together with the ALICE data.  From the $p_T$ spectra we can mainly extract the bulk viscosity effect. 
The $p_T$ spectra for (f), (g), and (h) are almost identical, which means that 
the bulk viscosity effect during hydrodynamic expansion in the three cases is the same. 
Also, compared with those in Fig.~\ref{fig:pT_bulk}, the slope of our computed $p_T$ spectra becomes 
flat and shows better agreement with experimental data. 
Introduction of the temperature dependence of $\eta/s$ reduces the average value of 
$\zeta/s$ during hydrodynamic expansion through Eq.~(\ref{eq:bulk}) (Fig.~\ref{fig:viscosity}). 

\begin{figure*}[t!]
 \begin{minipage}{0.5\hsize}
 \centering \hspace{1cm}
 \includegraphics[width=7.3cm]{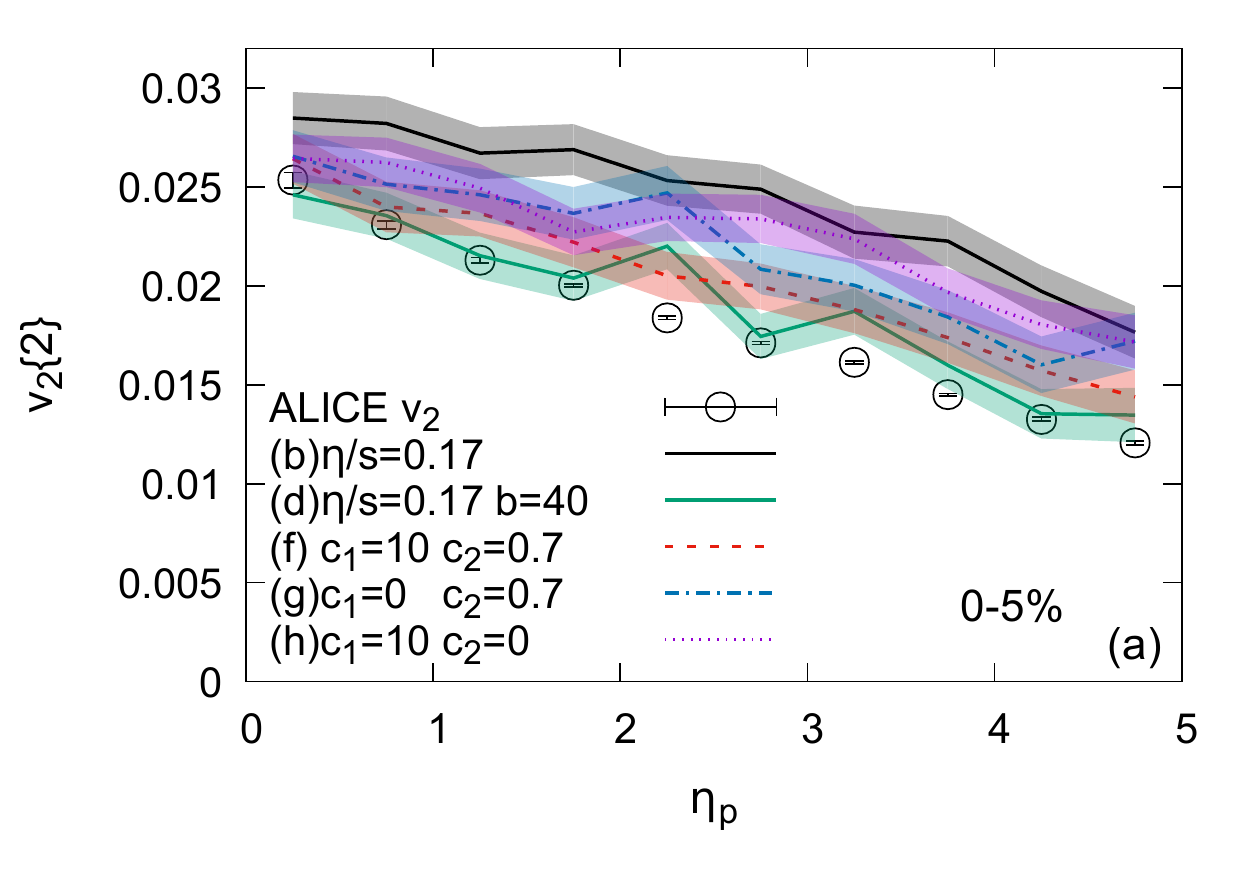}
 \end{minipage}
 \begin{minipage}{0.49\hsize}
 \centering \hspace{-1cm}
 \includegraphics[width=7.3cm]{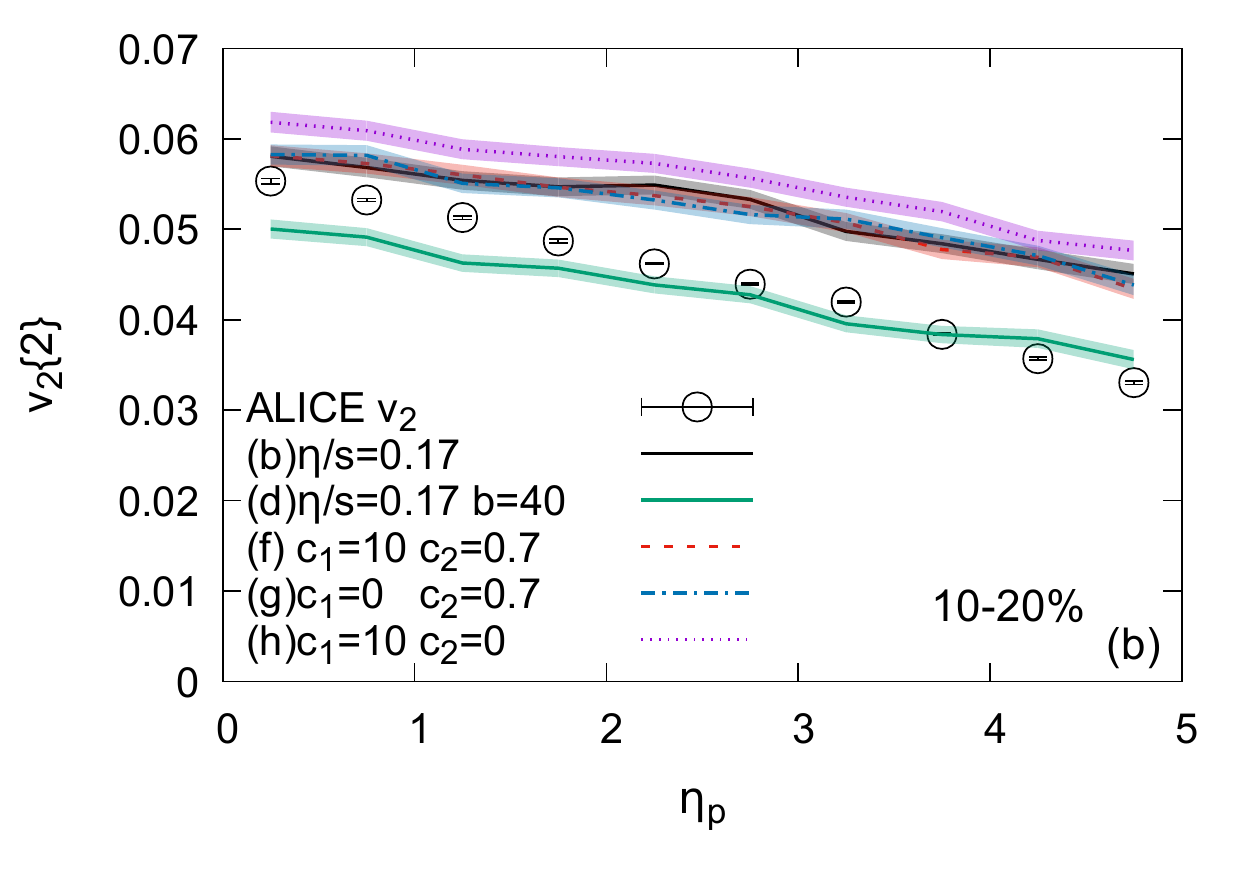}
 \end{minipage}
 \begin{minipage}{0.5\hsize}
 \centering \hspace{1cm}
 \includegraphics[width=7.3cm]{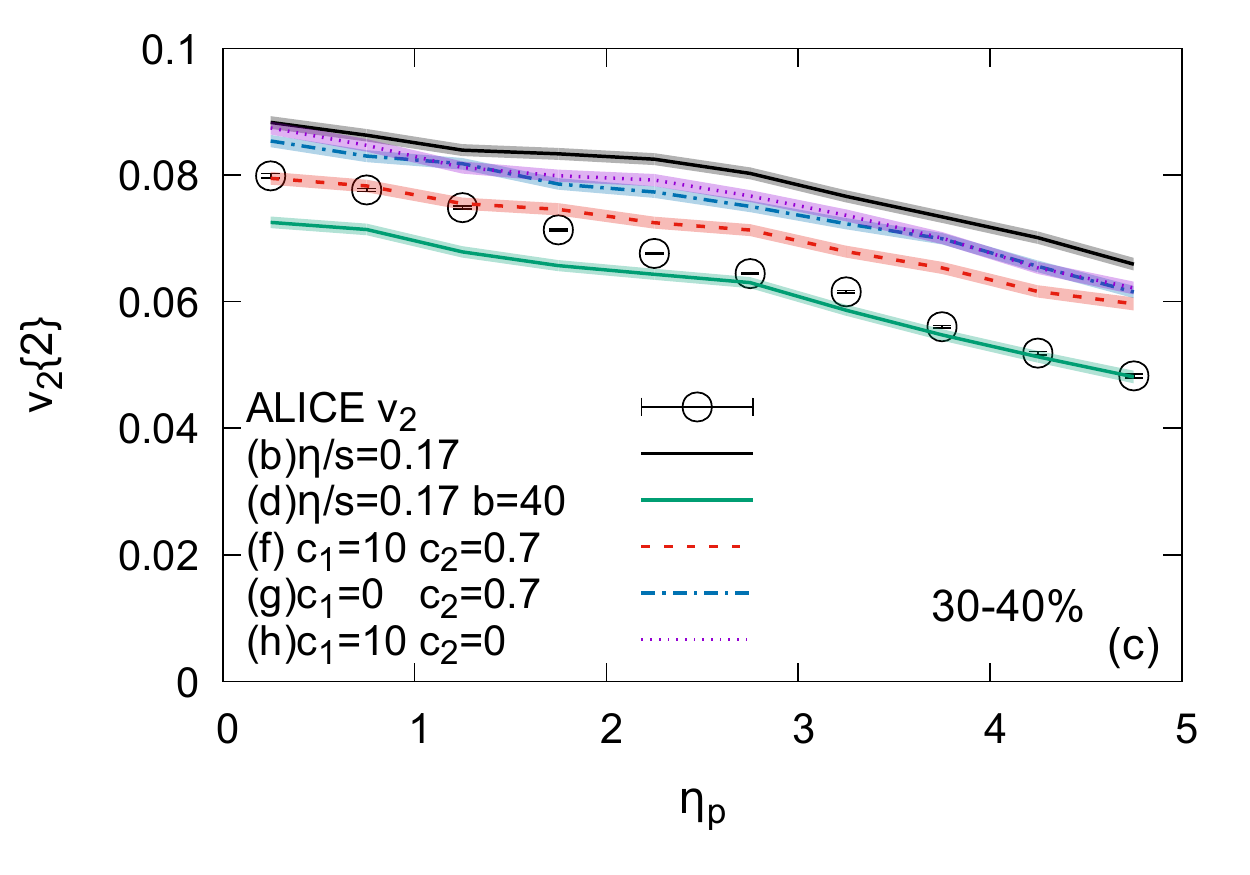}
 \end{minipage}
 \begin{minipage}{0.49\hsize}
 \centering \hspace{-1cm}
 \includegraphics[width=7.3cm]{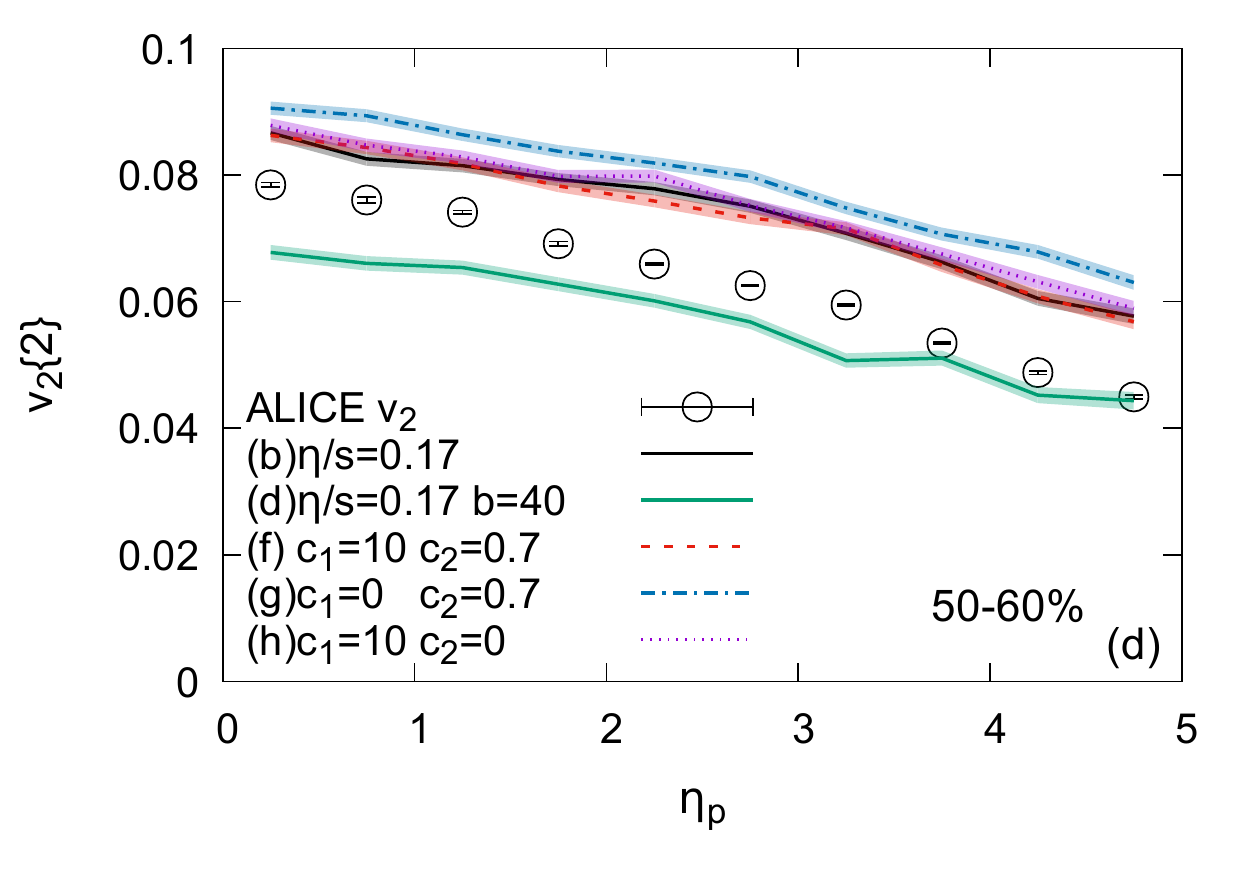}
 \end{minipage}
 \colorcaption{Calculated results of integrated $v_2$ as a function of pseudorapidity in centralities 0-5 \% (a),  
 10-20 \% (b), 30-40 \% (c), and 50-60 \% (d),  
 in the cases of (b) $\eta/s=0.17$ (the black solid line), 
 (d)$\eta/s=0.17, b=40$ (the green solid line), 
 (f) $c_1=10, c_2=0.7$ (the red dashed line), 
 (g) $c_1=0, c_2=0.7$ (the blue dashed-dotted line), and 
 (h) $c_1=10, c_2=0$ (the purple dotted line), together with the ALICE data (the open circles)  
 \cite{Adam:2016ows}.  
 \label{fig:v2eta}} 
\end{figure*}
\begin{figure*}[t!]
 \begin{minipage}{0.5\hsize}
 \centering \hspace{1cm}
 \includegraphics[width=7.3cm]{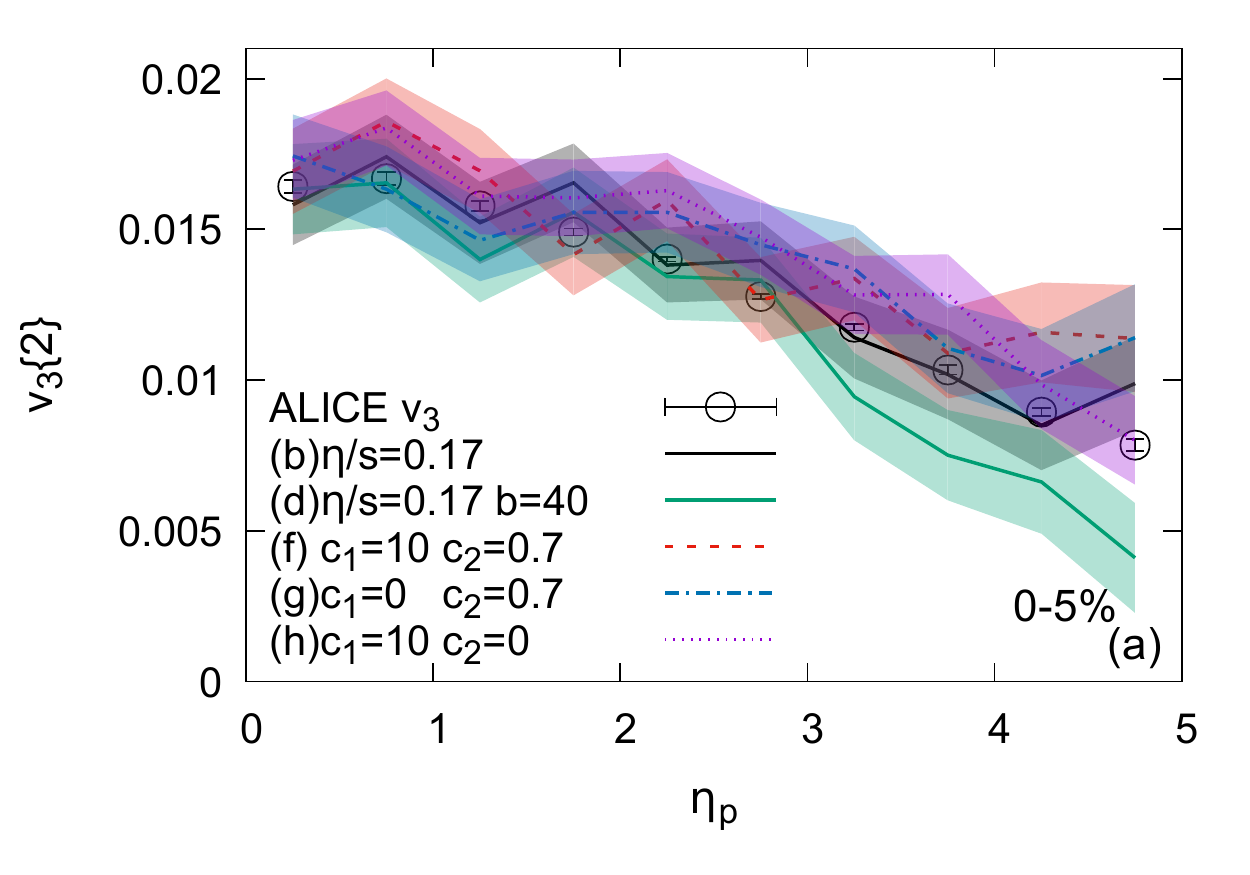}
 \end{minipage}
 \begin{minipage}{0.49\hsize}
 \centering \hspace{-1cm}
 \includegraphics[width=7.3cm]{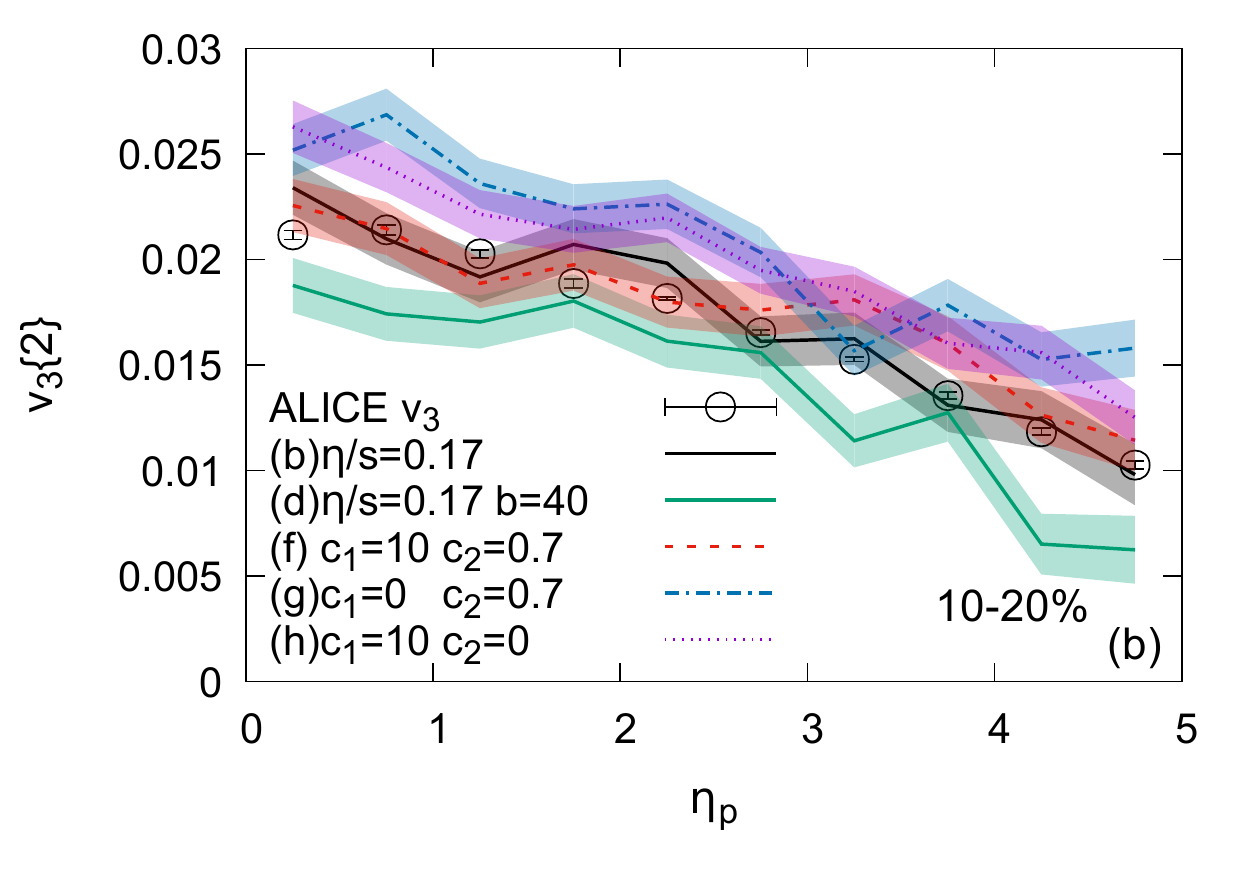}
 \end{minipage}
 \begin{minipage}{0.5\hsize}
 \centering \hspace{1cm}
 \includegraphics[width=7.3cm]{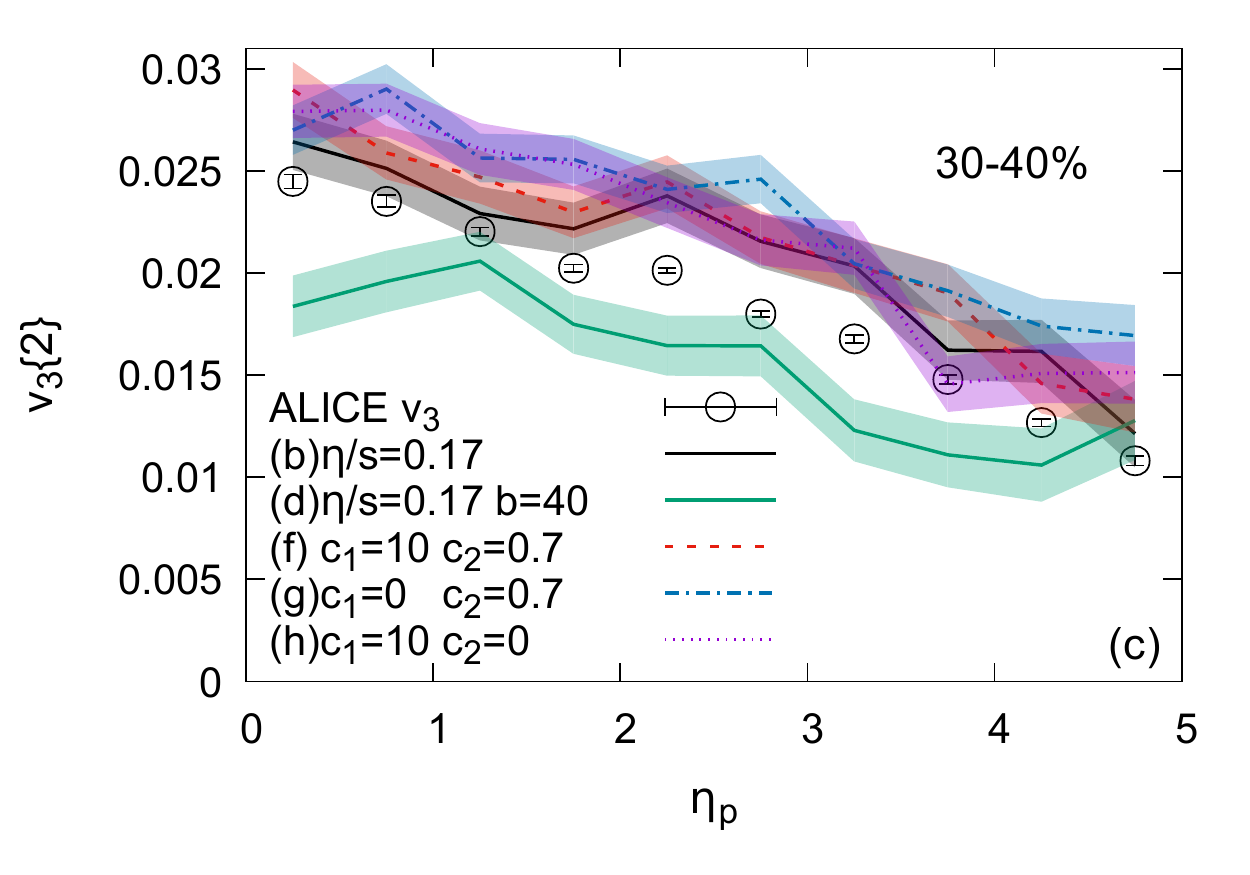}
 \end{minipage}
 \begin{minipage}{0.49\hsize}
 \centering \hspace{-1cm}
 \includegraphics[width=7.3cm]{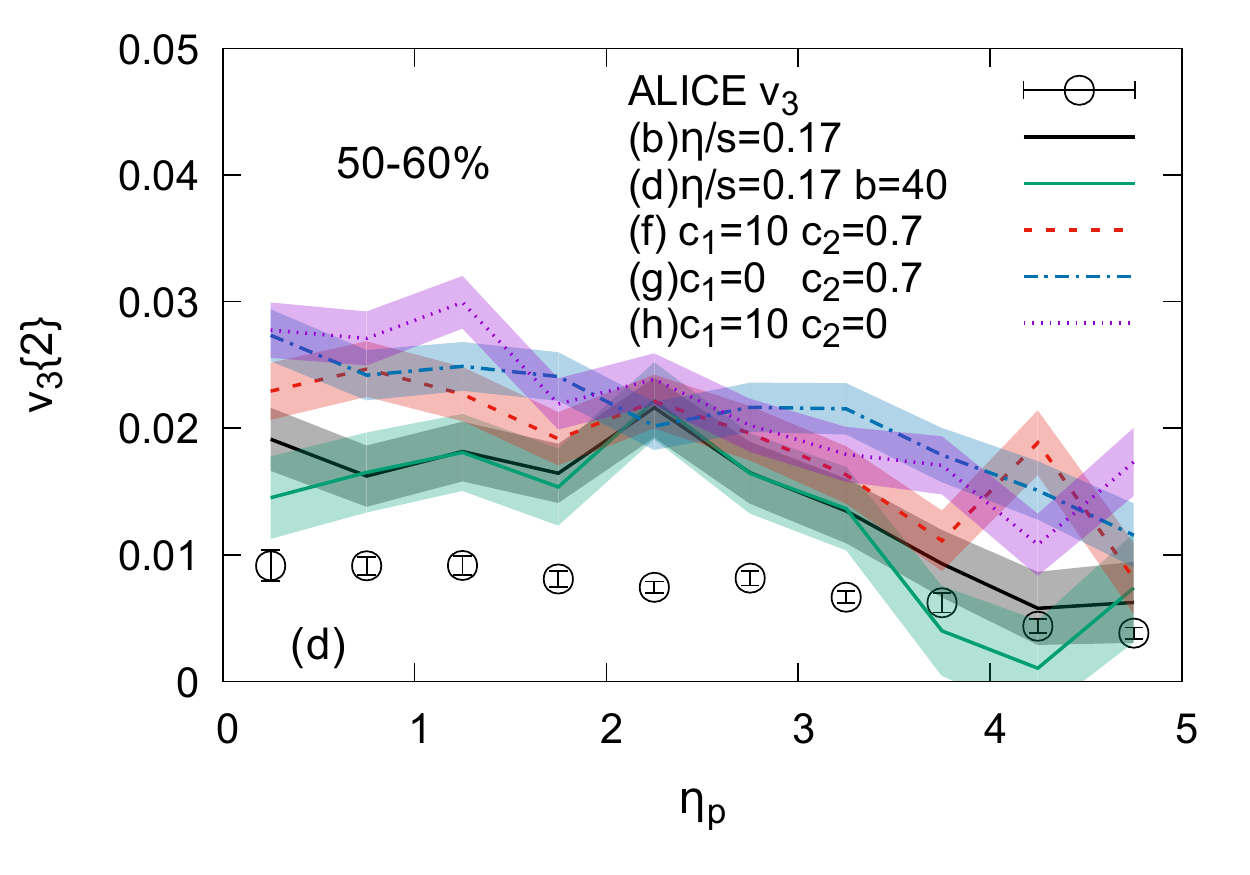}
 \end{minipage}
 \colorcaption{
Calculated results of integrated $v_3$ as a function of pseudorapidity in centralities 0-5 $\%$ (a), 
 10-20 \% (b), 30-40 $\%$ (c), and 50-60 \% (d) , 
  in the cases of (b) $\eta/s=0.17$ (the black solid line) 
 (d)$\eta/s=0.17,b=40$ (the green solid line), 
 (f) $c_1=10, c_2=0.7$ (the red dashed line), 
 (g) $c_1=0, c_2=0.7$ (the blue dashed-dotted line), and 
 (h) $c_1=10, c_2=0$ (the purple dotted line),   together 
 with the ALICE data \cite{Adam:2016ows}. 
 \label{fig:v3eta}} 
\end{figure*}

In Fig.~\ref{fig:meanpt}, we show the centrality dependence of mean 
$p_T$ of $\pi^+$, $K^+$ and $p$, together with the ALICE data \cite{Abelev:2013vea}.
All computational results show 
reasonable agreement with experimental data, except for the case of constant $\eta/s=0.17$ with vanishing bulk viscosity. 
In the case of (b), our calculated values of $\langle p_T\rangle$ are larger than experimental data, 
which implies that in our calculation radial flow grows stronger. 
In particular, a large deviation from experimental data exists in radial flow of protons. 
Centrality dependence of mean $p_T$ of our computational results shows steeper decrease with centrality, compared with experimental data. 
If bulk viscosity is included, our calculated results of (d), (f), (g), and (h) 
become close to experimental data, because bulk viscosity suppresses 
the growth of radial flow \cite{Rose:2014fba, Ryu:2015vwa}.
Our calculations of (d) show good agreement with experimental data up to 
centrality 10-20 $\%$; however, they deviate from the experimental data at peripheral 
collisions. 
In peripheral collisions the suppression due to bulk viscosity is too strong and 
the mean $p_T$ shows rapid decrease with centrality. 
We find that the mean $p_T$ is not sensitive to the differences of 
the temperature dependence of $\eta/s$ [cases (f), (g), and (h)], 
as expected  from $p_T$ spectra of $\pi^+$, $K^+$, and $p$. 

We examine the time evolution of the transverse flow in 0-5 \% centrality in Fig.~\ref{fig:vT-etaT}. 
Here we focus on four cases (d), (f), (g) and (h). 
Until $\tau \sim 8$ fm the order of the magnitude of $\langle v_T \rangle$ is the same as that 
of mean $p_T$: (h), (f), (g) and (d) in descending order. 
After  $\tau \sim 8$ fm $\langle v_T \rangle$ of case (d) 
becomes larger than that of case (g). It suggests that mean $p_T$ is determined mainly by 
the early stage of the expansion.

Figure \ref{fig:vn-shearT} shows the elliptic and triangular flows of 
charged hadrons as a function of $p_T$ in 0-5 $\%$ 
and 30-40 $\%$ centralities. 
From collective flows we can investigate both shear and bulk viscosities. 
The results of $p_T$ spectra and mean $p_T$ suggest that the cases (f), (g), and (h) have 
almost the same bulk viscosity effect during hydrodynamic expansion. 
Therefore we can understand the temperature dependent shear-viscosity effect on the collective flows. 
In both centralities, the elliptic flow of case (f)  is the 
smallest among three cases. 
In the case of (f) the average value of shear viscosity over fluid cells is largest, which 
leads the smallest amplitude of elliptic flow among them. 
There are small differences between cases (g) and (h) in the elliptic flows. 
This is because in the current parametrization of $\eta/s(T)$ 
the average values of shear viscosity are almost the same. 
Furthermore, compared with Fig.~\ref{fig:vn-shear}, enhancement of $v_2$ due to finite bulk viscosity 
is observed above $p_T > 2$ GeV. 
For the triangular flow we find the same tendency in 0-5 \% centrality, in spite of errors. 
In 30-40 \% centrality all (f), (g), and (h) cases show good agreement with the 
experimental data, which is realized by enhancement of $v_3$ due to finite bulk viscosity.

Figure \ref{fig:v2_centrality} shows centrality dependence of integrated $v_n$ ($n=2,3$). 
The results of case (b) are consistent with the ALICE data \cite{ALICE:2011ab}.  
However this agreement is accidentally achieved by larger $\langle p_T\rangle$ and smaller $v_2(p_T)$. 
The integrated $v_n$ is affected not only by $v_n(p_T)$ but also by $p_T$ spectra 
through integration over $0.2<p_T<5$ GeV. 
The computed $v_2$ for the case (d) is smaller than experimental data except for the central collision, 
because the mean $p_T$ is smaller than the experimental data especially in peripheral collisions. 
Cases (f), (g), and (h) show good agreement with the experimental data, which is 
a consequence of consistent with experimental data for $v_2(p_T)$ and $p_T$ spectra. 
In 50-60 \% centrality $v_2$ of case (g) is larger than experimental data, which suggests that 
in the centrality the shear viscosity of the hadronic phase is important. 

We examine the viscosity effects of elliptic flows as a function of pseudorapidity  for 
0-5 \%, 10-20 \%, 30-40 \%, and 50-60 \% centralities in Fig.~\ref{fig:v2eta}. 
If we input only the constant $\eta/s$ in the calculations, 
our results of  $v_2$ are larger than experimental data and 
the slope of $v_2(\eta_p)$ at peripheral collisions is gentler than that of the ALICE data \cite{Adam:2016ows}. 
On the other hand, if we add bulk viscosity [case (d)], the absolute value of $v_2(\eta_p)$ becomes 
small and approaches to the experimental data. 
Because in the computation of $v_2(\eta_p)$ $p_T$ integration is performed, 
the mean $p_T$ for cases (b) and (d) affects the amplitude of $v_2(\eta_p)$. 
For cases (f), (g), and (h), their centrality dependence of mean $p_T$ as well as 
behavior of $v_2(p_T)$ are almost the same, which means that 
there should be small differences among them in the behavior of $v_2(\eta_p)$. 
However we find an interesting viscosity effect in the centrality dependence of $v_2(\eta_p)$. 
For the three cases the average bulk viscosity effect during hydrodynamic expansion is almost the same. 
It means that different behavior of $v_2(\eta_p)$ among them originates mainly from different 
temperature dependence of shear viscosity. 
In 10-20 \% centrality, $v_2(\eta_p)$ of case (h) becomes the largest. In the centrality 
shear viscosity of the QGP phase is dominant. 
In 30-40 \% centrality, $v_2(\eta_p)$ of case (f) shows the smallest value, which suggests that 
shear viscosity of the QGP and hadronic phases important. 
In 50-60 \% centrality $v_2(\eta_p)$ of case (g) becomes the largest, which suggests that 
shear viscosity of the hadronic phase is important \cite{Molnar:2014zha, Denicol:2015nhu}. 
The centrality dependence of $v_2(\eta_p)$ reveals the detailed temperature dependence of $\eta/s$ and $\zeta/s$. 
The deviation between our results and the experimental data becomes large at forward 
rapidity.

\begin{figure*}[t]
 \begin{minipage}{0.5\hsize}
 \centering \hspace{1cm}
 \includegraphics[width=7.3cm]{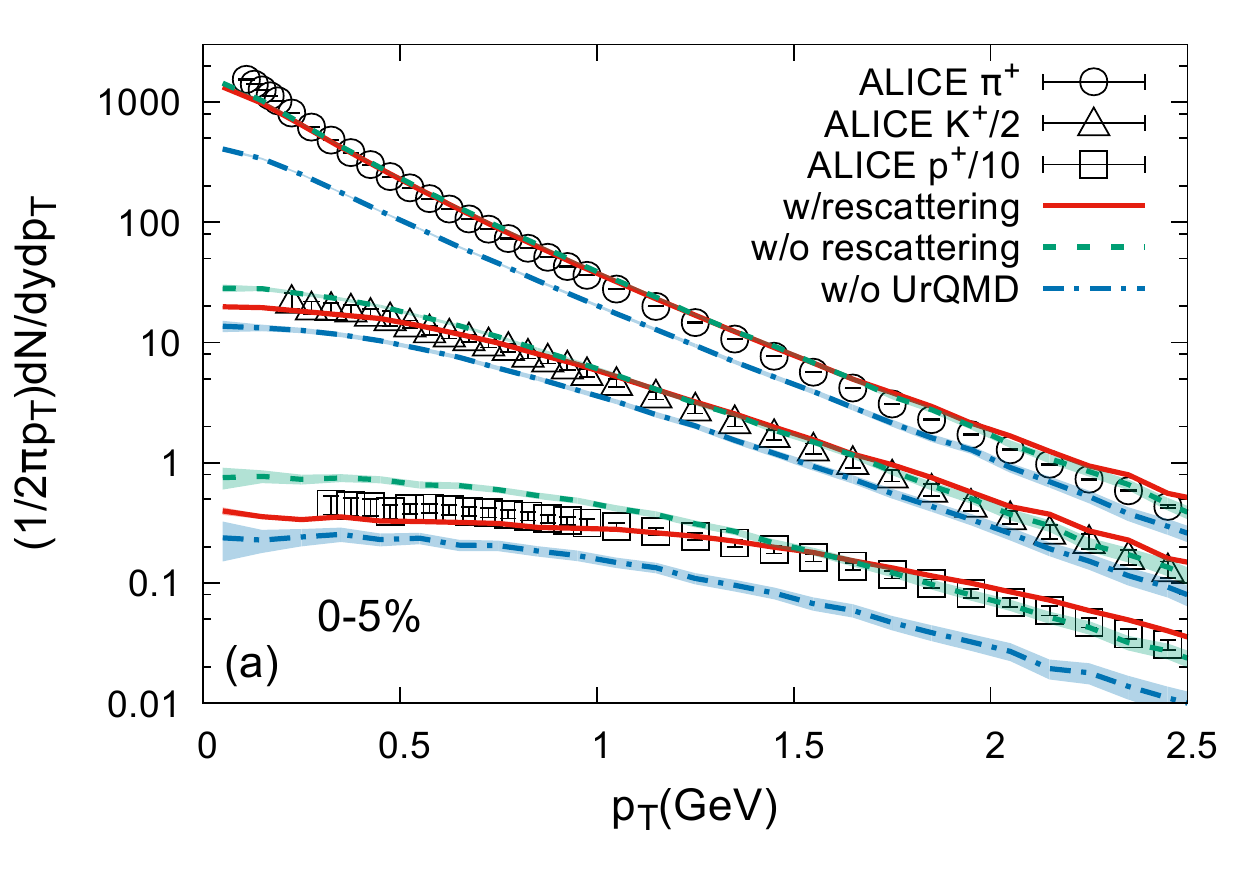} 
 \end{minipage}
 \begin{minipage}{0.49\hsize}
 \centering \hspace{-1cm}
 \includegraphics[width=7.3cm]{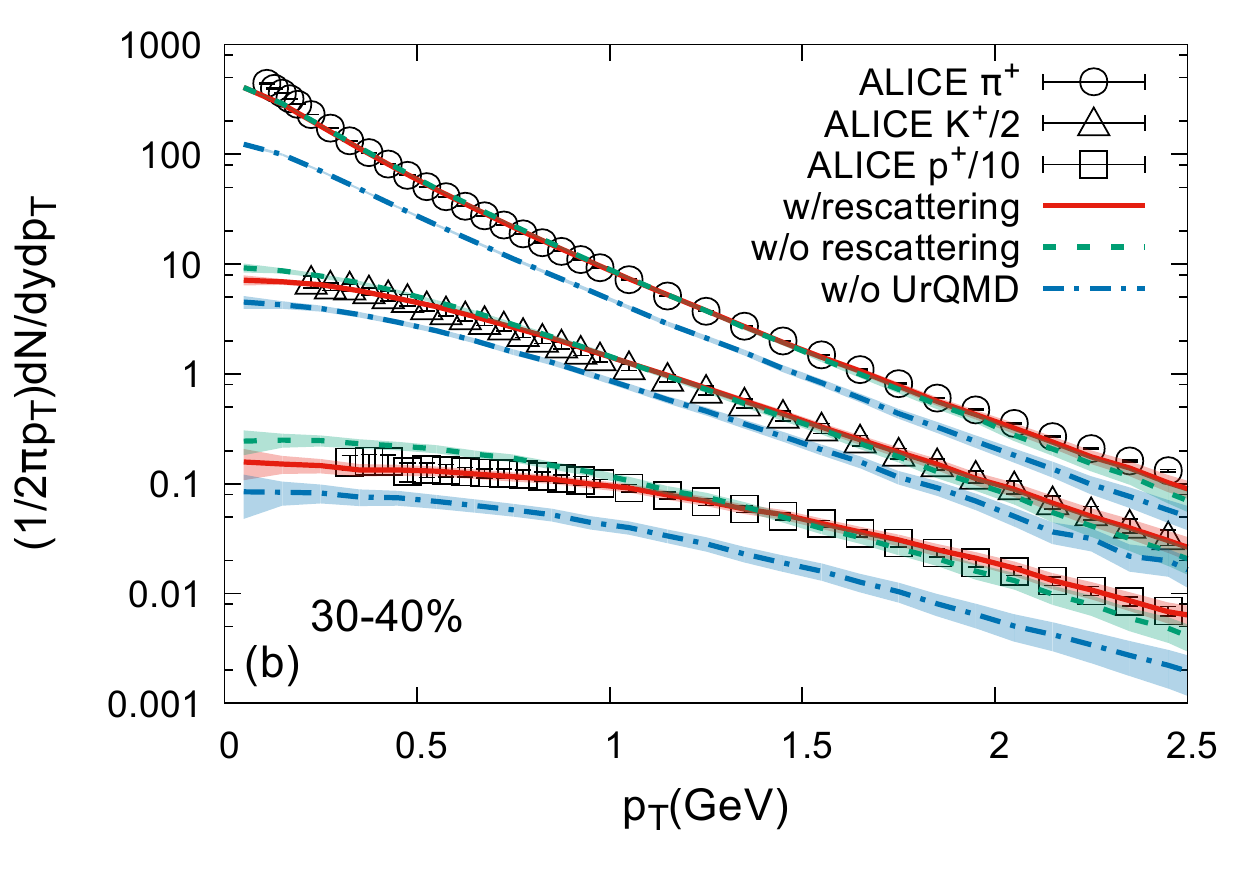}
 \end{minipage}
\colorcaption{
 The $p_T$ distributions for $\pi^+$, $K^+$, and $p$ 
 in 0-5 $\%$ (a)  and 
 30-40 $\%$ (b) centralities, 
 together with the ALICE data (the open circles)  \cite{Abelev:2013vea}. 
The red solid line, the green dashed line, and the blue dashed-dotted line 
stand for computed $p_T$ spectra with rescattering, without rescattering 
(only resonance decay), and without UrQMD (just from hydrodynamic evolution), 
respectively. \label{fig:pT_fsi} 
}
\end{figure*}

Here we make a comment on behavior of $v_2(\eta_p)$ at RHIC, which also shows rapid decrease at 
forward and backward $\eta_p$. It is understood by introduction of temperature dependent shear viscosity 
in the hadronic phase \cite{Denicol:2015nhu}. 
On the other hand, in our results for the LHC we do not find the clear difference between 
the slope of $v_2(\eta_p)$ of cases (b) and (h), because in our parametrization the average value of shear viscosity during hydrodynamic 
expansion between the two cases is almost the same. 
The shape of the $v_2(\eta_p)$ is determined  not only by the shear viscosity of the hadronic phase 
but also by the average value of shear viscosity of the fluid. 
Furthermore since $v_2(\eta_p)$ is evaluated by the correlation between particles at midrapidity and 
those at forward rapidity, decorrelation between them may need to be considered. 

\begin{figure*}[t]
 \begin{minipage}{0.5\hsize}
 \centering \hspace{1cm}
 \includegraphics[width=7.3cm]{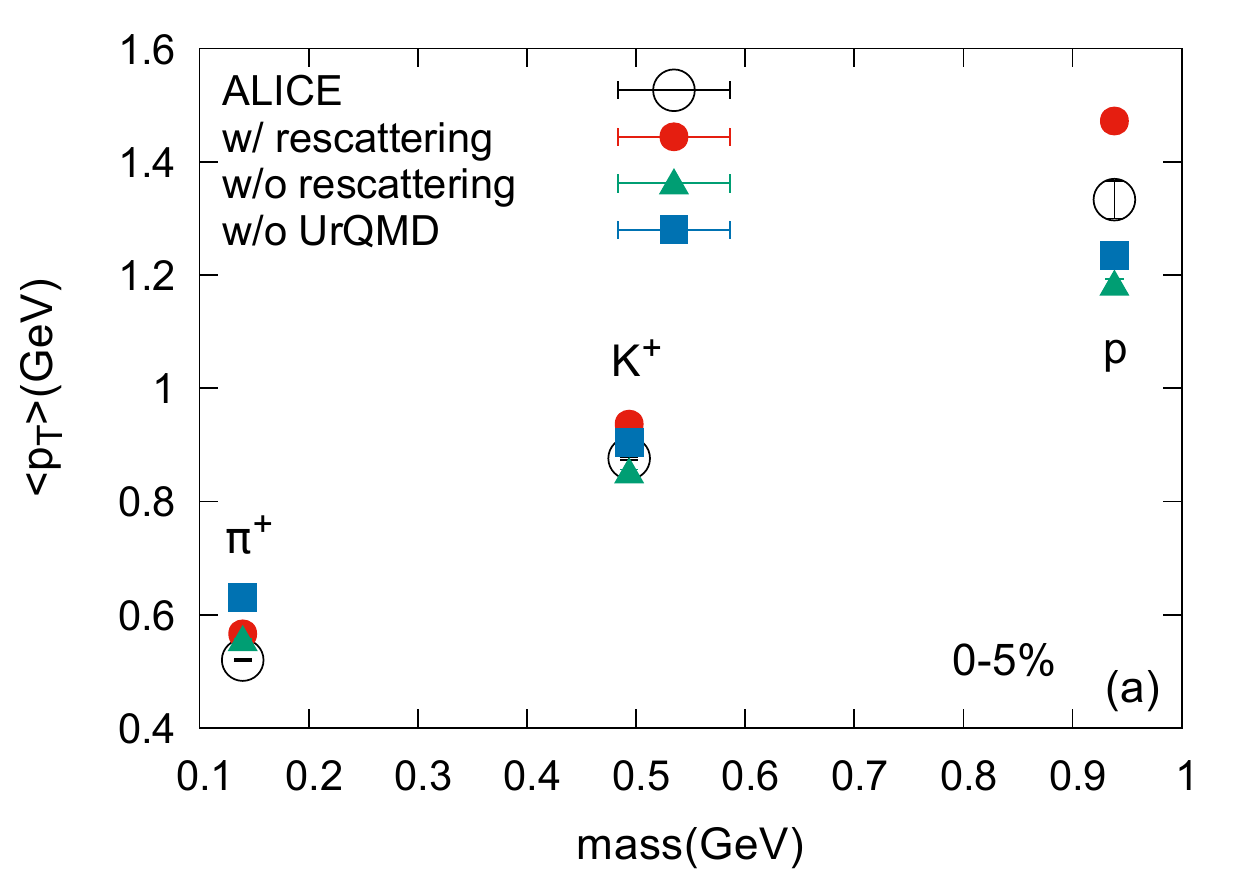} 
 \end{minipage}
 \begin{minipage}{0.49\hsize}
 \centering \hspace{-1cm}
 \includegraphics[width=7.3cm]{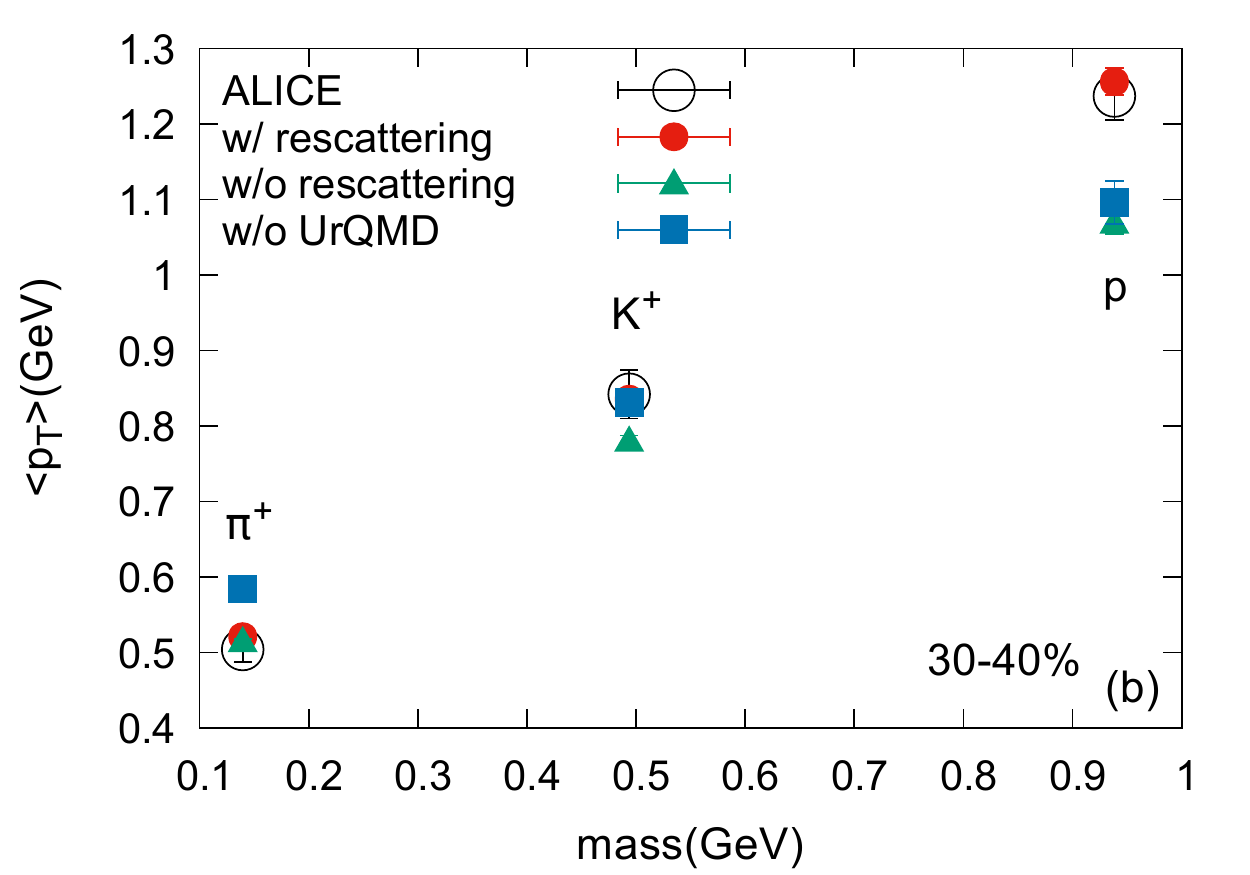}
 \end{minipage}
\colorcaption{The mean $p_T$ for $\pi^+$, $K^+$, and $p$ with rescattering 
(the red solid circles), without rescattering (the green solid triangles), 
and without UrQMD (the blue solid squares) in 0-5 \% (a) and 
30-40 \% centralities (b), 
together with the ALICE data (the open circles)  \cite{Abelev:2013vea}.
We evaluate the mean $p_T$ in the region of $p_T > 0$ GeV. 
\label{fig:mean-PT}
} 
\end{figure*}

In Fig.~\ref{fig:v3eta} we show $v_3$ of charged hadrons as a function of pseudorapidity 
in 0-5 \%, 10-20 \%, 30-40 \%, and 50-60 \% centralities. 
The errors in the triangular flows are larger than those in the elliptic flows. 
The case (b) shows good agreement with the experimental data; however, this agreement 
is realized from the combination of large mean $p_T$ (Fig.~\ref{fig:meanpt}) and 
small $v_3(p_T)$ (Fig.~\ref{fig:vn-shear}).  
The smaller value of $v_3(\eta_p)$ of case (d) comes from smaller mean $p_T$. 
Due to the large errors in the triangular flows cases (f), (g), and (h) are not distinguishable. 
To reach the conclusive results for the temperature-dependence of $\eta/s$ 
from $v_3(\eta_p)$, we need to perform calculations with more statistics. 
In 50-60 \% centrality $v_3(\eta_p)$ for all cases are larger than the experimental data. 

For a brief summary, the $p_T$ spectra are not 
sensitive to the parametrization $\eta/s(T)$, but their slopes depend on the 
average value of $\zeta/s(T)$ during 
the hydrodynamic expansion: they are steeper with larger $\zeta/s$. 
From the amplitude of mean $p_T$, we can extract the bulk viscosity effect. 
The collective flows $v_2$ and $v_3$ are affected by both shear and bulk viscosities. 
The average value of shear viscosity reduces the amplitude of $v_2$ and $v_3$, 
whereas the bulk viscosity reduces $v_2$ and $v_3$ at low $p_T$ and enhances them 
at high $p_T$, which is related to the change of $p_T$  slope. 
Our parametrization of (f), (g), and (h) shows the best agreement with the experimental data compared 
with other cases [(b) and (d)], which implies that the parametrization $\eta/s$ and $\zeta/s$ in (f), (g), and (h) is 
one of the most suitable choices. 
Furthermore we point out that the agreement with the experimental data found in $\langle v_n \rangle$ with 
constant $\eta/s$ is realized just from the combination of smaller $v_2(p_T)$ and larger $\langle p_T \rangle$. 
We find the effect of the temperature dependence of $\eta/s$ in the centrality dependence of $v_n$: 
in the central (peripheral) collisions, viscosities of the hadronic (QGP) phase becomes important. 

\begin{figure*}[t]
 \centering
 \begin{minipage}{0.5\hsize}
 \centering \hspace{1cm}
 \includegraphics[width=7.3cm]{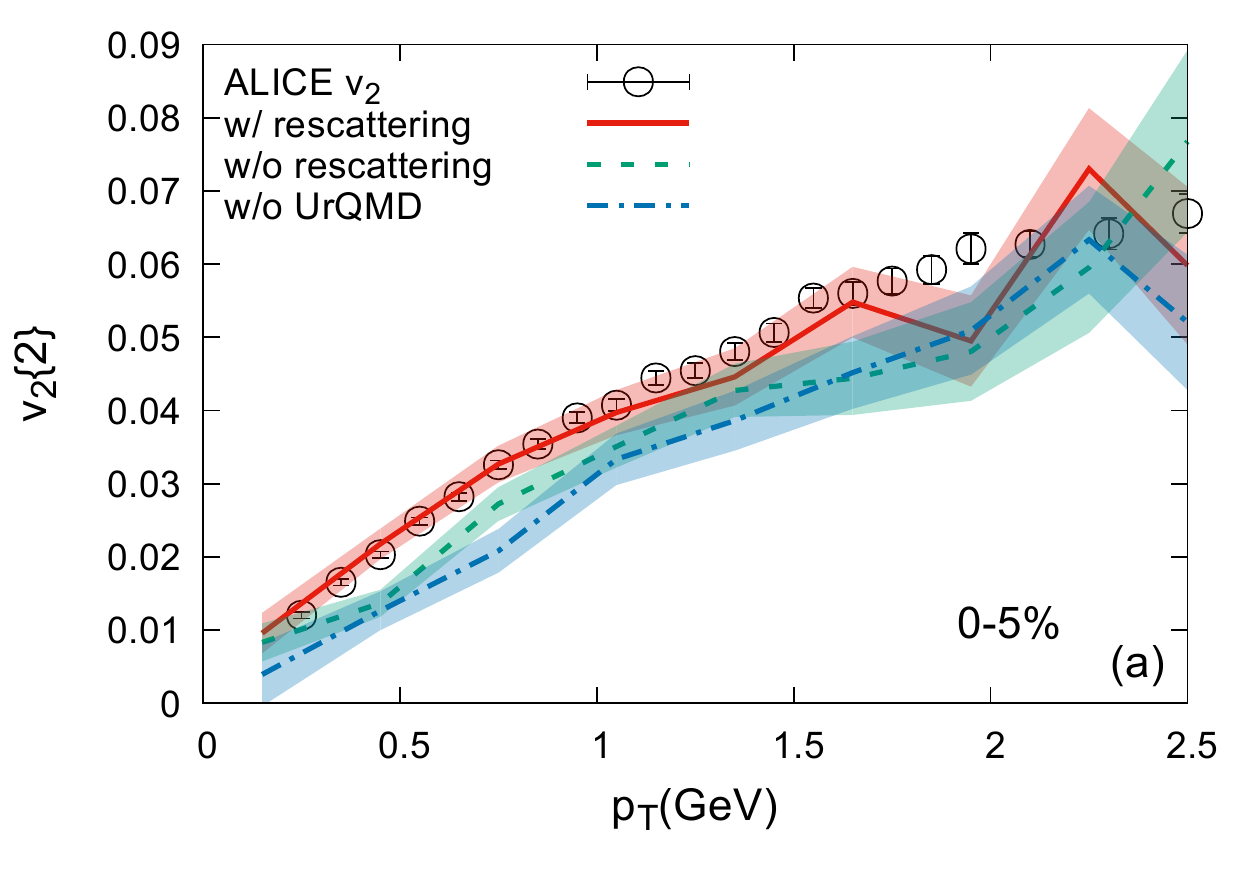}
 \end{minipage}
 \begin{minipage}{0.49\hsize}
 \centering \hspace{-1cm}
 \includegraphics[width=7.3cm]{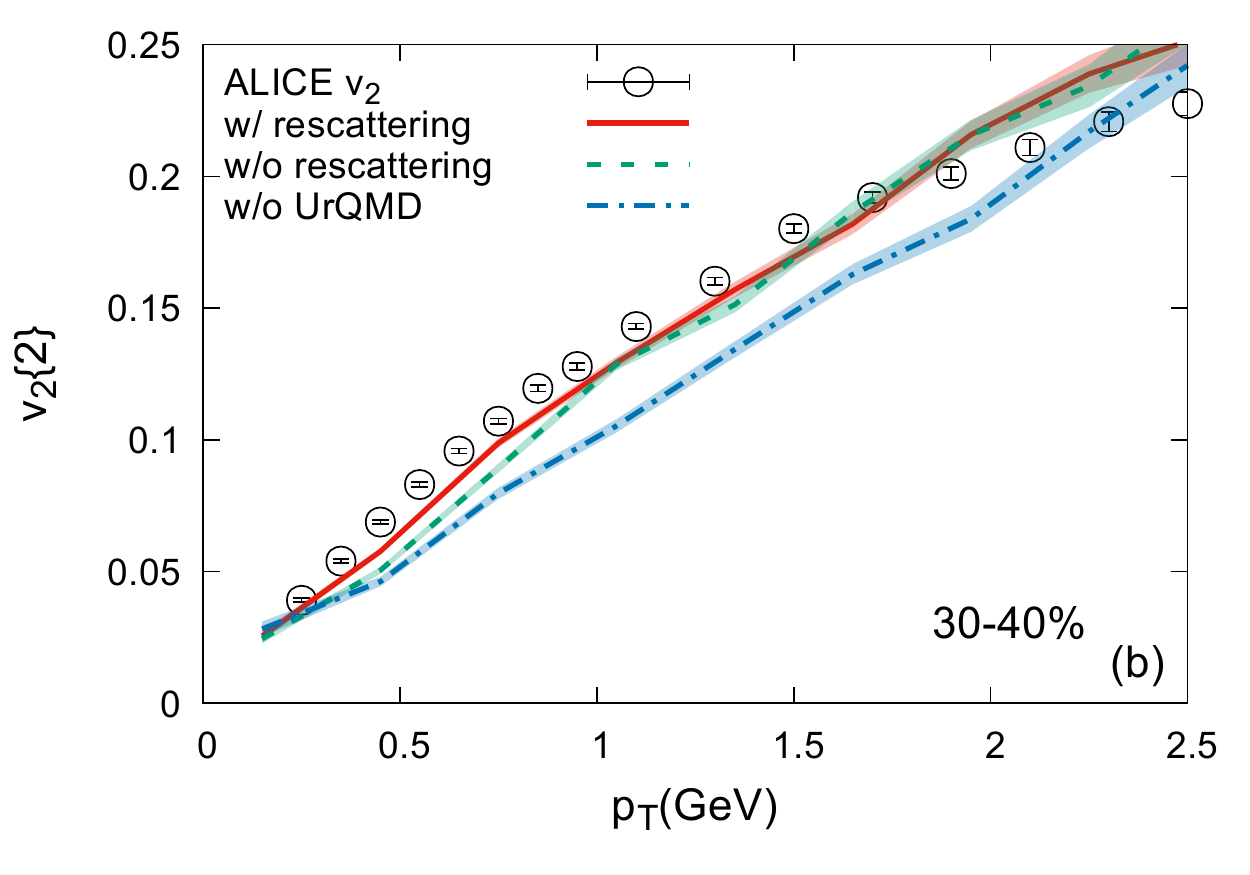}
 \end{minipage}
 \colorcaption{The elliptic flow of charged hadrons with rescattering (the red solid line), without rescattering 
 (the blue dashed line), and 
 without UrQMD (the blue dashed-dotted line) 
 as a function of $p_T$ in 0-5 \% (a) and 30-40 \% (b) centralities, together with the ALICE data \cite{ALICE:2011ab}. 
 \label{fig:v2-pt-fsi}} 
\end{figure*}
\begin{figure*}[t]
 \begin{minipage}{0.5\hsize}
 \centering \hspace{1cm}
 \includegraphics[width=7.3cm]{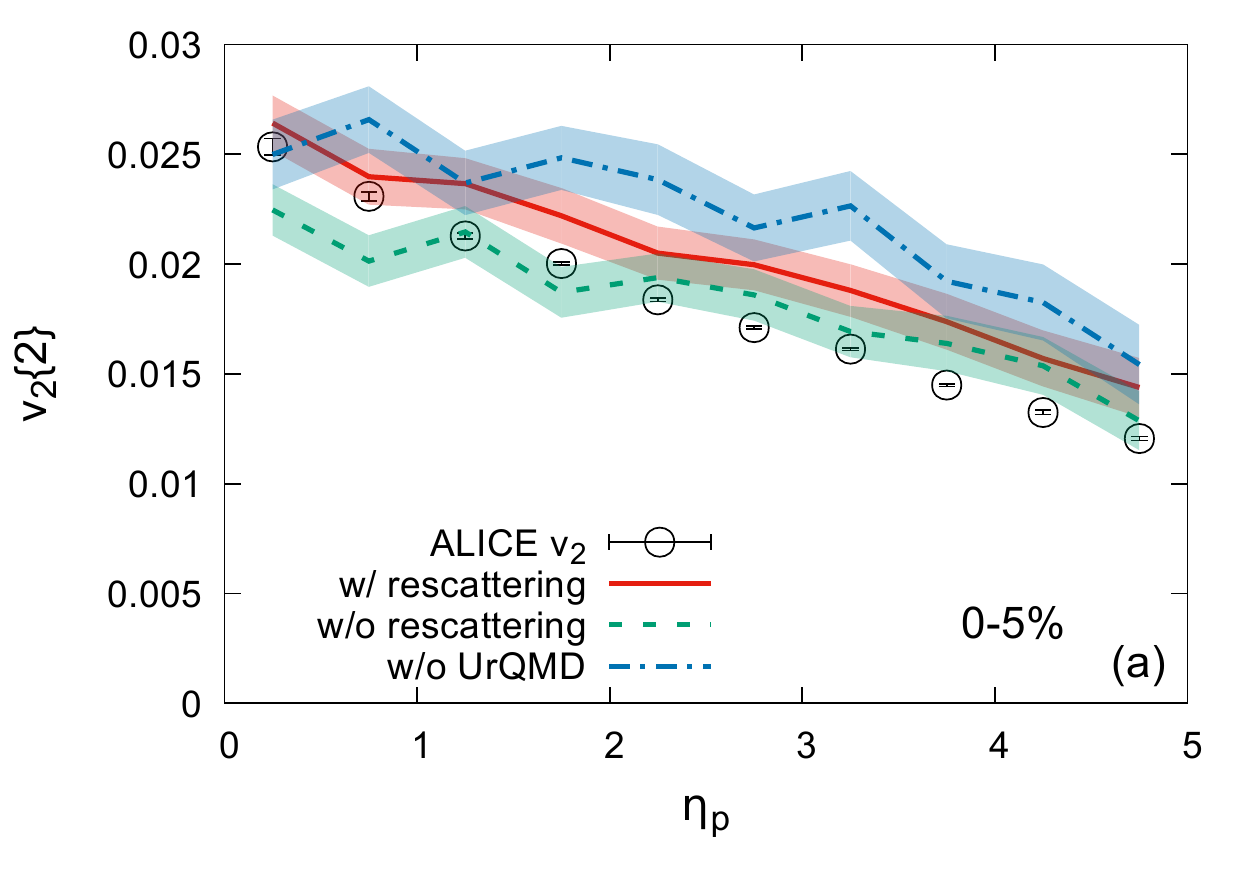}
 \end{minipage}
 \begin{minipage}{0.49\hsize}
 \centering \hspace{-1cm}
 \includegraphics[width=7.3cm]{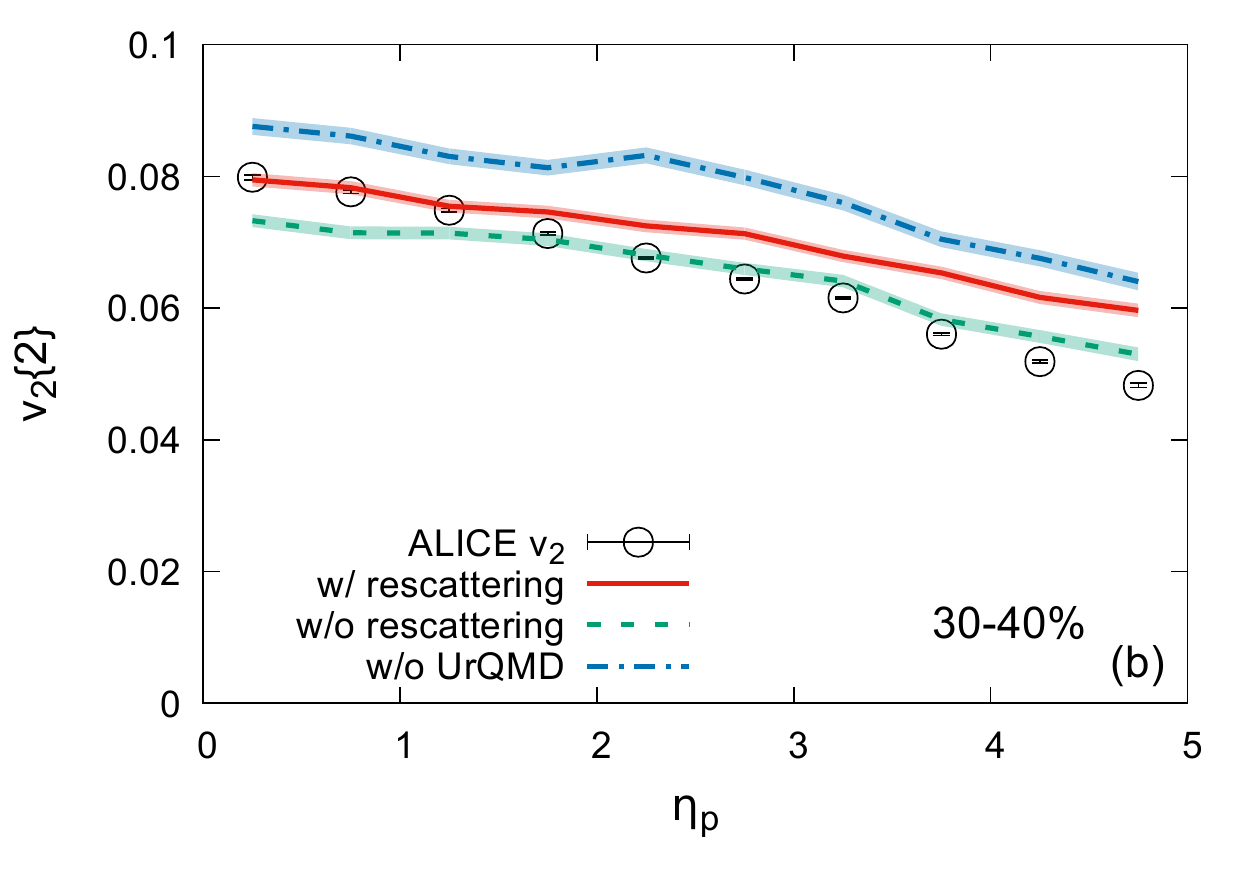}
 \end{minipage}
 \colorcaption{The integrated $v_2$ as a function of pseudorapidity 
 in centralities 0-5 $\%$ (a) and 30-40 $\%$ (b), together with the
 ALICE data \cite{Adam:2016ows}. 
 The red solid line stands for $v_2$ with rescattering, the green dashed line stands for $v_2$ 
 without rescattering, and the blue dashed-dotted line stands for $v_2$ without UrQMD. 
 \label{fig:v2-eta-fsi}} 
\end{figure*}

\subsection{Final state interactions \label{subsec:FSI}}

\begin{table*}[t!]
  \centering
      \caption{The values of mean $p_T$ of $\pi^+$, $K^+$, and $p$ with rescattering, without rescattering, and without 
      UrQMD in 0-5 \% and 30-40 \% centralities. \label{tab:meanpt}}
  \begin{tabular}{lllllll}
    \hline \hline
     &   & 0-5 $\%$  & &  & 30-40 $\%$ & \\
      & $\pi^+$ & $K^+$ & $p$ & $\pi^+$& $K^+$ & $p$ \\ \hline
  w/ rescattering & $0.560\pm0.002$ & $0.929\pm 0.005$ & $1.462\pm 0.013$  
           & $0.502\pm0.002$ & $0.800\pm 0.007$ & $1.225\pm 0.016$   \\ 
  w/o rescattering & $0.546\pm0.002$  & $0.837\pm 0.005$  & $1.168\pm 0.010$  &  
          $0.493\pm0.002$ & $0.749\pm0.006$ & $1.031\pm 0.013$ \\ 
  w/o UrQMD  & $0.617\pm0.001$ & $0.888\pm 0.002$ & $1.208\pm 0.005$ & $0.553\pm0.001$ & $0.794\pm 0.003$ & $1.070\pm 0.008$ \\ 
 experimental data & $0.517\pm 0.019$  & $0.876\pm 0.026$  &  $1.333\pm 0.033$  & $0.504\pm0.017$  &  $0.842\pm 0.032$  &  $1.237\pm 0.032$ \\
\hline
  \end{tabular}
\end{table*}

Finally we investigate the effect of the final state interactions on $p_T$ spectra and 
collective flows $v_2$ and $v_3$ in detail. 
Here we focus on case (f) ($c_1=10$, $c_2=0.7$ and $\eta/s$=0.17). 
Figure \ref{fig:pT_fsi}  shows the $p_T$ spectra of $\pi^+$, $K^+$, and $p$ in 0-5 \% 
(left panel) and 30-40 \% (right panel) centralities, together with the 
ALICE data \cite{Abelev:2013vea}. 
For the $p_T$ spectra just from hydrodynamics at the switching temperature ($T_{SW}=150$ MeV), 
the slopes are almost the same as those of the experimental data; however, 
their yields are much less than the experimental data. 
Once we include the resonance decay, the $p_T$ spectra move to 
close to the experimental data. In particular, for $\pi^+$ we observe a significant 
decay effect at low $p_T$. 
Furthermore, during final state interactions, 
particles earn the transverse momentum so that the slope of $p_T$ 
spectra including rescattering 
becomes flat compared with that without rescattering. 
We can clearly see the same tendency in $p_T$ spectra of protons. 

In Fig.~\ref{fig:mean-PT}, we show the values of mean $\langle p_T \rangle$ of $\pi^+$, $K^+$, and $p$ 
with rescattering, without rescattering, and without UrQMD 
in 0-5 \% (left panel) and 30-40 \% (right panel), together with the ALICE data \cite{Abelev:2013vea}. 
We also list the values of $\langle p_T \rangle$ in Table~\ref{tab:meanpt}.
In the case of $\pi^+$, the resonance decay at low $p_T$ is so large that the slope of $p_T$ is steeper. 
As a result $\langle p_T \rangle$ with rescattering and without rescattering becomes small, compared to 
that without UrQMD. 
On the other hand, for $\langle p_T \rangle$ of $p$ due to the rescattering in the final state interactions 
the slope of $p_T$ spectra becomes flat, which leads to growth of mean $\langle p_T \rangle$.

In Fig.~\ref{fig:v2-pt-fsi}, we investigate the effect of resonance decay and final state interactions 
from the elliptic flow of charged hadrons in 30-40 \% centrality. 
We find that most of the elliptic flow develops during the hydrodynamic evolution. 
It indicates that the elliptic flow reflects the features of QGP fluid such as  the EoS and transport 
coefficients without being smeared by the final state interactions. 
If the decay effect is included, the elliptic flow of charged hadrons becomes large,
close to the experimental data at low $p_T$. 
Particles through decay process tend to be  produced in the same direction as that of their parent   
particles \cite{Landau-classical}, which enhances the amplitude of the 
elliptic flow \cite{Greco:2004ex}.
On the other hand, we only find small effect from the rescattering on $v_2$, 
comparing $v_2$ with rescattering (the red solid line) and that without rescattering 
(the green dashed line). 
The change is small; however, elliptic flow with rescattering shows the best 
agreement with the experimental data. 

Finally, in Fig.~\ref{fig:v2-eta-fsi} we compare the elliptic flows of charged hadrons 
as a function of pseudorapidity with rescattering, without 
rescattering, and without UrQMD in 0-5 \% and 30-40 \% centralities. 
Because $\pi$ is dominant over the produced charged hadrons, the difference among 
them is explained by focusing $\pi$ in the final state interactions. 
Interestingly, the elliptic flow without UrQMD shows the largest values, though 
in Fig.~\ref{fig:v2-pt-fsi} $v_2(p_T)$ without UrQMD is the smallest in the whole 
$p_T$ region. From Fig.~\ref{fig:mean-PT} the value of mean $p_T$ without UrQMD of $\pi$ 
is the largest, which leads to the largest $v_2(\eta_p)$ through the $p_T$ integral. 
In other words,  mean $p_T$ becomes smaller by resonance decay 
so that $v_2(\eta_p)$ without rescattering becomes smaller than that 
without UrQMD. 
In addition, the resonance decay only changes the magnitude of $v_2(\eta_p)$, but it 
does not change the $\eta_p$ dependence of $v_2$. 
If we include the rescattering in the $v_2$, the value of it becomes slightly
larger than that without rescattering. 
This is also understood by the behavior of mean $p_T$ of $\pi$ in Fig.~\ref{fig:mean-PT}.

\section{Summary \label{sec:summary}}
We have investigated the temperature dependence of shear and bulk viscosities 
from comparison with ALICE data: single particle spectra and collective flows 
at Pb+Pb $\sqrt{s_{\rm NN}}=2.76$ collisions. 

First we studied  the constant shear viscosity by comparison between our calculated results 
and the ALICE data. 
The $p_T$ spectra of $\pi^+$, $K^+$, and $p$ are insensitive to the value of $\eta/s$, and in 0-5 \% and
10-20 \% centralities our computed $p_T$ spectra overestimate above $p_T > 1.5$ GeV. 
We find the clear $\eta/s$ dependence in $v_2$ and $v_3$;
{\it i.e.,} the larger $\eta/s$ is, the smaller $v_2$ and $v_3$ of charged hadrons are. 

For the finite bulk viscosity, we obtain the following results. 
The slope of $p_T$ spectra of $\pi^+$, $K^+$, and $p$ becomes steep in the finite bulk viscosity, 
which suggests a small mean $p_T$. The elliptic flow $v_2$ becomes small at low $p_T$, whereas 
above $p_T > 2$ GeV it becomes large. The triangular flow $v_3$ is enhanced for the larger bulk viscosity in 
30-40 \% centrality. 
Furthermore we find the bulk viscosity effect as the enhancement of the profile function around 
the critical temperature which may affect physical observables. 

Furthermore we have discussed consequences of the temperature dependence of $\eta/s$ and $\zeta/s$. 
In the parametrization of $\eta/s(T)$ and $\zeta/s(T)$, $p_T$ spectra are not 
sensitive to the parametrization $\eta/s$, but their slopes depend on the average value of $\zeta/s$ during 
the hydrodynamic expansion. They are steeper with larger $\zeta/s$. 
From the amplitude of mean $p_T$, we can extract the bulk-viscosity effect. 
The collective flows $v_2$ and $v_3$ are affected by both shear and bulk viscosities. 
The shear viscosity reduces the amplitude of $v_2$ and $v_3$. 
On the other hand, the bulk viscosity reduces $v_2$ and $v_3$ at low $p_T$ and enhances them 
at high $p_T$, which is related to the change of $p_T$  slope. 
Our parametrization of (f), (g), and (h) shows the best agreement with the experimental data compared 
with other cases [(b) and (d)], which implies that the parametrization $\eta/s$ and $\zeta/s$ in (f), (g), and (h) is 
one of the most suitable choices. 
Furthermore we point out that the agreement with the experimental data found in $\langle v_n \rangle$ with 
constant $\eta/s$ is realized just from the combination of smaller $v_2(p_T)$ and larger $\langle p_T \rangle$. 
We find the effect of the temperature dependence of $\eta/s$ in the centrality dependence of $v_n$. 
In the central (peripheral) collisions, viscosities of the hadronic (QGP) phase becomes important. 

Finally we have investigated the effect of the final state interactions. 
The resonance decays increase yields and the slope of $p_T$ 
spectra becomes flat during final state interactions. 
Besides, the mean $p_T$ of $\pi^+$ becomes small, whereas that of $p$ becomes large. 
For elliptic flow as a function of $p_T$, most of the elliptic flow 
develops during the hydrodynamic evolution, 
though the elliptic flow still continues to grow a little  through resonance decays. 
It indicates that the elliptic flow reflects the feature of QGP fluid such as the 
EoS and transport coefficients. 
In $v_2(\eta_p)$, through integration of $p_T$, $v_2(\eta_p)$ without UrQMD 
shows the largest value and the resonance decay only changes the magnitude of $v_2(\eta_p)$. 
From the comprehensive analyses of centrality dependence of 
single particle spectra and collective flow, we can extract the detailed information of 
the QGP bulk property, without the information being smeared by the final state interactions. 

\begin{acknowledgments}
We would like to thank Marcus Bluhm for providing the code of the lattice QCD-based equations of state 
as a function of pressure at vanishing net-baryon density. 
We acknowledge valuable discussions on construction of the hybrid model with 
Jonah E. Bernhard,  Weiyao Ke, and  J. Scott Moreland. 
We are grateful for the hospitality of the members of Duke University. 
Finally we would like to thank Steffen Bass and Berndt Mueller for encouraging us to pursue 
this work. 
The work of C.N. is supported by the JSPS KAKENHI Grant-in-Aid for Scientific Research (S) Grant No. JP26220707 and 
the JSPS KAKENHI Grant-in-Aid for Scientific Research (C) Grant No. JP17K05438. 
\end{acknowledgments}


\input{rsub-v3.bbl}

\end{document}

%% file: rsub-v3.bbl
%